\RequirePackage{ifpdf}
\pdfoutput=1
\documentclass[11pt,a4paper]{article}
\usepackage{jheppub}
\usepackage{graphicx}
\usepackage{amssymb}
\usepackage{amsmath}
\usepackage{mathtools}
\usepackage{mathbbol}
\usepackage{amsthm}
\usepackage{slashed}
\usepackage{color,rotating}
\usepackage{bbm}
\usepackage{array}
\usepackage{colonequals}
\usepackage[utf8]{inputenc}
\usepackage{subcaption}
\usepackage{xcolor}
\usepackage{stackrel}
\DeclareMathOperator{\tr}{tr}
\allowdisplaybreaks

\usepackage[normalem]{ulem} 
\newcommand{\ds}{{\slashed{\nabla}}}
\newcommand{\Tr}{{\rm Tr}}

\newcommand{\be}{\begin{equation}}
\newcommand{\ee}{\end{equation}}
\newcommand{\bi}{\begin{itemize}}
\newcommand{\ei}{\end{itemize}}

\newcommand{\p}{\partial}

\newcommand{\gb}{\bar{g}}
\newcommand{\bpsi}{\bar{\psi}} 
\newcommand{\btheta}{\bar{\theta}}
\newcommand{\bchi}{\bar{\chi}}
\newcommand{\hh}{\widehat{h}}

\newcommand{\Db}{\bar{D}}
\newcommand{\Cb}{\bar{C}}
\newcommand{\Rb}{\bar{R}}

\newcommand{\talpha}{\tilde{\alpha}}
\newcommand{\snabla}{\slashed{\nabla}}
\newcommand{\bnabla}{\bar{\nabla}}

\newcommand{\unit}{{\bf 1}}

\newcommand{\cM}{\ensuremath{\mathcal{M}}}
\newcommand{\cO}{\ensuremath{\mathcal{O}}}
\newcommand{\cR}{\ensuremath{\mathcal{R}}}

\newcommand{\bB}{{\bf B}}

\title{Asymptotically Safe Gravity-Fermion systems \\ on curved backgrounds}
\author[a]{Jesse Daas,}
\author[a]{Wouter Oosters,}
\author[a]{Frank Saueressig\,\href{https://orcid.org/0000-0002-2492-8271}{\protect \includegraphics[scale=.07]{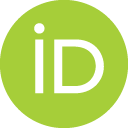}}}
\author[a]{and Jian Wang}

\affiliation[a]{Institute for Mathematics, Astrophysics and Particle Physics (IMAPP), \\
Radboud University Nijmegen, Heyendaalseweg 135, 6525 AJ Nijmegen, The Netherlands}

\emailAdd{j.daas@science.ru.nl}
\emailAdd{woosters@science.ru.nl}
\emailAdd{f.saueressig@science.ru.nl}
\emailAdd{jian.wang@science.ru.nl}

\abstract{We set up a consistent background field formalism for studying the renormalization group (RG) flow of gravity coupled to $N_f$ Dirac fermions on maximally symmetric backgrounds. Based on Wetterich’s equation we perform a detailed study of the resulting fixed point structure in a projection including the Einstein-Hilbert action, the fermion anomalous dimension, and a specific coupling of the fermion bilinears to the spacetime curvature. The latter constitutes a mass-type term which breaks chiral symmetry explicitly. Our analysis identifies two infinite families of interacting RG fixed points which are viable candidates to provide a high-energy completion through the asymptotic safety mechanism. The fixed points exist for all values of $N_f$ outside of a small window situated at low values $N_f$ and become weakly coupled in the large $N_f$-limit. Symmetry-wise, they correspond to ``quasi-chiral'' and ``non-chiral''  fixed points. The former come with enhanced predictive power, fixing one of the couplings via the asymptotic safety condition. Moreover, the interplay of the fixed points allows for cross-overs from the non-chiral to the chiral fixed point, giving a dynamical mechanism for restoring the symmetry approximately at intermediate scales. Our discussion of chiral symmetry breaking effects provides strong indications that the topology of spacetime plays a crucial role when analyzing whether quantum gravity admits light chiral fermions.}


\begin{document}
%======================================================
\maketitle
%======================================================
\section{Introduction}
\label{sect:intro}
%======================================================
Asymptotic Safety \cite{Percacci:2017fkn,Reuter:2019byg} constitutes a powerful mechanism for providing a consistent and predictive high-energy completion of a quantum field theory. The construction hinges on the presence of a suitable (interacting)  fixed point of the theories renormalization group (RG) flow. Building on the initial proposal by Weinberg \cite{Weinberg:1976xy,Weinberg:1980gg}, functional renormalization group methods pioneered in \cite{Wetterich:1992yh,Morris:1993qb,Reuter:1993kw,Reuter:1996cp}, have provided substantial evidence supporting the existence of such a fixed point, the so-called Reuter fixed point, in the context of gravity \cite{Souma:1999at,Reuter:2001ag, Lauscher:2001ya,Litim:2003vp,Machado:2007ea, Benedetti:2009rx,Machado:2009ph,Benedetti:2009gn,Manrique:2009uh,Manrique:2010am,Groh:2010ta,Eichhorn:2010tb,Manrique:2010mq, Manrique:2011jc,Benedetti:2012dx,Rechenberger:2012dt,Christiansen:2012rx,Dietz:2012ic,Ohta:2013uca,Falls:2013bv,Falls:2014tra,Christiansen:2014raa,Becker:2014qya,Demmel:2014hla,Christiansen:2015rva,Morris:2015oca,Ohta:2015efa,Ohta:2015fcu,Gies:2015tca, Demmel:2015oqa, Biemans:2016rvp,Gies:2016con,Denz:2016qks,Platania:2017djo,Houthoff:2017oam,Falls:2017lst, Knorr:2017fus,Eichhorn:2017egq,Christiansen:2017bsy,deBrito:2018jxt,Falls:2018ylp,Knorr:2018kog,Kluth:2020bdv,Falls:2020qhj,Knorr:2020ckv} 
and also a wide range of gravity-matter systems \cite{Narain:2009fy,Shaposhnikov:2009pv,Dona:2013qba,Dona:2014pla,Meibohm:2015twa,Dona:2015tnf,Oda:2015sma,Eichhorn:2016vvy,Wetterich:2016uxm,Biemans:2017zca,Becker:2017tcx,Christiansen:2017cxa,Hamada:2017rvn,Eichhorn:2017als,Eichhorn:2017ylw,Eichhorn:2017sok,Alkofer:2018fxj,Eichhorn:2018akn,Eichhorn:2018ydy,Eichhorn:2018nda,Pawlowski:2018ixd,Alkofer:2018baq,Knorr:2019atm,Burger:2019upn,Eichhorn:2019yzm,Reichert:2019car,Daas:2020dyo,Eichhorn:2020kca,Eichhorn:2020sbo}. 
See also \cite{Niedermaier:2006wt,Codello:2008vh,Reuter:2012id,Eichhorn:2018yfc,Pawlowski:2020qer,Bonanno:2020bil} for reviews and \cite{Nagy:2012ef,Pereira:2019dbn,Reichert:2020mja} for recent lecture notes. In particular, there has been a significant effort in developing the form-factor program, analyzing approximations which retain an arbitrary momentum dependence at the level of the effective action \cite{Knorr:2019atm,Draper:2020bop}, reconstructing the graviton propagator \cite{Bosma:2019aiu,Bonanno:2021squ,Knorr:2021niv}, and refining the computational toolbox in arbitrary backgrounds \cite{Benedetti:2010nr,Codello:2012kq,Becker:2020mjl,Knorr:2021slg}.

An important step towards a realistic theory of quantum gravity is the inclusion of matter degrees of freedom. Within the asymptotic safety program this step is conceptually straightforward. The corresponding matter fields are simply added to the theory. Subsequently, one analyzes the resulting RG flow, searching for fixed points which could provide potential high-energy completions. This opens the perspective on a \emph{new standard model of particle physics} in which the matter degrees of freedom mandated by the standard model of particle physics are supplemented by the graviton as an additional force carrier. Tentative studies suggest that the  RG flow of this setting indeed supports interacting fixed points which could render the construction asymptotically safe \cite{Dona:2013qba,Biemans:2017zca,Alkofer:2018fxj}. Remarkably, these fixed points come with the potential of fixing some of the free parameters of the standard model as, e.g., the value of the electromagnetic coupling \cite{Harst:2011zx,Eichhorn:2017lry}, the Higgs mass \cite{Shaposhnikov:2009pv,Pawlowski:2018ixd}, or the ratio of quark masses \cite{Eichhorn:2018whv}.

An important element in the construction of the \emph{new standard model of particle physics} is the inclusion of fermions. Within the asymptotic safety program, this question has already been looked into in a series of works \cite{Eichhorn:2011pc,Dona:2012am,Dona:2013qba,Meibohm:2016mkp,Eichhorn:2017eht,Eichhorn:2016esv,Eichhorn:2016vvy,Biemans:2017zca,Alkofer:2018fxj,Alkofer:2018baq,Gies:2018jnv,deBrito:2019epw,Daas:2020dyo,Gies:2021upb}. In particular, non-minimal couplings between fermions and gravity have been considered in \cite{Eichhorn:2016vvy,Eichhorn:2018nda}. This entails the effect of ``gravitational catalysis'' where the coupling to the spacetime curvature essentially provides a mass to the fermions \cite{Gies:2018jnv,Gies:2021upb}. Up to now, fermions in Asymptotic Safety have predominantly been investigated using a hybrid approach where the beta functions in the matter sector are computed in a flat background. While this leads to significant technical simplifications and is conceptually close to computations carried out in the context of particle physics, the specific background also comes with features and choices that are non-generic when considering more general curved backgrounds, see, e.g., \cite{Dona:2012am} for an exemplary exposition.

This provides a clear motivation for investigating the RG flow of gravity coupled to fermionic matter beyond these flat background studies. Building on our earlier work  \cite{Daas:2020dyo} we investigate this question in a truncation comprising the Einstein-Hilbert action, the fermion anomalous dimension and a coupling between the fermion bilinears and the spacetime curvature. The latter provides a prototypical example for an interaction breaking chiral symmetry explicitly. We provide a detailed derivation of the beta functions on a background-sphere and carry out a thorough analysis of the resulting fixed point structure. An important feature of our setting is that it can be restricted to several well-motivated subtruncations, comprising
\begin{itemize}
	\item[i)] fermions minimally coupled to the Einstein-Hilbert action,
	\item[ii)] fermions coupled to the Einstein-Hilbert action including the fermion anomalous dimension,
	\item[iii)] the setting i) complemented by the coupling of the fermion-bilinear to the spacetime curvature,
	\item[iv)] the full system including all couplings and anomalous dimensions.
\end{itemize}
This feature allows to explicitly test the robustness of a fixed point structure under a refinement of the approximation. In this way the influence of various factors (as the matter anomalous dimension) can be investigated systematically. Moreover, mechanisms leading to new classes of fixed points, which are not visible in simpler approximations, are readily understood based on explicit examples. 

As a key result, we identify continuous families of interacting renormalization group fixed points, so-called non-Gaussian fixed points (NGFPs), which are robust under the refinement strategy. Within our approximation, the fixed points come with varying degrees of predictive power. The existence of two of these families are tied to the inclusion of the non-minimal interaction term coupling the fermion-bilinear to the spacetime curvature. Moreover, our analysis shows that the role of chiral symmetry in a curved background is actually more subtle than suggested by flat-background computations since the change of the background topology unlocks new symmetry-breaking ingredients, also see \cite{Hamada:2020mug} for a related discussion.

Our work is organized as follows. Sect.\ \ref{sect:fermions} introduces the Wetterich equation together with the truncation ansatz studied in this article. In particular, the fate of chiral symmetry is discussed in detail. The beta functions encoding the scale-dependence of the couplings and anomalous dimensions contained in our ansatz are reported in Sect.\ \ref{sect:beta}. We perform a detailed analysis of the fixed point structure encoded in this system in Sect.\ \ref{sect:fp}. The complexity of the approximation is increased from subsection to subsection, giving control on the robustness of the fixed point structure. The RG flow controlled by the interplay of these fixed points is investigated in Sect.\ \ref{sect:chiralsymmetry}. Sect.\ \ref{sect:summary} contains our conclusion. The technical details of our investigation, including a brief introduction to the spin-base formalism employed in our computation, have been relegated to two appendices.

%======================================================
\section{Gravitational RG flows including fermions in a curved background}
\label{sect:fermions}
%======================================================
We start by outlining the general framework underlying our computation. Technical details about the spin-base formalism have been relegated to App.\ \ref{App.A}.
%======================================================
\subsection{The Wetterich equation and its projection}
\label{ssect:2.1}
%======================================================
Starting from the pioneering work \cite{Reuter:1996cp}, the functional renormalization group equation (FRGE) for the effective average action $\Gamma_k$ \cite{Wetterich:1992yh,Morris:1993qb,Reuter:1993kw} has played a key role in exploring the existence and predictive power of the Reuter fixed points in the context of gravity and gravity-matter systems. The Wetterich equation \cite{Wetterich:1992yh,Morris:1993qb,Reuter:1993kw}
\be\label{FRGE}
k \p_k \Gamma_k = \frac{1}{2} {\rm STr} \left[ \left( \Gamma_k^{(2)} + \cR_k \right)^{-1} k \p_k \cR_k \right]
\ee
encodes the change of $\Gamma_k$ when integrating out quantum fluctuations with momenta $p^2 \approx k^2$ where $k$ corresponds to a coarse-graining scale. Here STr contains a sum over all fluctuation fields, an integral over loop-momenta, and a minus sign for fermionic degrees of freedom. The Hessian $\Gamma_k^{(2)}$ is the second functional derivative of $\Gamma_k$ with respect to the fluctuations. The regulator $\cR_k$ suppresses fluctuations with $p^2 \lesssim k^2$ by a $k$-dependent mass-term. By construction $\cR_k$ falls off sufficiently fast for $p^2 \gg k^2$, ensuring that the trace does not give rise to UV-divergences. The interplay of the regulator appearing in the numerator and denominator ensures that the flow of $\Gamma_k$ is driven by integrating out momenta close to the scale $k$ \cite{Berges:2002ew,Dupuis:2020fhh}.

In this work, we are interested in studying the RG flow of gravity  supplemented by $N_f$ Dirac fermions $\psi$ in a four-dimensional spacetime. The gravitational degrees of freedom are encoded in the spacetime metric $g_{\mu\nu}$ which is decomposed into a fixed background metric $\gb_{\mu\nu}$ and fluctuations $h_{\mu\nu}$ by performing a linear split
\be
g_{\mu\nu} = \gb_{\mu\nu} + h_{\mu\nu} \, . 
\ee 
In the sequel quantities constructed from the background metric will be marked by a bar.
The metric fluctuations are further decompose into their trace $h$ and traceless part $\hh_{\mu\nu}$,
\be
 h_{\mu\nu} = \hh_{\mu\nu} + \frac{1}{4} \gb_{\mu\nu} h \, , \qquad \gb^{\mu\nu} \hh_{\mu\nu} = 0 \, . 
\ee
Similarly, we decompose the fermions according to
\be\label{Fermion linear split}
\psi = \theta + \chi \, , \qquad \bpsi = \btheta + \bchi \, ,
\ee
where $\theta$ and $\chi$ are the background and fluctuations, respectively. We define $\bpsi \equiv \psi^\dagger h$ with the internal spinor metric $h$ satisfying $h^\dagger = - h$, c.f.\ App.\ \ref{App.A} for more details on our spinor conventions.

We then approximate the effective average action by
\be\label{Gans}
\Gamma_k[g,\psi,\bpsi,\gb] =  \Gamma_k^{\rm grav}[g,\gb]  + \Gamma_k^{\rm fermion}[g,\psi,\bpsi,\gb] + \Gamma^{\rm gf}_k[g,\gb] + S^{\rm ghost}[g,C,\bar{C}, \gb] \, . 
\ee
The gravitational part is approximated by the Einstein-Hilbert action,
\be
\Gamma_k^{\rm grav}[g,\gb] = \frac{1}{16 \pi G_k} \int d^4x \sqrt{g} \left[ - R + 2 \Lambda_k \right] \, ,
\ee
with $G_k$ and $\Lambda_k$ being the running Newton's coupling and cosmological constant, respectively.\footnote{We stress that physics should be extracted from the endpoint of an RG-trajectory at $k = 0$ \cite{Knorr:2019atm,Bonanno:2020bil}. This entails that the Newton's constant and cosmological constant appearing in physical processes is scale-independent.} The gravitational sector is supplemented by the harmonic gauge condition
\be
\Gamma_k^{\rm gf}[g,\gb] = \frac{1}{32 \pi G_k} \int d^4x \sqrt{\gb} \, \gb^{\mu\nu} F_\mu F_\nu \, , \qquad F_\mu = \Db^\nu h_{\mu\nu} - \frac{1}{2} \Db_\mu \, h \, , 
\ee
and the resulting (classical) ghost action
\be
S^{\rm ghost}[g,C,\bar{C}, \gb] = - \sqrt{2} \int d^4x \sqrt{\gb} \, \Cb_\mu \, \cM^\mu{}_\nu \, C^\nu \, , 
\ee
For $g = \gb$, one has $\cM^\mu{}_\nu = \delta_\nu^\mu \Db^2 + \Rb^\mu{}_\nu$.
The fermionic part of the action is taken to be
\be\label{Gfermion}
\Gamma_k^{\rm fermion}[g,\psi,\bpsi;\gb] = \int d^4x \sqrt{g} \, Z^\psi_k \, \left\{\, \bpsi \left[ i \slashed{\nabla} + m \gamma_5 \right]  \psi + \talpha_k R \, \bpsi \gamma_5 \psi \right\} \, . 
\ee
Here $Z^\psi_k$ denotes the wave function renormalization associated with the fermionic fields, $\slashed{\nabla} = \gamma^\mu \nabla_\mu$ is the covariant derivative containing the spin connection and $m$ denotes the fermion mass. In addition, our setting includes the scale-dependent coupling $\talpha_k$ multiplying a mass-type term where the mass of the fermions is provided by the Ricci curvature $R$. The conventions on the $\gamma$-matrices are chosen such that the kinetic term squares to the Klein-Gordon equation while, at the same time, satisfying the identity \eqref{commutator spin connection}.

We compute the flow of $G_k$, $\Lambda_k$, $\talpha_k$ and the scale-dependence of $Z^\psi_k$ in the background field approximation, evaluating the FRGE \eqref{FRGE} at zeroth order in the fluctuation fields. The information about the $k$-dependence of the couplings is then encoded in the coefficients multiplying the interaction monomials
\be\label{Imono}
I_1 = \int d^4x \sqrt{\gb} \, , \quad 
I_2 = \int d^4x \sqrt{\gb} \Rb \, , \quad 
I_3 = \int d^4x \sqrt{\gb} \btheta \, i\slashed{\nabla} \, \theta \, , \quad 
I_4 = \int d^4x \sqrt{\gb} \Rb \, \btheta \gamma_5 \theta \, . 
\ee 
The computation of these coefficients can significantly be simplified by a clever choice of background fields which allow to disentangle these contributions. In practice, we work with $\gb_{\mu\nu}$ being a one-parameter family of metrics on the Euclidean four-sphere, parameterized by the radius of the spheres. The background curvature tensors then satisfy
\be\label{S4curvature}
\Rb_{\mu\nu} = \frac{1}{4} \gb_{\mu\nu} \, \Rb \, , \qquad \Rb_{\mu\nu\rho\sigma} = \frac{\Rb}{12}\left(\gb_{\mu\rho} \gb_{\nu\sigma} - \gb_{\mu\sigma} \gb_{\nu\rho} \right) \, , \qquad \Db_\mu \Rb = 0 \, . 
\ee
In addition, we impose that the background spinor satisfies the generalized eigenvalue equation
\be\label{thetabackground}
\nabla_\mu \, \theta = i \sqrt{\frac{\Rb}{48}} \, \gamma_\mu \, \theta \, . 
\ee
Noting that the spectrum of the Dirac operator on a four-sphere is given by \cite{Camporesi:1995fb}
\be
{\rm spec}\{\slashed{\nabla}\} = \{\pm i \sqrt{\frac{\Rb}{12}} \left(l+2\right) \} \, , \quad l = 0,1,2,\cdots \, ,
\ee
it is easily verified that the choice \eqref{thetabackground} corresponds to the lowest eigenvalue of $\slashed{\nabla}$. Notably, this is the only eigenspinor of the Dirac operator which satisfies a generalized eigenvalue equation of the form \eqref{thetabackground}, see \eqref{Only possible c value}. At this stage, the following important remark is in order. Structurally, eq.\ \eqref{thetabackground} is quite different from a flat background where eigenspinors of the Dirac operator are obtained in Fourier space, $i \snabla \theta = \slashed{p} \, \theta$. In this case $\snabla$ has a continuous spectrum and it is admissible to work with a background spinor which is covariantly constant, $\nabla_\mu \theta =0$. Since the spectrum of $\snabla$ on the sphere does not admit a zero mode, it is clear that this choice is not admissible on spherically symmetric backgrounds for purely geometrical reasons, though.

The completion of our setup requires specifying the regulator $\cR_k$. In the gravitational sector, we generate the matrix elements via the substitution rule
\be\label{cutrule}
\Box \mapsto \Box + R_k(\Box) \, . 
\ee
In this sector, we identify the coarse-graining operator to be the Laplacian constructed from the background metric, $\Box^{\rm grav} = - \gb^{\mu\nu} \Db_\mu \Db_\nu$. In the classification \cite{Codello:2008vh}, this corresponds to a regulator of type I. Applying \eqref{cutrule} to the graviton propagator \eqref{gravprop} yields the matrix elements of $\cR_k$ associated with the graviton
\be
\begin{split}
\left[ \left. \cR_k \right|_{\hh\hh}\right]_{\mu\nu}{}^{\rho\sigma} = & \, \frac{1}{32 \pi G_k} \, R_k(\Box) \left( \frac{1}{2} \left( \delta_\mu^\rho \delta_\nu^\sigma + \delta_\nu^\rho \delta_\mu^\sigma \right) - \frac{1}{4} \gb_{\mu\nu} \gb^{\rho\sigma} \right) \, , \\  
\left. \cR_k \right|_{hh} = & \, -\frac{1}{128 \pi G_k} \, R_k(\Box) \, .
\end{split} 
\ee
%
%where $\unit_{\mu\nu}{}^{\rho\sigma} \equiv \frac{1}{2} \left( \delta_\mu^\rho \delta_\nu^\sigma + \delta_\nu^\rho \delta_\mu^\sigma \right) - \frac{1}{4} \gb_{\mu\nu} \gb^{\rho\sigma}$ is the unit on the space of traceless, symmetric tensors.
 The scalar regulator function $R_k(\Box)$ is taken to be of Litim-type \cite{Litim:2000ci,Litim:2001up}, $R_k^{\rm Litim}(z) = (k^2 - z) \Theta(k^2-z)$, where $\Theta(x)$ is the Heaviside step function. Following the discussion \cite{Dona:2012am}, the fermionic sector admits two canonical choices for $\Box$:
\be\label{coarsegrain1}
\Box^{\rm I} \equiv - \gb^{\mu\nu} \bnabla_\mu \bnabla_\nu \, , \qquad \Box^{\rm II} \equiv - \bar{\snabla}^2 \, . 
\ee
These two choices are related by the Lichnerowicz formula
\be\label{eq:nabla2}
- \bar{\snabla}^2 = - \gb^{\mu\nu} \bnabla_\mu \bnabla_\nu + \frac{1}{4} \Rb \, , 
\ee
i.e., the two coarse-graining operators differ by a covariantly constant endomorphism only. Hence, they can be combined into a single expression,
\be\label{coarsegrainingop}
\Box^\psi \equiv - \bar{\snabla}^2 + \beta \Rb \, , 
\ee
where $\beta = 0$ and $\beta = -1/4$ corresponds to $\Box^\psi$ being the squared Dirac operator and the Laplacian, respectively. Motivated by the structure of the mass term appearing in \eqref{Gfermion}, we construct the regulator in the fermionic sector by replacing $m$ with a $k$-dependent regulator of dimension one
\be\label{Rkpsi}
\left. \cR_k \right|_{\bpsi\psi} = k \, Z^\psi_k \gamma^5 \left(1 - \sqrt{\Box^{\psi}}/k \right) \Theta\left(1 - \sqrt{\Box^{\psi}}/k \right) \, . 
\ee
This specific form of the regulator dresses the fermion kinetic term according to $\pm Z_k^\psi \sqrt{p^2} \rightarrow \pm Z_k^\psi \sqrt{p^2 + k^2 (r_k^\psi)^2}$ with $r_k^\psi \equiv  \left(1 - \sqrt{\Box^{\psi}}/k \right) \Theta\left(1 - \sqrt{\Box^{\psi}}/k \right)$ being the dimensionless profile function in the fermionic sector. Hence, it suppresses fluctuation modes with $p^2 \ll k^2$ irrespective of the corresponding sign of the eigenvalue. This completes the setup underlying the RG analysis of our work.
%------------------------------------------------------
\subsection{Chiral symmetry on spherically symmetric backgrounds}
\label{ssect:2.2}
%======================================================
Before giving the results of our computation, it is instructive to discuss the role of chiral symmetry breaking in the setting described in the previous subsection. By definition, chiral transformations rotate the left-handed and the right-handed components of a spinor by an independent phase, 
\be\label{chiralsym1}
\psi \rightarrow e^{i \theta \gamma_5} \, \psi \, , \qquad 
\bpsi \rightarrow  \bpsi \, e^{i \theta \gamma_5} \, ,
\ee
with identical transformations for the background and fluctuation fields. The standard model of particle physics indicates that there are sectors where chiral symmetry must be unbroken at the typical energy scales associated with quantum gravity effects, typically assumed to be set by the Planck scale $M_{\rm Pl} \approx 10^{18}$ GeV. Any phenomenologically viable theory of quantum gravity must therefore admit ``light'' chiral fermions \cite{Eichhorn:2011pc}. The symmetry may then provide a valuable testbed for possible high-energy completions of gravity-matter systems in general and via the asymptotic safety mechanism in particular.

Applying the transformation \eqref{chiralsym1} to our ansatz for $\Gamma_k$ shows that the gravitational sector and the fermion kinetic term is invariant under chiral transformations. The mass term as well as the interaction coupling the fermion-bilinear to the spacetime curvature break this symmetry explicitly. On the basis that gravity couples to Dirac fermions, independently of their handedness, one expects that there is a consistent solution of the flow equation where the couplings of the chiral symmetry breaking terms are zero for all values of the coarse-graining scale $k$. 

At this point, we point out though that, besides the chiral symmetry breaking terms in the action, our setup of the flow equation contains two additional sources for chiral symmetry breaking. Firstly, the regulator $\cR_k^\psi$ introduced in eq.\ \eqref{Rkpsi} is modeled according to a fermion mass term. As a result it induces a chiral symmetry breaking component in the flow.\footnote{Similarly to the case of split-symmetry breaking, this effect may be controlled by a modified ward-identity.} The flat background computation recently presented in \cite{deBrito:2020dta} followed an alternative route, taking the regulator to be proportional to the fermion kinetic term, i.e., $\cR_k(p^2)|_{\bpsi\psi} = Z_k^\psi \slashed{p} \left(\sqrt{k^2/p^2}  - 1\right) \Theta(1-p^2/k^2)$.  This suggests a chiral symmetry preserving extension to general backgrounds given by
\be
\cR_k(\Box^\psi)|_{\bpsi\psi} = Z_k^\psi \, \bar{\snabla} \, \left(\sqrt{k^2/\Box^\psi}  - 1\right) \Theta(1-\Box^\psi/k^2) \, .
\ee
 While it would be interesting to compare our results to a flow computed based on this choice, this is beyond the present study and may be subject of subsequent work.
 
 A second source of chiral symmetry breaking emanates from the background spinor formalism. Explicitly, the generalized eigenvalue equation \eqref{thetabackground} does not adhere to chiral symmetry. This may easily be seen by evaluating the fermion kinetic term for this specific choice of background
 \be\label{chiralsym2}
\left. \int d^4x \sqrt{\gb} \,  \bpsi \, i \bar{\snabla} \,   \psi \right|_{\bchi=\chi=0} = - \, \int d^4x \sqrt{\gb} \, \sqrt{\frac{\Rb}{3}} \,  \btheta \theta \, . 
 \ee
 The resulting term has the structure of a mass term, signaling the breaking of chiral symmetry by the choice of background. This property is actually not limited to the specific choice of mode made in \eqref{thetabackground}. Any background spinor which is an eigenvalue of the Dirac operator on the background sphere, satisfying $\snabla \theta = \lambda \theta$, induces chiral symmetry breaking terms based on the argument \eqref{chiralsym2}. Since any spinor can be expanded in this eigenbasis, it is difficult to have a non-trivial fermionic background field which preserves the transformation law \eqref{chiralsym1} on a spherically symmetric background.

%======================================================
\section{Beta functions}
\label{sect:beta}
%======================================================
We now give the result for the beta-functions encoding the scale-dependence of the couplings $G_k$, $\Lambda_k$, and $\tilde{\alpha}_k$. The technical details underlying this computation have been relegated to App.\ \ref{App.C}. Further details on working with spinors on a spherically symmetric background and the graviton-fermion vertices can be found in App.\ \ref{App.A}.

%======================================================
\subsection{Beta functions}
\label{sect:beta2}
%======================================================
The scale-dependence of $G_k$, $\Lambda_k$ and $\tilde{\alpha}_k$ together with the fermion wave-function renormalization $Z_k^\psi$ is obtained by substituting the ansatz \eqref{Gans} into the Wetterich equation \eqref{FRGE}, compute the resulting operator traces, and comparing the coefficients of the interaction terms \eqref{Imono} appearing on the left- and right-hand side.
The result is conveniently expressed in terms of the dimensionless couplings
\be\label{dimless}
g_k \equiv G_k \, k^{-2} \, , \qquad \lambda_k \equiv \Lambda_k k^2 \, , \qquad \alpha_k \equiv \tilde{\alpha}_k \, k \, , 
\ee
supplemented by the anomalous dimension of Newton's coupling, $\eta_N$ and the  fermionic fields
\be\label{anomdef}
\eta_N \equiv G_k^{-1}  \, k\p_k G_k \, , \qquad \eta_{\psi} \equiv - \, k \p_k \ln Z_k^\psi \, . 
\ee
The scale-dependence of the couplings \eqref{dimless} is then encoded in an autonomous system of coupled differential equations
\be
k\p_k g_k = \beta_g(g,\lambda,\alpha) \,, \quad
k\p_k \lambda_k = \beta_\lambda(g,\lambda,\alpha) \, , \quad
k\p_k \alpha_k = \beta_\alpha(g,\lambda,\alpha) \, .
\ee 
Besides their dependence on the couplings, the beta functions also depend on the parameter $\beta$ specifying the coarse-graining operator \eqref{coarsegrainingop}.
The beta functions in the gravitational sector are given by\footnote{We correct a misprint in the earlier work \cite{Daas:2020dyo} where the factor of $g$ multiplying the fermionic contribution to $\beta_\lambda$ is missing.}
\be\label{betagrav}
\begin{split}
\beta_g = & \, (2+\eta_N) \, g \, , \\
\beta_\lambda = & \left(\eta_N - 2 \right) \lambda + \frac{g}{4\pi} \left[  \left( 10 - \frac{5}{3} \eta_N \right) \frac{1}{(1-2\lambda)} - 8 \right] - \frac{N_f \, g}{12} \left(4 + (8-3 \pi) \eta_\psi \right)\, . 
\end{split}
\ee
The explicit expression for the anomalous dimension $\eta_N$ takes the form
\be\label{etaN}
\eta_N = \frac{g \left(B_1^{\rm grav} + N_f (B^{\rm ferm}_1 + B^{\rm ferm}_2 \, \eta_\psi) \right)}{1 - g \, B_2^{\rm grav}} \, , 
\ee
where
\be\label{Bresults}
\begin{split}
B_1^{\rm grav} = & \, - \frac{1}{3\pi} \left[ \frac{9}{(1-2\lambda)^2} - \frac{5}{(1-2\lambda)} + 7 \right] \, , \\
B_2^{\rm grav} = & \, \frac{1}{12\pi} \left[ \frac{6}{(1-2\lambda)^2} - \frac{5}{(1-2\lambda)} \right]\, \, , \\
B^{\rm ferm}_1 = & \, - \frac{1}{6\pi} \left[ (2 -  \pi) (1+6\beta)
  + 12 \alpha 
\right] \, , \\
B^{\rm ferm}_2 = & \,  \frac{1}{12\pi} \left[ 
 (2-\pi)  - 24   (3-\pi)  \, \alpha 
+ 6  (8-3 \pi)  \beta
\right] \, .
\end{split}
\ee

The running of $\alpha_k$ and the fermion anomalous dimension turn out to be cubic and quadratic in $\alpha$, respectively, and can be parameterized as
\be\label{betaalpha}
\beta_\alpha = A_0 + (A_1+1 + \eta_\psi)\,\alpha + A_2\,\alpha^2 + A_3\,\alpha^3 \, , 
\ee
and
\be\label{etapsi}
\eta_\psi = C_0 + C_1 \, \alpha + C_2 \, \alpha^2 \, . 
\ee
The coefficients $A_i$ and $C_i$ depend on $g$, $\lambda$ and $\eta_\psi$ and take the form
\be\label{Ai}
A_i = \frac{g}{\pi}\Big[\frac{A_i^1 + \Tilde{A}_i^1 \, \eta_\psi}{(1-2\lambda)} + \frac{A_i^2 + \Tilde{A}_i^2 \, \eta_\psi}{(1-2\lambda)^2} + \frac{A_i^3}{(1-2\lambda)^3}\Big] \, , \quad
%\ee
%
%and
%
% \be\label{Ci}
C_i = \frac{g}{\pi}\Big[ \, \frac{C_i^1 + \Tilde{C}_i^1 \, \eta_\psi}{(1-2\lambda)} + \frac{C_i^2}{(1-2\lambda)^2} \, \Big] \, . 
\ee 
The explicit computation gives the values tabulated in Table \ref{tab.1}. 
The terms proportional to $\eta_\psi$ can be understood as non-perturbative contributions resulting from resuming certain perturbative contributions to the flow. The results for the anomalous dimensions are given in implicit form. The corresponding system of linear equations is readily solved to obtain the anomalous dimensions as a function of the scale-dependent couplings $g,\lambda$ and $\alpha$.
\begin{table}[p!]
	\renewcommand{\arraystretch}{1.2}
	\begin{center}
	\begin{tabular}{cl}
		coefficient & numerical value \\
	\hline
\; $A_0^1$ \; & $-\frac{3}{32}$ \\
	$A_1^1$ & $-\frac{7}{6} + \frac{7}{16}\pi - \frac{1}{2}\beta$\\
	$A_2^1$ & $\frac{47}{12} - \frac{5}{4}\pi + (\frac{45}{4} - \frac{45}{16}\pi)\beta$\\
	$A_3^1$ & $\frac{9}{10}$\\
	$A_0^2$ & $\frac{3}{8} - \frac{15}{128}\pi + \frac{1}{32}\eta_N$\\
	$A_1^2$ & $\frac{107}{30} + \frac{1}{32}\pi + (\frac{1}{30}-\frac{13}{64}\pi)\eta_N -\frac{1}{4}\beta - (\frac{39}{40}-\frac{21}{64}\pi) \, \beta \, \eta_N$\\
	$A_2^2$ & $\frac{169}{120} - \frac{1}{2}\pi + (\frac{101}{280}-\frac{1}{8}\pi)\eta_N + (\frac{9}{2}-\frac{9}{8}\pi)\beta - (\frac{61}{20} - \frac{15}{16}\pi) \, \beta \, \eta_N$\\
	$A_3^2$ & $-\frac{17}{10} + (\frac{79}{28}-\frac{27}{32}\pi)\eta_N$\\
	$A_0^3$ & $\frac{7}{20}-\frac{3}{32}\pi - (\frac{179}{1120} - \frac{3}{64}\pi)\eta_N$\\
	$A_1^3$ & $-\frac{67}{30} -\frac{1}{8}\pi + (\frac{47}{210} + \frac{1}{32}\pi)\eta_N$\\
	$A_2^3$ & $-\frac{17}{105}+(\frac{143}{630}-\frac{1}{16}\pi)\eta_N$\\ \hline
	$\Tilde{A}_0^1$ & $-\frac{9}{32} + \frac{3}{32} \pi$\\
	$\Tilde{A}_0^2$ & $-\frac{5}{32} + \frac{3}{64} \pi$\\
	$\Tilde{A}_1^1$ & $\frac{21}{16} - \frac{7}{16} \pi - \frac{1}{8}\beta$ \\
	$\Tilde{A}_1^2$ & $\frac{1}{12} - \frac{1}{32} \pi$\\
	$\Tilde{A}_2^1$ & $\frac{43}{40} - \frac{7}{4} \pi - \left(\frac{9}{4} - \frac{9}{16} \pi\right)\beta$ \\
	$\Tilde{A}_2^2$ & $\frac{49}{60} - \frac{1}{4} \pi$\\
	$\Tilde{A}_3^1$ & $\frac{163}{10} - \frac{21}{4} \pi$ \\
	\hline	
	$C_0^1$ & $\frac{3}{4} -\frac{9}{32} \pi$ \\
	$C_0^2$ & $-\frac{75}{16} - \frac{9}{64} \pi + (\frac{23}{40} + \frac{3}{32} \pi) \, \eta_N$ \\
	$C_1^1$ & $\frac{21}{4} - \frac{45}{32}\pi$ \\
	$C_1^2$ & $\frac{21}{10} - \frac{9}{16} \pi - (\frac{277}{280} - \frac{9}{32} \pi ) \, \eta_N$ \\
	$C_2^1$ & $-\frac{39}{10} + \frac{9}{8}\pi$\\
	$C_2^2$ & $-\frac{23}{10} + \frac{3}{8}\pi + (\frac{41}{140} - \frac{3}{64}\pi) \, \eta_N$\\
	\hline
	$\Tilde{C}_0^1$ & $-\frac{27}{32} + \frac{9}{32}\pi$\\
	$\Tilde{C}_1^1$ & $-\frac{15}{8} + \frac{9}{16}\pi$ \\
	$\Tilde{C}_2^1$ & $-\frac{1}{2} + \frac{3}{16}\pi$ \\ \hline
\end{tabular}
\end{center}
\caption{\label{tab.1} List of the numerical coefficients appearing in the parameterization \eqref{Ai}. The coefficients not listed in the table vanish. The coefficients $A_i^j$ agree with the ones published in \cite{Daas:2020dyo}.}
\end{table}

The beta functions \eqref{betagrav} and \eqref{betaalpha} together with the explicit expressions for the anomalous dimensions \eqref{etaN} and \eqref{etapsi} constitute the main result of this section. They extend the computation presented in \cite{Daas:2020dyo} by including the effect of the wave-function renormalization in the fermionic sector. This extension gives valuable information about the robustness of the fixed points and RG flows resulting from the ansatz given in Sect.\ \ref{ssect:2.1}.

%======================================================
\subsection{Structural Properties of the Beta Functions}
\label{sect:beta3}
%======================================================
At this stage, it is instructive to take limits of our beta functions and compare to earlier results in the literature. Setting the matter contributions to zero, one recovers the beta functions of the Einstein-Hilbert truncation \cite{Reuter:1996cp,Reuter:2019byg}. 

The ambiguity in the coarse-graining operator \eqref{coarsegrainingop} raises the natural question if the parameter $\beta$ can affect the RG flow at a qualitative level. This is conveniently studied by restricting the full setup to the case of minimally coupled fermions, setting $\alpha_k = 0$ and switching off the fermion wave-function renormalization $\eta_\psi = 0$. The equation $\beta_g = 0$ then admits solutions with non-vanishing $g_*$ if $\eta_N^* = -2$. Assuming that its numerator is positive, eq.\ \eqref{etaN} then entails that $g_* > 0$ requires 
\be\label{fpposcond}
B_1^{\rm grav} + B^{\rm ferm} < 0 \, . 
\ee
Substituting the explicit expressions given in \eqref{Bresults} entails that $g_*$ changes sign if
\be
  \left[ \frac{9}{(1-2\lambda)^2} - \frac{5}{(1-2\lambda)} + 7 \right] 
 + \frac{N_f}{2} \bigg[ 
2 -  \pi 
+ \left( 12-6\pi \right) \beta \bigg] = 0 \, . 
\ee
For $\beta < - \frac{1}{6}$ the straight bracket multiplying $N_f$ is positive. If the first term is positive and finite, this entails that \eqref{fpposcond} is satisfied for all values of $N_f$. Conversely, $\beta > - \frac{1}{6}$ entails that (again for $\lambda < 1/2$ kept fixed) there is a critical value 
\be
N_f^{\rm crit} =  2 \left[ \frac{9}{(1-2\lambda)^2} - \frac{5}{(1-2\lambda)} + 7 \right]/\bigg[ \pi - 2 + \left( 6\pi - 12 \right) \beta \bigg] > 0 \, ,
\ee 
where the condition \eqref{fpposcond} is violated and $g_*$ moves to negative values.
This mechanism, leading to a upper bound on the number of fermions for which $g_* > 0$, was first discussed in detail in \cite{Dona:2012am}. In a flat background one has $\slashed{p}^2 = p^2$, corresponding to the case $\beta = 0$. Computations on a spherical symmetric background can be based on coarse-graining with the square of the Dirac operator ($\beta = -1/4)$ which does not lead to a violation of \eqref{fpposcond}. From the structure of $B^{\rm ferm}$ found in \eqref{Bresults}, it is clear that both the inclusion of $\alpha$ and $\eta_\psi$ may crucially affect the condition \eqref{fpposcond}. Following \cite{Daas:2020dyo}, this observation consititutes one motivation to study RG flows including the non-minimal fermion interaction terms \eqref{Gfermion}.

From the viewpoint of chiral symmetry, an intriguing property of $\beta_\alpha$ is the non-vanishing coefficient $A_0$. Baring miraculous cancellations, this entails that any interacting fixed point with $g_* \not = 0$ also comes with a non-zero value $\alpha_*$. This entails that any interacting RG fixed point found in the present computation comes with a (possibly small) chiral symmetry breaking component associated with the coupling of the spinor bilinear to the spacetime curvature. At this point it is instructive to scrutinize this property in more detail. For this purpose, we compute the coefficient $A_0$ with the right-hand side of \eqref{thetabackground} artificially set to zero
\be\label{A00res}
\begin{split}
&\left. A_0^1 \right|_{\nabla_\mu \theta = 0}= \frac{3}{32} \, , \\
&\left. A_0^2 \right|_{\nabla_\mu \theta = 0} = \frac{1}{2} - \frac{15}{128} \pi + \left( \frac{23}{160} - \frac{3}{64} \pi \right) \eta_N  \, , \\
&\left. A_0^3 \right|_{\nabla_\mu \theta = 0} = \frac{7}{20} - \frac{3}{32} \pi - \left( \frac{179}{1120} - \frac{3}{64} \pi \right) \eta_N \, . 
\end{split}
\ee
In a sense, this may correspond to a hybrid-type of computation where the flow of the matter couplings is evaluated on a flat background while the gravitational sector is computed utilizing the background sphere. The contribution of the non-trivial background background spinor $\Delta A_0^i \equiv  A_0^i - \left. A_0^i \right|_{\nabla_\mu \theta = 0} $ can then be isolated by taking the difference between the result \eqref{A00res} and the coefficients given in Table \ref{tab.1}
\be
\Delta A_0^1 = - \frac{3}{16} \, , \qquad
\Delta A_0^2  = - \frac{1}{8} - \left( \frac{9}{80} - \frac{3}{64} \pi \right) \eta_N \, , \qquad
\Delta A_0^3  = 0 \, . 
\ee
The vanishing of $\Delta A_0^3$ indicates that this term receives contributions from the regulator only. In general the $\Delta A_0^i$ are non-vanishing though. This establishes that even in the case where no additional chiral symmetry breaking terms are induces (e.g., by a mass-type regulator in the fermionic sector), the background spinor equation will induce a non-trivial coefficient $A_0$. This makes the flow qualitatively different from the one computed for matter sectors on a flat background.

%======================================================
\section{Fixed Point structure for gravity-fermion systems}
\label{sect:fp}
%======================================================
The main result of the previous section is the explicit form of the beta functions encoding the scale-dependence of the dimensionless couplings $g_k, \lambda_k$, and $\alpha_k$ as well as the fermion anomalous dimension $\eta_\psi$. These beta functions depend parameterically on the number of fermions $N_f$ and the parameterization of the coarse-graining operator $\beta$. The explicit expressions \eqref{betagrav} and \eqref{betaalpha} then  serve as our starting point for studying the fixed point structure of the system. Clearly, there is a Gaussian fixed point (GFP), situated at $\{g_*^{\rm GFP}, \lambda_*^{\rm GFP}, \alpha_*^{\rm GFP}\} =\{0,0,0\}$ which is present at all levels of our approximation. Thus our goal is to identify the non-Gaussian fixed points (NGFPs) which could provide a high-energy completion of the gravity-fermion system via the asymptotic safety mechanism.
The interplay of the fixed points will be studied in Sect.\ \ref{sect:chiralsymmetry}.
 
Let us start by introducing the general framework. We denote the dimensionless couplings collectively by $u^i = \{g,\lambda,\alpha\}$. At a fixed point $u^i_*$ all beta functions vanish simultaneously, 
\be
\beta_{u^j}(u_*^i) = 0 \, , \qquad \forall \, i \, . 
\ee 
 This entails that the dimensionless coupling constants remain finite if the underlying RG trajectory is dragged into a fixed point as $k \rightarrow \infty$. The set of these asymptotically safe trajectories spans the UV-critical hypersurface of the fixed point. Whether a given direction is UV-attractive (UV-relevant) or UV-repulsive (UV-irrelevant) is conveniently studied by linearizing the beta functions at the fixed point. Defining the stability matrix associated with a given fixed point $u_*$,
\be\label{stabmat}
\bB^i{}_j \equiv \left. \p_{u^j} \beta_{u^i} \right|_{u = u_*} \, , 
\ee
eigendirections $V_I$ with stability coefficients $\theta_I$, $\bB V_I = - \theta_I V_I$, are UV-attractive if ${\rm Re} \, \theta_I > 0$ and UV-repulsive if ${\rm Re} \, \theta_I < 0$. Since $\bB$ is not necessarily symmetric, the stability coefficients can be complex. Notably, the $\theta_I$ constitute observable quantities. Thus they provide valuable probes for judging the robustness of approximations based on truncating the action functional.  

In the following subsections we will gradually build up the analysis, starting with minimally coupled fermions in Sect.\ \ref{ssect:4.1}. The effect of the chiral symmetry breaking fermion-curvature coupling, initially studied in \cite{Daas:2020dyo} and also \cite{Eichhorn:2018nda,deBrito:2020dta}, is reviewed in Sect.\ \ref{ssect:4.3}. We analyze the fixed point structure of the full system in Sect.\ \ref{ssect:4.4} and conclude with a discussion of the chiral symmetry breaking terms in $\beta_\alpha$ in section \ref{ssect:4.42}. Our main result are the two families of RG fixed points characterized in Figs.\ \ref{FullSysA} and \ref{FullSysB}, which are robust under the extension of the projection space and satisfy the criterion of being almost Gaussian in the sense that quantum fluctuations do not provide ``large corrections'' to the classical values of the critical exponents.

%======================================================
\subsection{Fermions minimally coupled to gravity}
\label{ssect:4.1}
%======================================================
We start by studying the fixed point structure arising from minimally coupled fermions, restricting ourselves to the subsystem where $\alpha_k = 0$.
%======================================================
\subsubsection{Minimally coupled fermions without the fermion anomalous dimension}
\label{ssect:4.4.1}
%======================================================
The first level of the analysis also switches off the fermion wave function renormalization by setting $\eta_\psi = 0$. In this case the beta functions \eqref{betagrav} simplify to
\be
\begin{split}
	\beta_g = & \, (2+\eta_N) \, g \, , \\
	\beta_\lambda = & \left(\eta_N - 2 \right) \lambda + \frac{g}{4\pi} \left[  \left( 10 - \frac{5}{3} \eta_N \right) \frac{1}{(1-2\lambda)} - 8 \right] - \frac{N_f \, g}{3}\, .
\end{split}
\label{betafnctsEH}
\ee
\\
The explicit expression for $\eta_N$ takes the form
\be
\eta_N = \frac{2g\left(22 - 36\lambda + 56\lambda^2 - N_f\ (\pi-2) \ (1+ 6\beta)\ (1-2\lambda)^2 \right)}{g\ (1+10\lambda) - 12\pi \ (1-2\lambda)^2}.
\label{explicitAdimsEH}
\ee
The Einstein-Hilbert truncation without matter fields is recovered by setting $N_f = 0$. In this case the flow is independent of $\beta$ and one recovers the well-known Reuter fixed point 
\be\label{ReuterFP}
g_* = 0.707 \, , \quad \lambda_* = 0.193 \, , \quad \theta_{1,2} = 1.48 \pm  3.04i \, . 
\ee

\begin{figure}[t!]
	\includegraphics[width=0.48\textwidth]{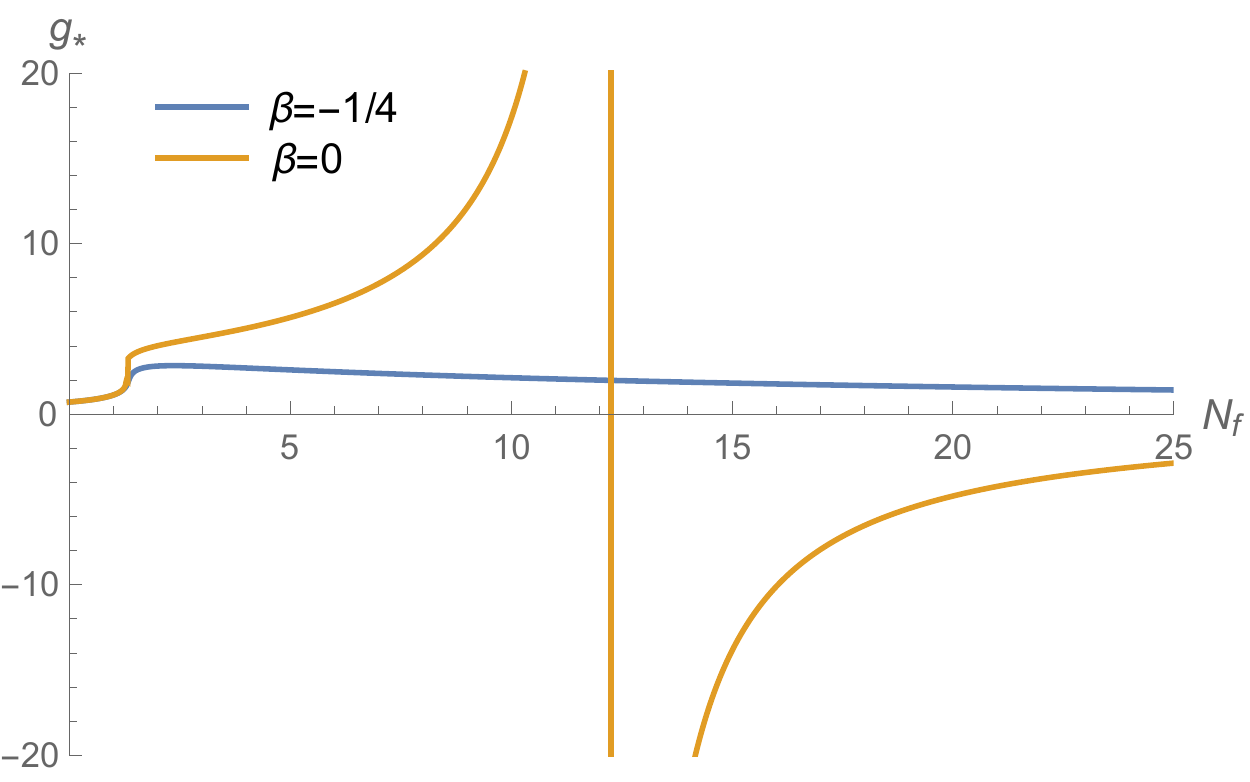} \, 
	\includegraphics[width=0.48\textwidth]{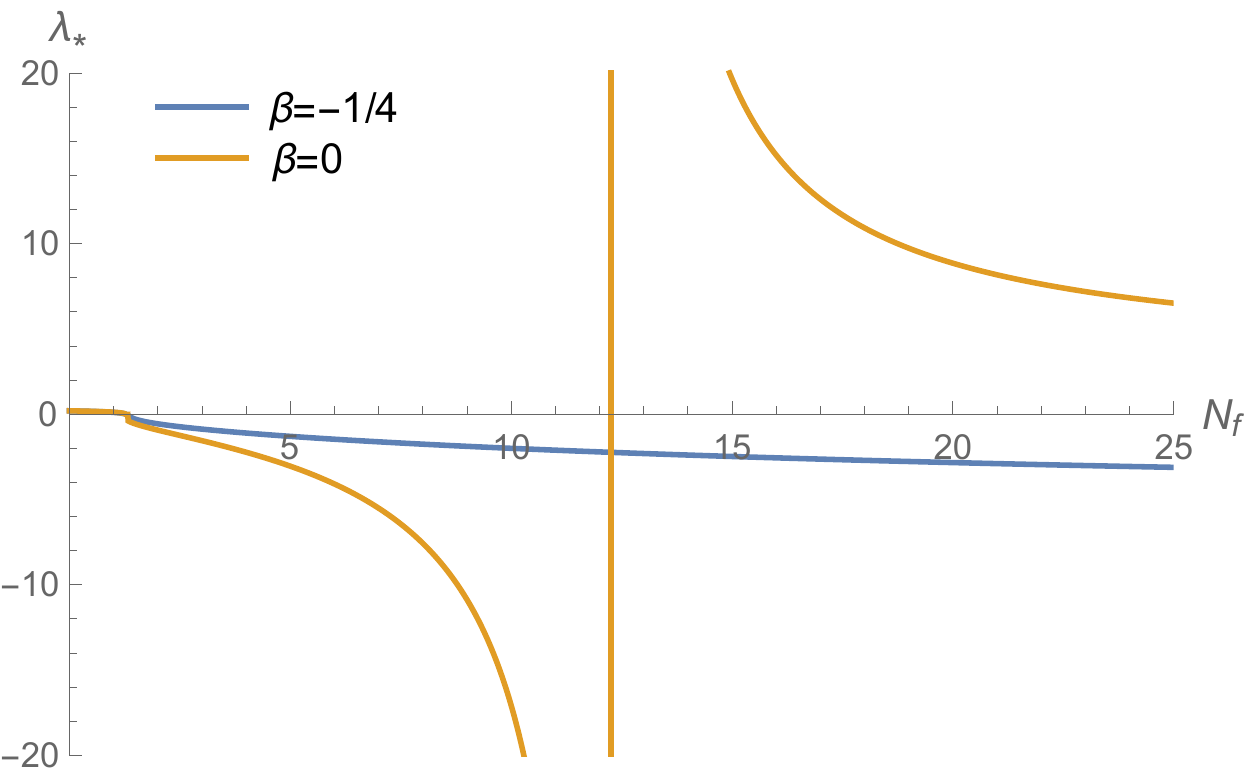} \\
	\includegraphics[width=0.48\textwidth]{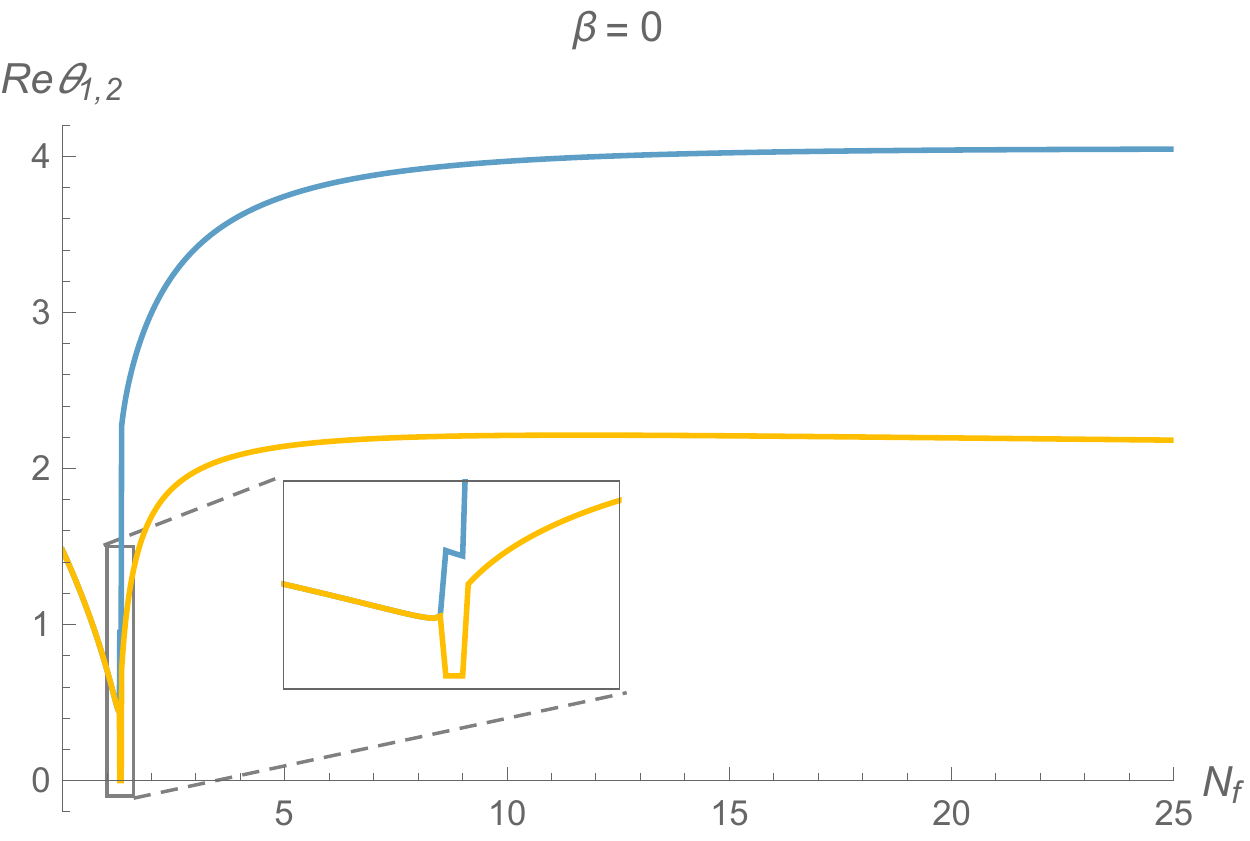} \, 
	\includegraphics[width=0.48\textwidth]{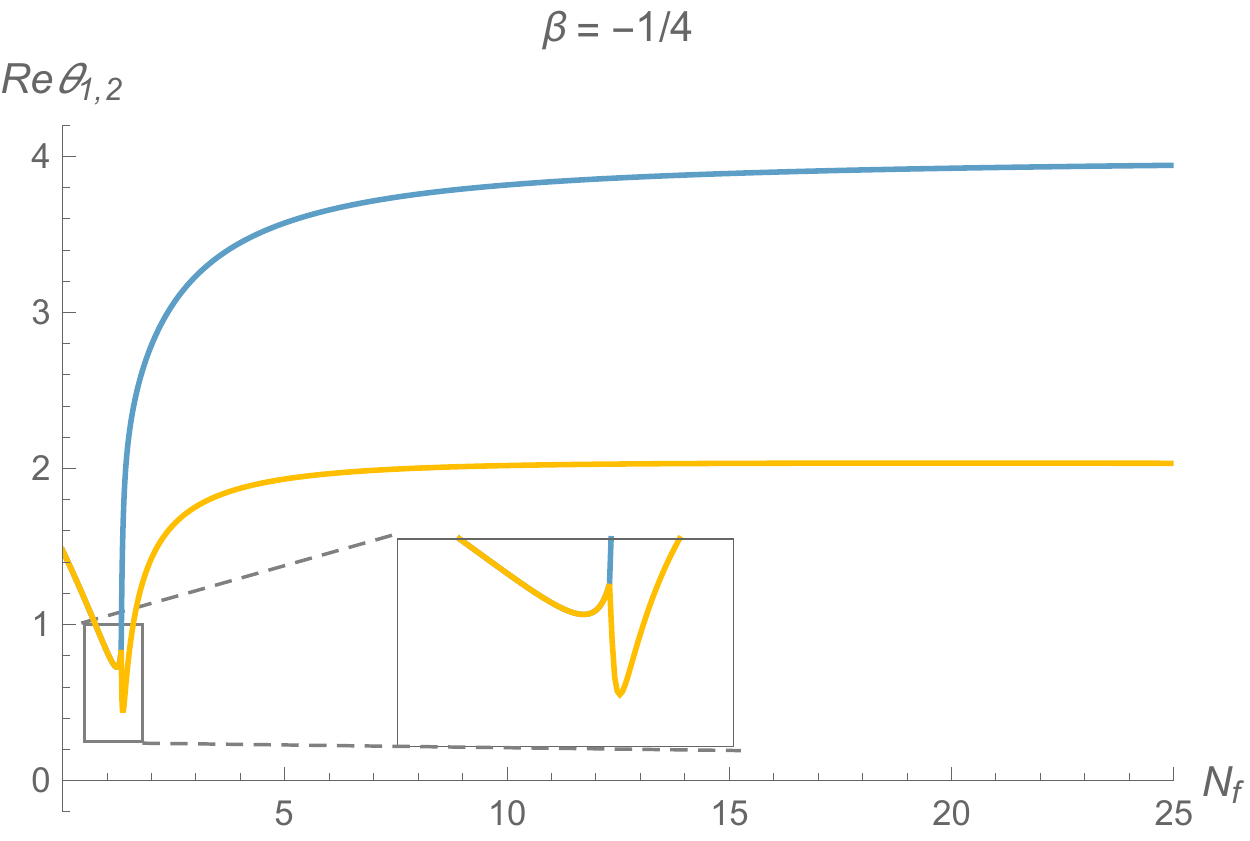}
	\caption{\label{EHFPs} The position and critical exponents of the NGFP arising from \eqref{betafnctsEH} as function of $N_f$. The blue and orange lines correspond to $\beta = -1/4$ (coarse-graining by the Laplacian) and $\beta = 0$ (coarse-graining by the squared Dirac operator). Notably, the NGFP exists for all values $N_f$. For $\beta = 0$ there is a critical number of fermions $N_f^{\text{crit}} = 12.26$ where the NGFP transition from $g_* > 0$ to $g_* < 0$. The real part of the critical exponents as a function of $N_f$ is shown in the bottom row. For low values $N_f$, the stability coefficients $\theta_{1,2}$ are a complex pair. The critical exponents turn real at $N_f \approx 1.33$ ($\beta = -1/4$) and $N_f \approx 1.29$ ($\beta = 0$), respectively. This transition is magnified in the insets shown in the lower row.} 
\end{figure}
 The inclusion of fermions results in a deformation of the gravitational fixed point. Notably, the deformed fixed points exist for all values of $N_f$ for both values of $\beta$. This can be seen in the top line of Fig.\ \ref{EHFPs}. A remarkable feature is that the $\beta = 0$-case exhibits a critical number of fermions $N_f^{\text{crit}} = 12.26$ where the NGFP transitions from $g_* > 0$ to $g_* < 0$. From the perspective of the background field approximation NGFPs situated at $g_* < 0$ are problematic for phenomenologically admissible high-energy completions of gravity: from \eqref{betafnctsEH} one concludes that $\beta_g = 0$ when $g=0$. As a consequence the RG flow of Newton's coupling can not change its sign. At low energy, gravity being attractive requires a positive value $G_0$ which can not be reached from a NGFP situated at $g_* < 0$.\footnote{Also see our discussion in section \ref{sect:beta3} and ref.\ \cite{Meibohm:2015twa} for arguments bypassing this logic at the level of fluctuation field computations.} On this basis one could then conclude that not all NGFPs are located such that it may serve as a building block for a theory with a phenomenologically viable high energy completion.

The critical exponents of the NGFPs are shown in the lower row of Fig.\ \ref{EHFPs}. Starting from the gravitational fixed point \eqref{ReuterFP} and increasing $N_f$, one observes that they become real at $N_f \approx 1.33$ ($\beta = -1/4$) and $N_f \approx 1.29$ ($\beta = 0$), respectively.  The fact that the critical exponents vary mildly with $N_f$ indicates that the even the inclusion of a large number of matter fields does not crucially alter the stability properties of the fixed point. This provides a clear indication that the NGFPs seen in the present setting are essentially gravity-dominated. Remarkably, the transition from $g_* > 0$ to $g_* < 0$ leaves no imprint on the critical exponents, justifying the deformation of the gravitational fixed point persists for all values of $N_f$ also in the peculiar case of $\beta = 0$.

Intriguingly, the properties of the NGFPs for large numbers of fermion fields can be studied analytically. Using a large-$N_f$-expansion, the position of the fixed point is given by
\be\label{NGFPNf1}
\begin{split}
g_* \simeq & \, - \frac{12 \pi}{N_f (\pi -2) \xi } \, , \qquad
\lambda_* \simeq \, \frac{\pi}{(\pi-2)\xi} + \frac{\lambda^{(1)}_*(\xi)}{N_f}  \, . 
\end{split}
\ee
Here we abbreviated $\xi \equiv 1+6\beta$ and the $\simeq$-symbol indicates that the right-hand side is an expansion in $1/N_f$ with terms of order $\cO(1/N_f^2)$ being neglected. The coefficient $\lambda^{(1)}_*(\xi)$ is positive for all values $\beta$. Its explicit form is tabulated in Table \ref{Tab.Nfcoeff}. The large-$N_f$ expansion of the critical exponents has a similar structure. Computing the stability matrix, substituting the general solution for $g_*$, and $\lambda_*$, and subsequently performing the large-$N_f$ expansion of the critical exponents yields
\be\label{critexp1}
\begin{split}
\theta_1 \simeq & \, 4 + \frac{\theta_1^{(1)}}{N_f}  \, , \qquad
\theta_2 \simeq  \, 2 + \frac{\theta_2^{(1)}}{N_f} \, . 
\end{split}
\ee
The coefficients at subleading order in $N_f$ again depend on $\xi$ and are given in the lower block of Table \ref{Tab.Nfcoeff}. Evaluating the coefficients for the specific values of $\beta$, one sees that these asymptotics entailed by eqs.\ \eqref{NGFPNf1} and \eqref{critexp1} matches the one shown in Fig.\ \ref{EHFPs}. 

A remarkable property of eqs.\ \eqref{NGFPNf1} and \eqref{critexp1} is that in the large-$N_f$ limit $g_* \propto 1/N_f$ becomes perturbatively small. At the same time, the critical exponents do not match the ones of the free theory, indicating that we are still dealing with a non-Gaussian fixed point in this limit. This opens a window for studying the gravity-matter fixed points using standard perturbative techniques. 

%======================================================
\subsubsection{Minimally coupled fermions including the fermion anomalous dimension}
\label{ssect:4.4.2}
%======================================================
We now refine the analysis of the previous subsection by including the fermion anomalous dimension $\eta_\psi$ while still keeping $\alpha_k = 0$. This extension of the truncation gives important information on the robustness of the fixed point structure uncovered in the system with $\eta_\psi=0$. The explicit form of the beta functions is
\be
\begin{split}
	\beta_g = & \, (2+\eta_N) \, g \, , \\
	\beta_\lambda = & \left(\eta_N - 2 \right) \lambda + \frac{g}{4\pi} \left[  \left( 10 - \frac{5}{3} \eta_N \right) \frac{1}{(1-2\lambda)} - 8 \right] - \frac{N_f \, g}{12}(4 + (8 - 3\pi) \, \epsilon \, \eta_\psi)\, .
\end{split}
\label{betafnctsEHAdim}
\ee
The explicit expressions for $\eta_N$ and $\eta_\psi$ are obtained from eqs.\ \eqref{etaN} and \eqref{etapsi} by taking $\alpha = 0$ and solving the system of linear equations
\be\label{adim}
\begin{split}
	\eta_N = & \frac{g \left[ \left( B_1^{\rm grav} + N_f  B_1^{\rm ferm}\right) \big( \tilde{C}_0^1 g (1-2\lambda) - \pi (1-2\lambda)^2 \big) - g \, N_f \, B_2^{\rm ferm}\left(C_0^{2a} + C_0^1 (1-2\lambda)\right)\right]}{g^2 N_f  B_2^{\rm ferm} C_0^{2b}- B_2^{\rm grav}  g\big( \tilde{C}_0^1 g (1-2\lambda) - \pi (1-2\lambda)^2 \big) + \tilde{C}_0^1 \, g \, (1-2\lambda) - \pi \, (1-2\lambda)^2}\, , \\
	\eta_\psi = & \,  \frac{g \left[ g \, B_1^{\rm grav} \, C_0^{2b}  + (1- gB_2^{\rm grav})(C_0^1 (1-2\lambda) +C_0^{2a})  + N_f \, g \, B_1^{\rm ferm} \, C_0^{2b} \right]}{(1 - g B_2^{\rm grav}) (1 - 2 \lambda) ( \pi(1-2\lambda) - g \, \tilde{C}_0^1) - N_f \, g^2 \, B_2^{\rm ferm} \, C_0^{2b}} \, .
\end{split}
\ee
The constants $C_0^i$ and $\tilde{C}_0^i$ are listed in Table \ref{tab.1} and we decomposed $C_0^2 = C_0^{2a} + C_0^{2b} \, \eta_N$ such that
\be
C_0^{2a} \equiv - \frac{75}{16} - \frac{9}{64} \pi \, , \qquad C_0^{2b} \equiv \frac{23}{40} + \frac{3}{32} \pi \, . 
\ee
The parameter $\epsilon$ introduced in \eqref{betafnctsEHAdim} distinguishes between the cases where the fermion anomalous dimension is included ($\epsilon = 1$) or switched off ($\epsilon=0$).

\begin{figure}[t!]
	\includegraphics[width=0.47\textwidth]{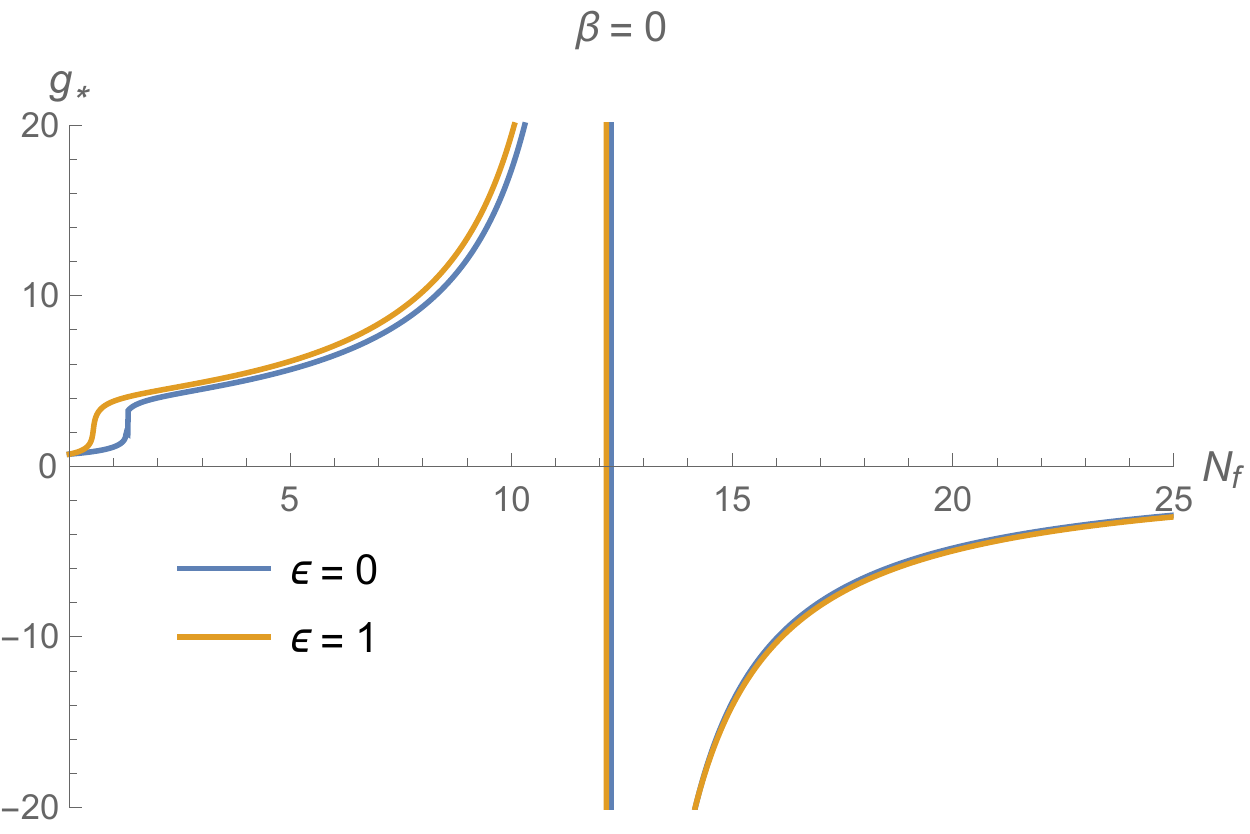}  \, 
	\includegraphics[width=0.47\textwidth]{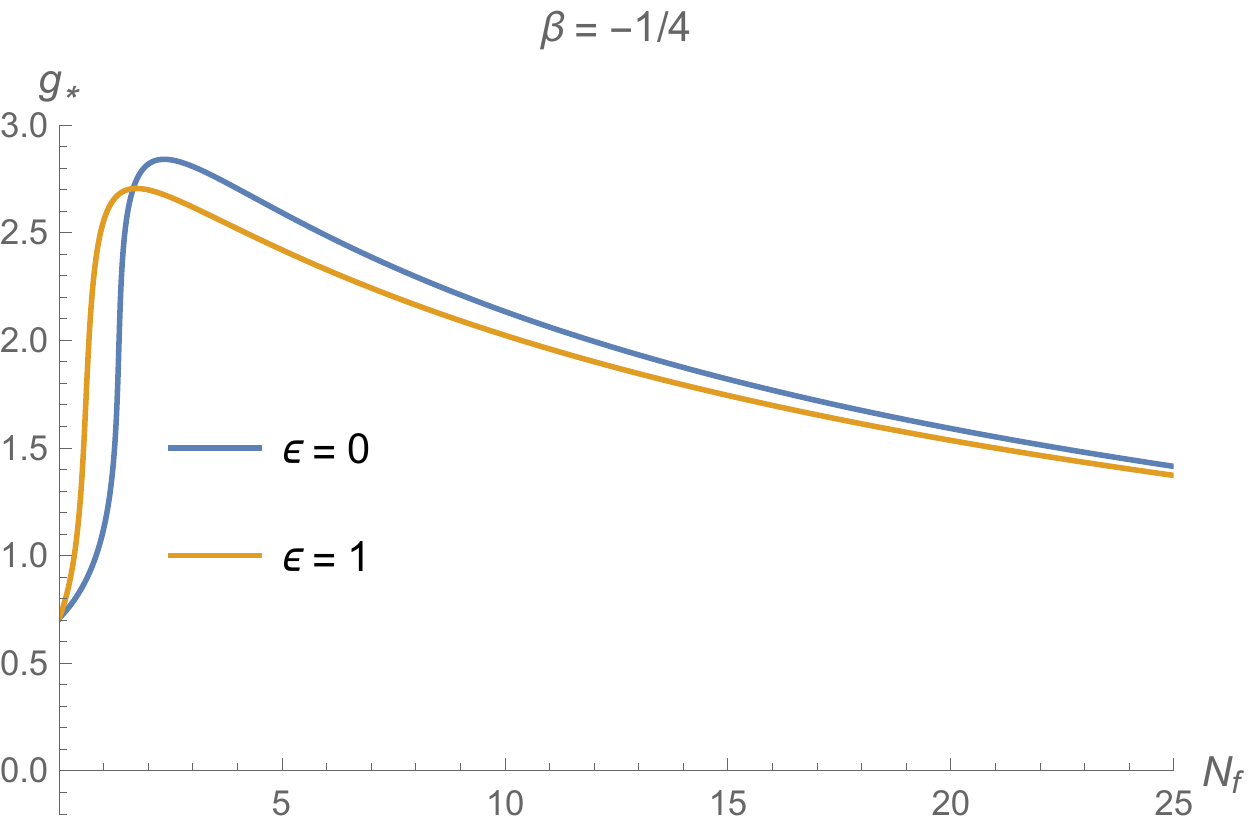} \\
	\includegraphics[width=0.47\textwidth]{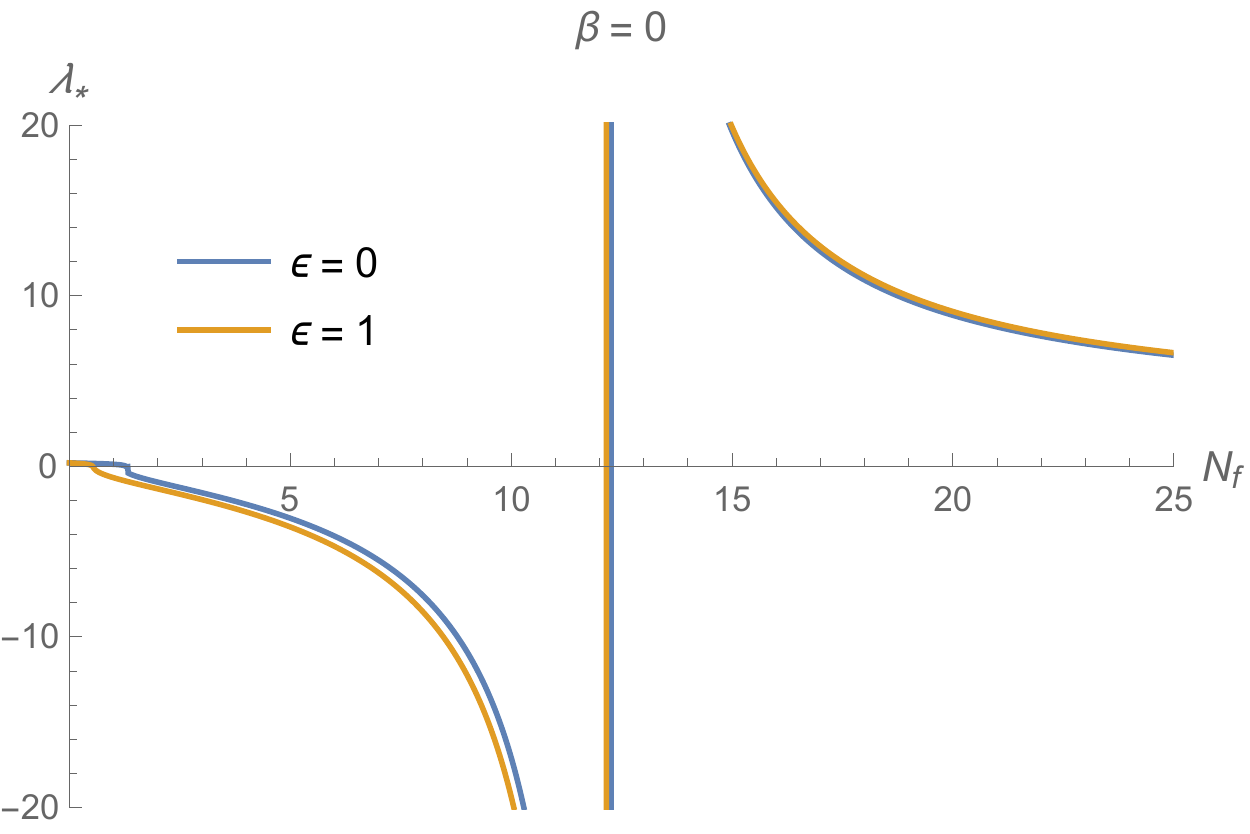} \, 
	\includegraphics[width=0.47\textwidth]{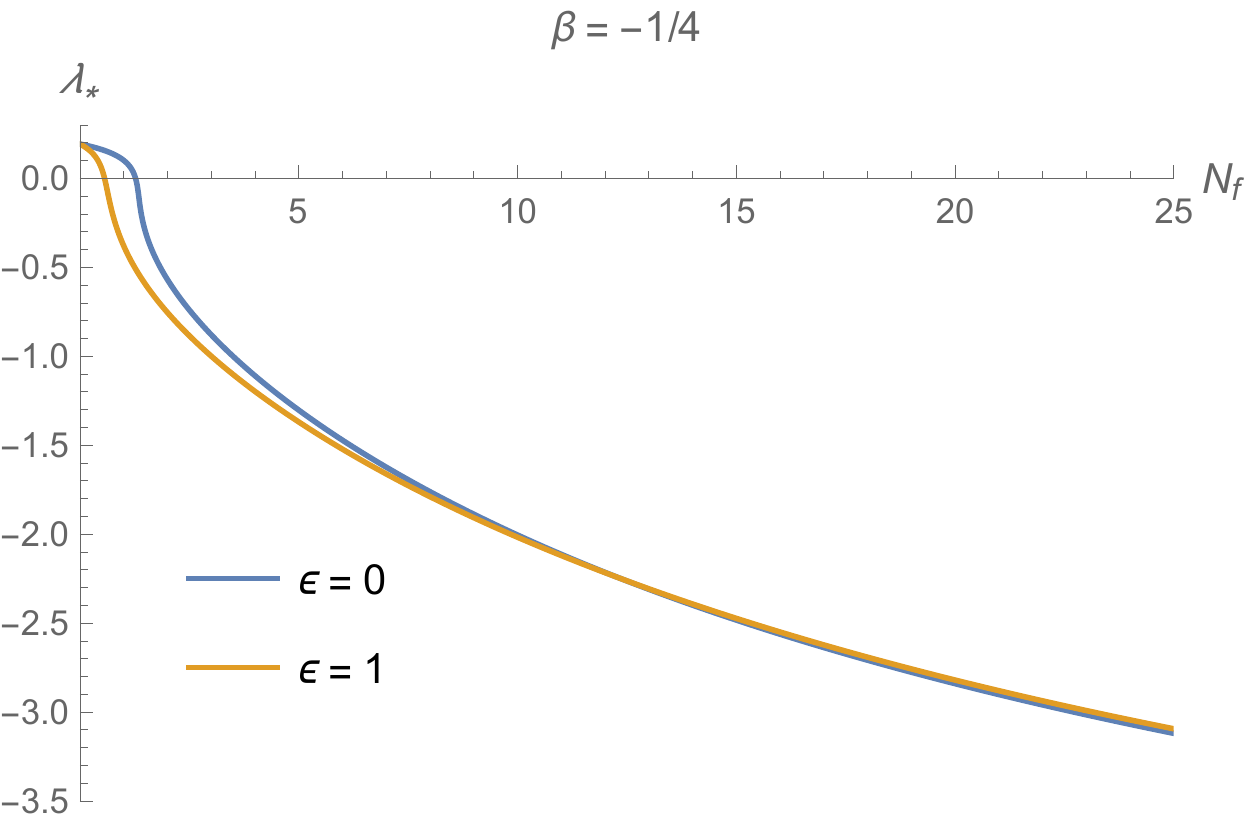} \\
	\caption{\label{EHwithAdim} Corrections to the position of the NGFPs resulting from the inclusion of the fermion anomalous dimension ($\epsilon = 1$, orange lines). The postion of the NGFP for $\eta_\psi = 0$ is given by the blue lines. Notably, the corrections provided by the fermion anomalous dimension are small. They are most pronounced in small values of $N_f$.} 
\end{figure}

The impact of the fermion anomalous dimension on the position of the NGFP is shown in Fig.\ \ref{EHwithAdim} for both choices of the coarse-graining scheme.
Notably, the fixed point positions receive only minor corrections once the fermionic wave function is included. These corrections are most pronounced in the range $0.7 \lesssim N_f \lesssim 1.3$ for $\beta = 0$ whereas for $\beta = -1/4$ the corrections are over a wider range although they get significantly smaller as $N_f$ increases. For $\beta = 0$ there is again a critical number of fermions, albeit at a slightly larger value. 

\begin{figure}[t]
		\includegraphics[width=0.47\textwidth]{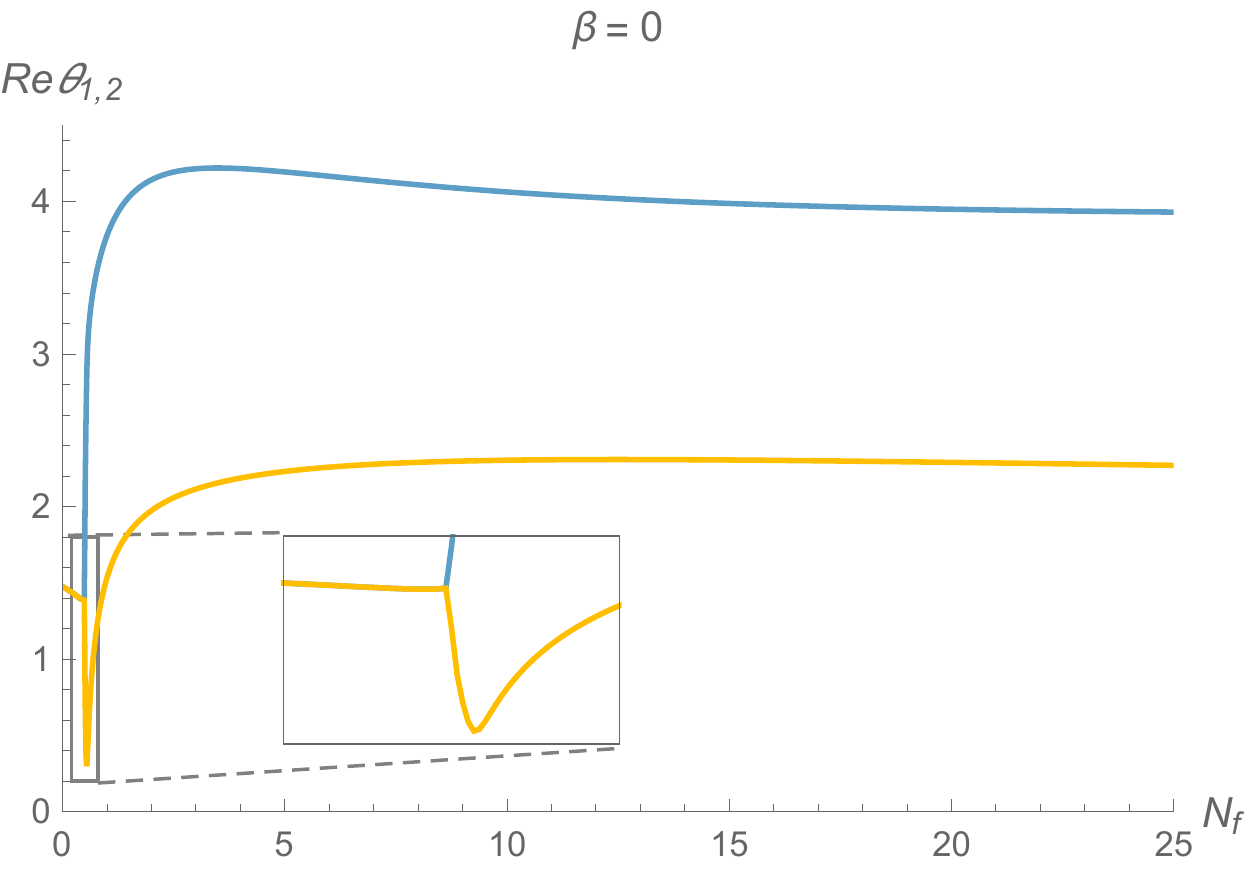} \, 
	\includegraphics[width=0.47\textwidth]{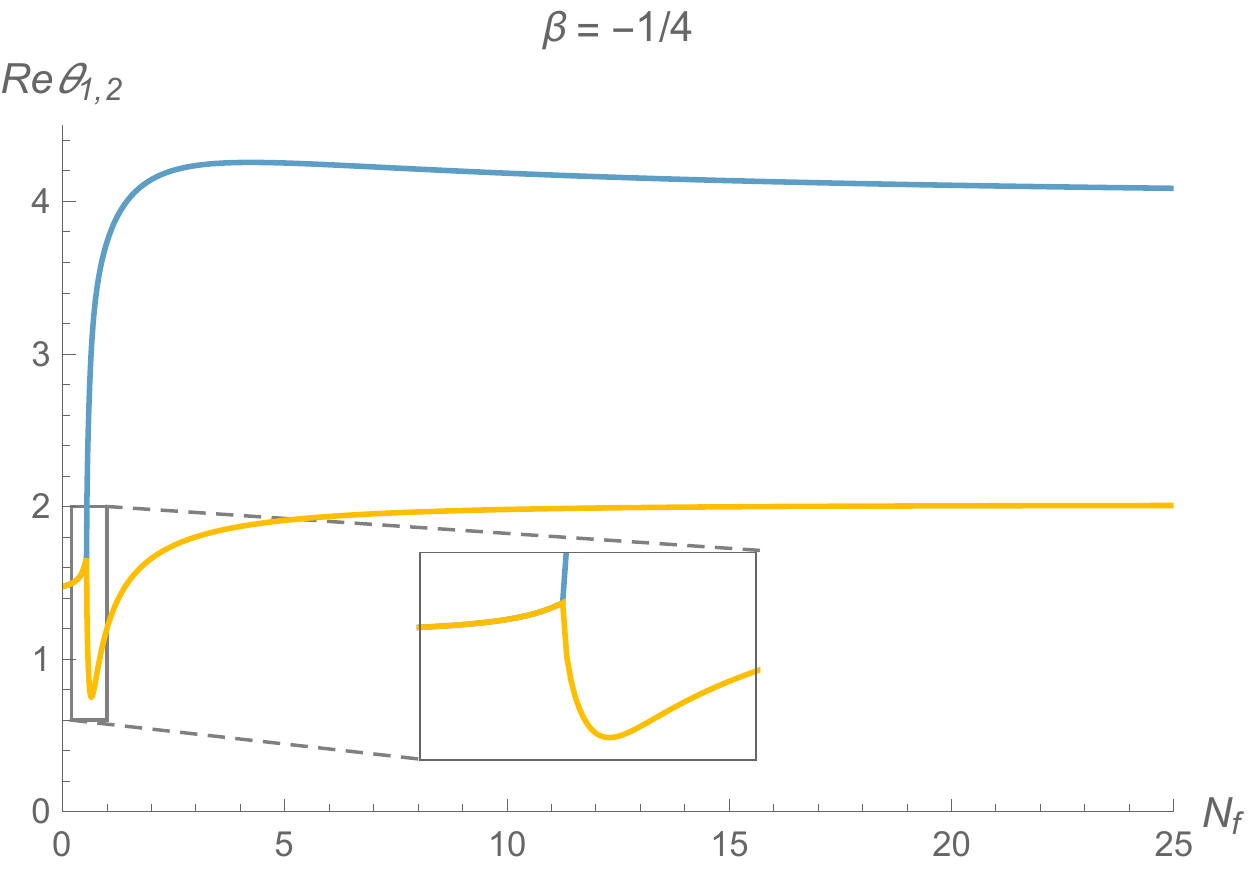} \\
	\includegraphics[width=0.47\textwidth]{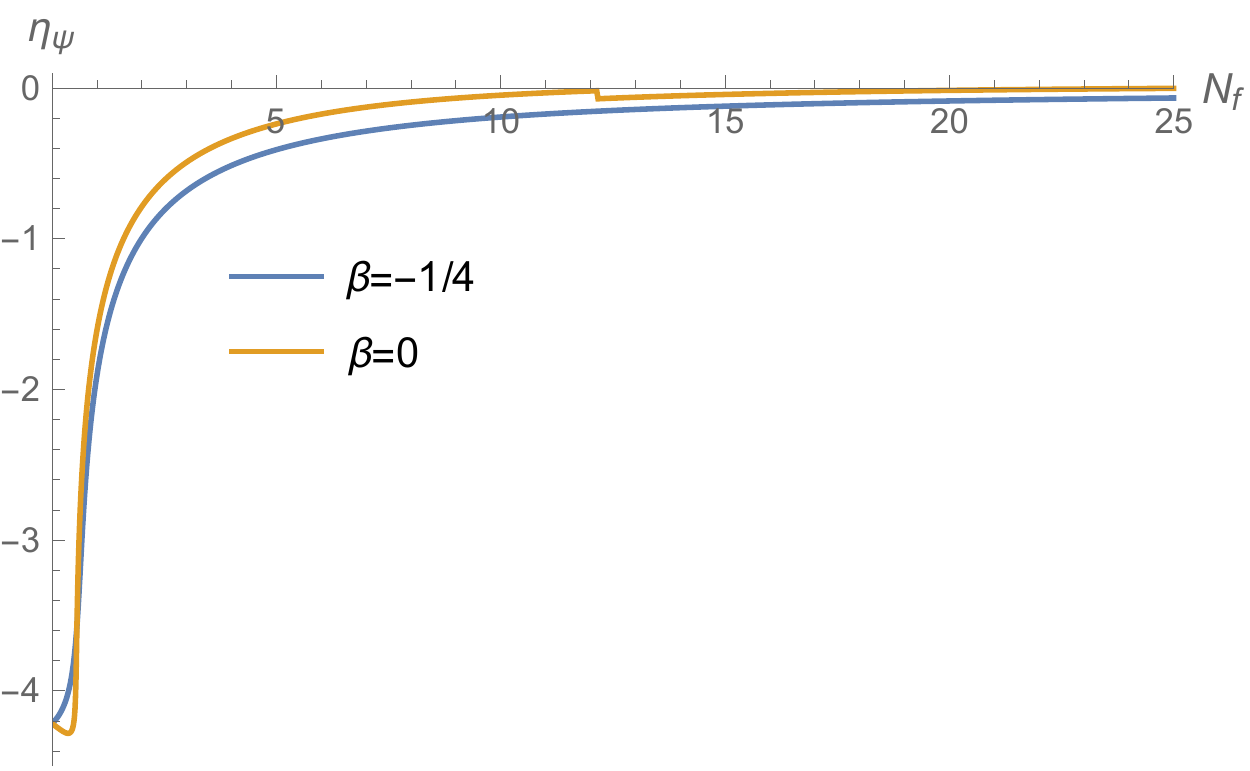}
	\centering
	\caption{\label{EHAdim} Illustration of the critical exponents (top line) and fermion anomalous dimension (bottom line) for $\beta = 0$ and $\beta = -1/4$, respectively. The bifurcation point appearing for low values of $N_f$ signals the transition from complex to real critical exponents. The sign change in $g_*$ and $\lambda_*$ that occur when $\beta = 0$ induces a tiny discontinuity where the fermion anomalous dimension jumps by $\Delta \eta_\psi = 0.03$ without changing its sign.} 
\end{figure}

The  the critical exponents of the NGFPs together with the corresponding values for $\eta_\psi$ are shown in Fig.\ \ref{EHAdim}.
The comparison between Fig.\ \ref{EHFPs} and \ref{EHAdim} shows that the critical exponents obtained for $\epsilon = 0$ and $\epsilon=1$ exhibit the same qualitative behavior. Again there is a transition from complex to real critical exponents occurring, now at $N_f = 0.5$ ($\beta = 0$) and $N_f = 0.55$ ($\beta = -1/4$). Moreover, the transition from $g_* > 0$ to $g_* < 0$ ($\beta = 0$) does not reflect in $\theta_{1,2}$ which are again continuous at $N_f^{\rm crit}$. The fermion anomalous dimension shown in the bottom diagram reveals that $\eta_\psi$ is bounded and vanishes for large values of $N_f$. The largest absolute values are found for small values of $N_f$, consistent with the observation that the corrections arising from the inclusion of the  fermionic wave function renormalization are most pronounced in this regime.

Eqs.\ \eqref{betafnctsEHAdim} and \eqref{adim} again allow to study the properties of the fixed points in a large-$N_f$ expansion analytically. This expansion takes a form similar to the one found in the minimally coupled case. The fixed point is located at
\be\label{largeNfpospsi}
g_*\simeq -\frac{12 \pi }{(\pi -2) \xi }\frac{1}{N_f} \, , \qquad
\lambda_* \simeq  \frac{\pi }{(\pi -2) \xi }+\frac{\lambda_*^{(1,\eta_\psi)}}{N_f} \, . 
\ee
The coefficient $\lambda_*^{(1,\eta_\psi)}$ is again a function of $\xi$ and listed in the second row of Table \ref{Tab.Nfcoeff}. The large-$N_f$ behavior of the stability coefficients is again obtained by constructing the general stability matrix, substituting the position of the fixed point and then performing the $N_f$-expansion for the resulting eigenvalues. This results in
\be\label{critexp1eta}
\begin{split}
	\theta_1 \simeq & \, 4 + \frac{\theta_1^{(1,\eta_\psi)}}{N_f}  \, , \qquad
	\theta_2 \simeq  \, 2 + \frac{\theta_2^{(1,\eta_\psi)}}{N_f} \, . 
\end{split}
\ee
Using that $g_* \propto 1/N_f$ and $\lambda_* \propto \text{constant}$, the asymptotics of $\eta_\psi$ follows from \eqref{adim}
\begin{equation}\label{psiasym}
\eta_\psi\simeq \frac{3 (\pi -2) (1628+195 \pi ) \xi +180 \pi  (8-3 \pi )}{80 (\pi  (\xi -2)-2 \xi )^2}\frac{1}{N_f} \, . 
\end{equation}
The sign of the prefactor depends on $\xi$. For $\xi \ge 0.105$, $\eta_\psi$ approaches zero from above while for $\xi \le 0.105$ it is negative for all values $N_f$. Again the asymptotics \eqref{largeNfpospsi}, \eqref{critexp1eta} and \eqref{psiasym} matches the one found numerically in Figs.\ \ref{EHwithAdim} and \ref{EHAdim}. 

\begin{table}[t!]
	\renewcommand{\arraystretch}{1.2}
	\begin{tabular}{lll}
		\hline \hline
		$\lambda_*^{(1)}$ & eq.\ \eqref{NGFPNf1} & $\frac{56 \pi^3 -2(\pi-2)\pi^2 \xi + 19 (\pi-2)^2 \pi \xi^2 - 4 (\pi-2)^3 \xi^3}{(\pi-2)^2 (\pi (\xi-2) - 2 \xi)^2 \xi^2}$ \\
		\hline
		$\lambda_*^{(1,\eta_\psi)}$ & eq.\ \eqref{largeNfpospsi} & 
		\parbox{12cm}{$\frac{1}{320 (\pi -2)^2 (\pi  (\xi -2)-2 \xi )^2 \xi ^2} \Big[-1280 (\pi -2)^3 \xi ^3 \\
			-(\pi -2) \pi  (\pi  (9 \pi  (328+195 \pi )-85040)+168448) \xi ^2\\
			+4 \pi  (\pi  (\pi  (45 \pi  (9 \pi -97)+9401)+9710)-29304) \xi +80 \pi ^2 (\pi  (71+27 \pi )+216)\Big] $} 
		\\
	\hline \hline
$\theta_1^{(1)}$ & eq.\ \eqref{critexp1} & $\frac{80 (\pi-2) \xi}{(\pi (\xi-2)  - 2 \xi)^2}$ \\
$\theta_2^{(1)}$ & eq.\ \eqref{critexp1} & $- \frac{2(4\xi^2 - 4 \pi \xi(37+\xi) + \pi^2 (40+74\xi + \xi^2))}{(\pi(\xi-2) - 2 \xi)^3}$ \\
\hline
$\theta_1^{(1,\eta_\psi)}$ & eq.\ \eqref{critexp1eta} & 
	\parbox{12cm}{$\frac{800 (\pi -2)^2 \xi ^2+(\pi -2) \pi  (9 \pi  (674+75 \pi )-22576) \xi -45 \pi ^2 (8-3 \pi )^2}{10 (\pi  (\xi -2)-2 \xi )^3} $} 
 \\
$\theta_2^{(1,\eta_\psi)}$ & eq.\ \eqref{critexp1eta} &
	\parbox{12cm}{$\frac{-1}{80 (\pi  (\xi -2)-2 \xi )^3} \Big[(\pi -2) (\pi  (12004+2295 \pi )-48224) \xi ^2\\
    -2 (\pi  (\pi  (1170 \pi -20497)+28250)+35928) \xi +20 \pi  (\pi  (27 \pi -805)+3132)\Big] $} 
 \\
\hline \hline		
	\end{tabular}
	\caption{\label{Tab.Nfcoeff} Summary of the numerical coefficients determining the leading corrections in a large-$N_f$ expansion of the fixed point position (top rows) and stability coefficients (bottom rows).}
\end{table}

\emph{In summary, we find that the gravitational fixed point admits a deformation by the fermion number $N_f$. This deformation exists for all values $N_f$. The critical exponents vary mildly with $N_f$, indicating that the fixed point is ``gravity-dominated''. For $\beta > -\frac{1}{6}$ there is a critical number of fermions for which the value of the background Newton's coupling $g_*$ transits to negative values. In general, the inclusion of the fermion anomalous dimension leads to small corrections in the fixed point properties and in particular to the critical exponents. The one-parameter family of NGFPs admits a large $N_f$ expansion where $g_* = 0$ to leading order. Moreover, the leading term in the $N_f$-expansion of the stability coefficients are universal in the sense that they are independent of our choice of coarse-graining operator.}
%======================================================
\subsection{Including non-minimally coupled gravity-fermion interactions}
\label{ssect:4.3}
%======================================================
We now consider the system where $\alpha_k \neq 0$ and $\eta_\psi = 0$. This subsystem includes the gravity-mediated chiral symmetry breaking term while neglecting the effect of the fermion wave function renormalization. Since this case has been discussed in \cite{Daas:2020dyo} already, the present exposition will be brief. 

The explicit form of the beta functions for this subsystem is readily obtained from \eqref{betagrav}, \eqref{etaN}, and \eqref{betaalpha} and reads  
\be
\begin{split}
	\beta_g = & \, (2+\eta_N) \, g \, , \\
	\beta_\lambda = & \left(\eta_N - 2 \right) \lambda + \frac{g}{4\pi} \left[  \left( 10 - \frac{5}{3} \eta_N \right) \frac{1}{(1-2\lambda)} - 8 \right] - \frac{N_f \, g}{3}\, , \\
	\beta_\alpha = & A_0 + (A_1 + 1)\alpha + A_2 \alpha^2 + A_3 \alpha^3\, .
\end{split}
\label{betafnctsNMC}
\ee
The anomalous dimension $\eta_N$ has the explicit form
\be
\eta_N = \frac{2g\left(22 - 36 \lambda + 56 \lambda^2 - N_f\left( \pi + 6\pi\beta -2(1+6\alpha + 6\beta) \right) (1-2\lambda)^2 \right)}{g(1+10\lambda) - 12\pi (1-2\lambda)^2} \, .
\ee
The coefficients $A_i$ are parametrized as
\be
A_i = \frac{g}{\pi}\left[\frac{A_i^1}{(1-2\lambda)} + \frac{A_i^2}{(1-2\lambda)^2} + \frac{A_i^3}{(1-2\lambda)^2} \right] \, ,
\ee
with the numerical values $A_i^j$ are listed in Table \ref{tab.1}. 

Examining the fixed point structure of this system leads to the following observations:
\begin{enumerate}
	\item Including the new fermionic term proportional to $\alpha$ gives rise to a new beta function $\beta_\alpha$. This is cubic in $\alpha$ for a fixed $\lambda$, $g$ and thus guarantees that there is at least one real solution to the equation $\beta_\alpha = 0$.
	\item As already discussed in Sect.\ \ref{sect:beta3}, the coefficient $A_0$ in $\beta_\alpha$ is generically non-zero. Thus $\alpha = 0$ is not a root of $\beta_\alpha$. As a consequence, any fixed point discussed in the previous section has to generalize to a fixed point where $\alpha_* \not= 0$. 
	\item When investigating the transition from the case of pure gravity, $N_f = 0$, to the inclusion of a small number of fermions, $N_f \ll 1$, one finds that the fixed point from the minimally coupled case splits into 3 families of NGFPs that are distinguished by their value of $\alpha_*$. In addition to these families there is a fourth one coming in from $\alpha_* \to -\infty$.
	\item When $N_f$ is increased to $N_f \approx 3$ the NGFPs from $\alpha_* \to -\infty$ and one family emanating from the pure-gravity fixed point annihilate.  The remaining two families of NGFPs extend to arbitrary values of $N_f$. These solutions will be named NGFP$^A$ and NGFP$^B$ (Family $A$ and Family $B$).
\end{enumerate}

\begin{figure}[t]
	\includegraphics[width=0.48\textwidth]{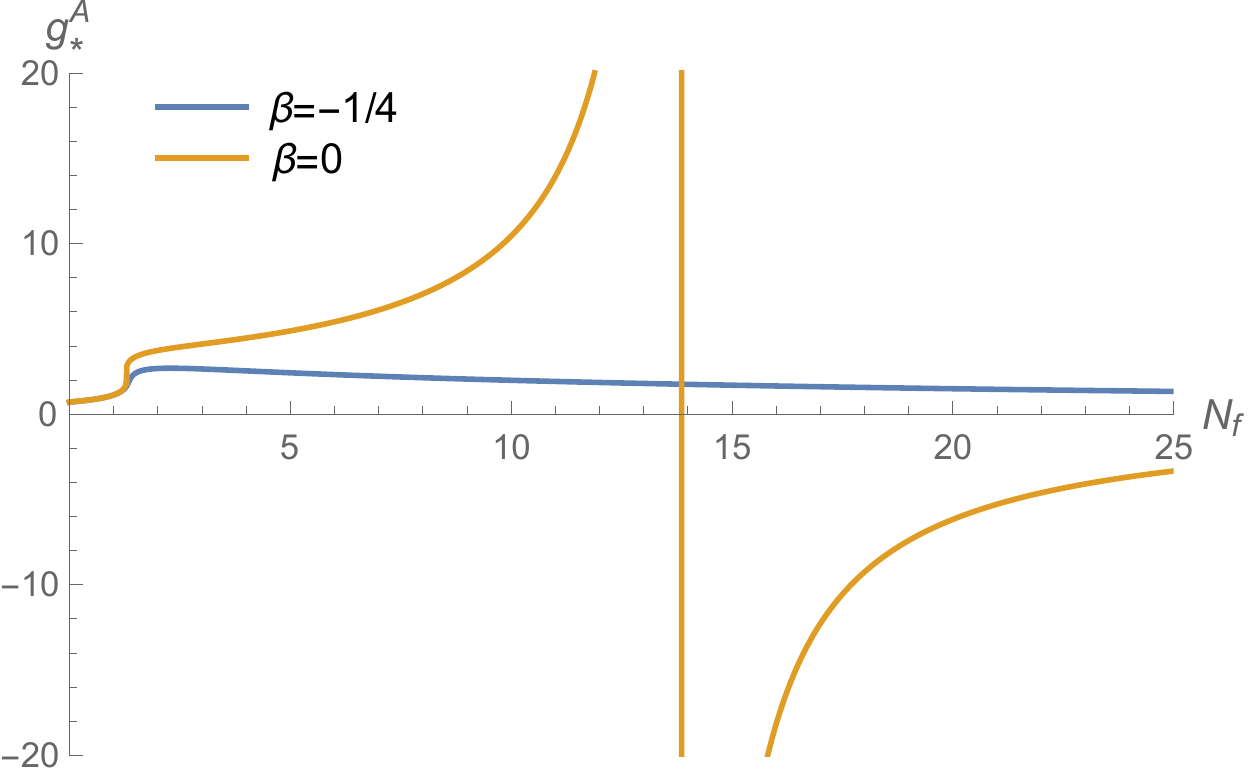}
	\includegraphics[width=0.48\textwidth]{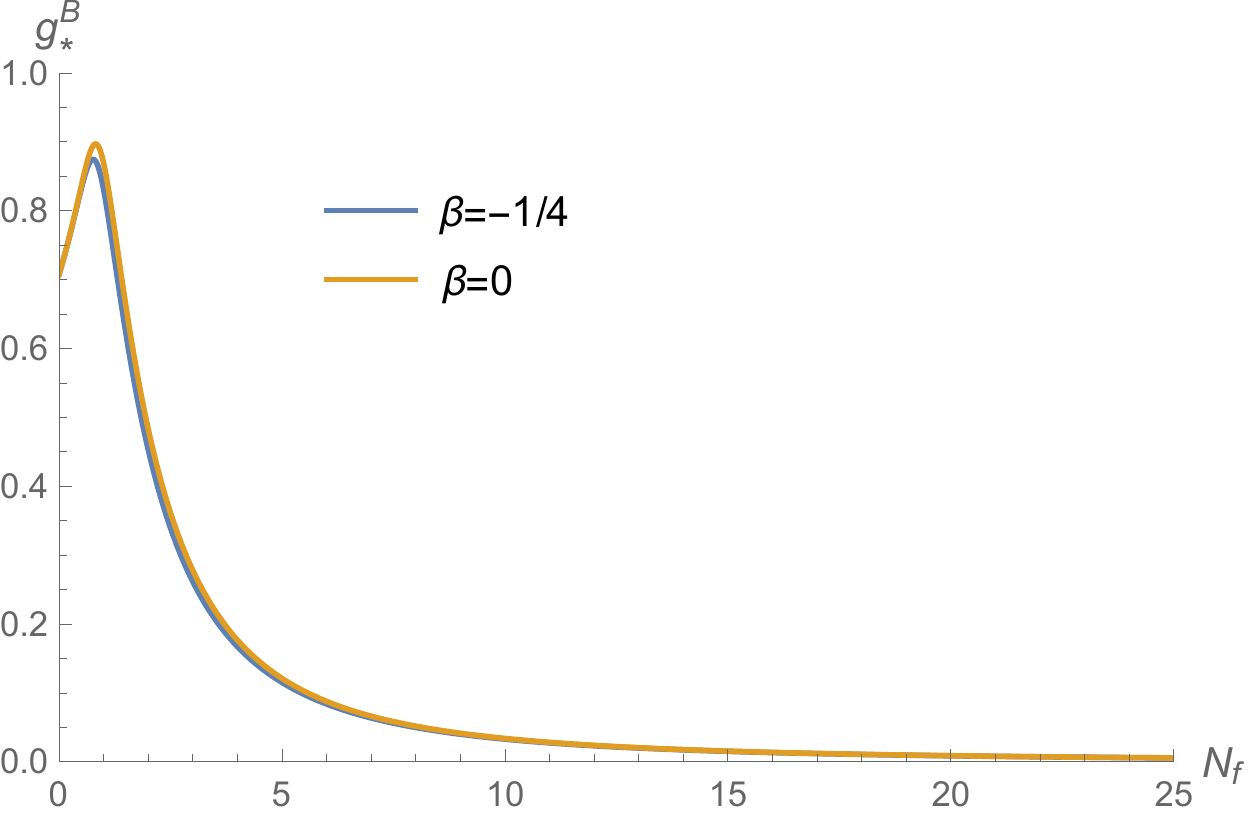}
	\includegraphics[width=0.48\textwidth]{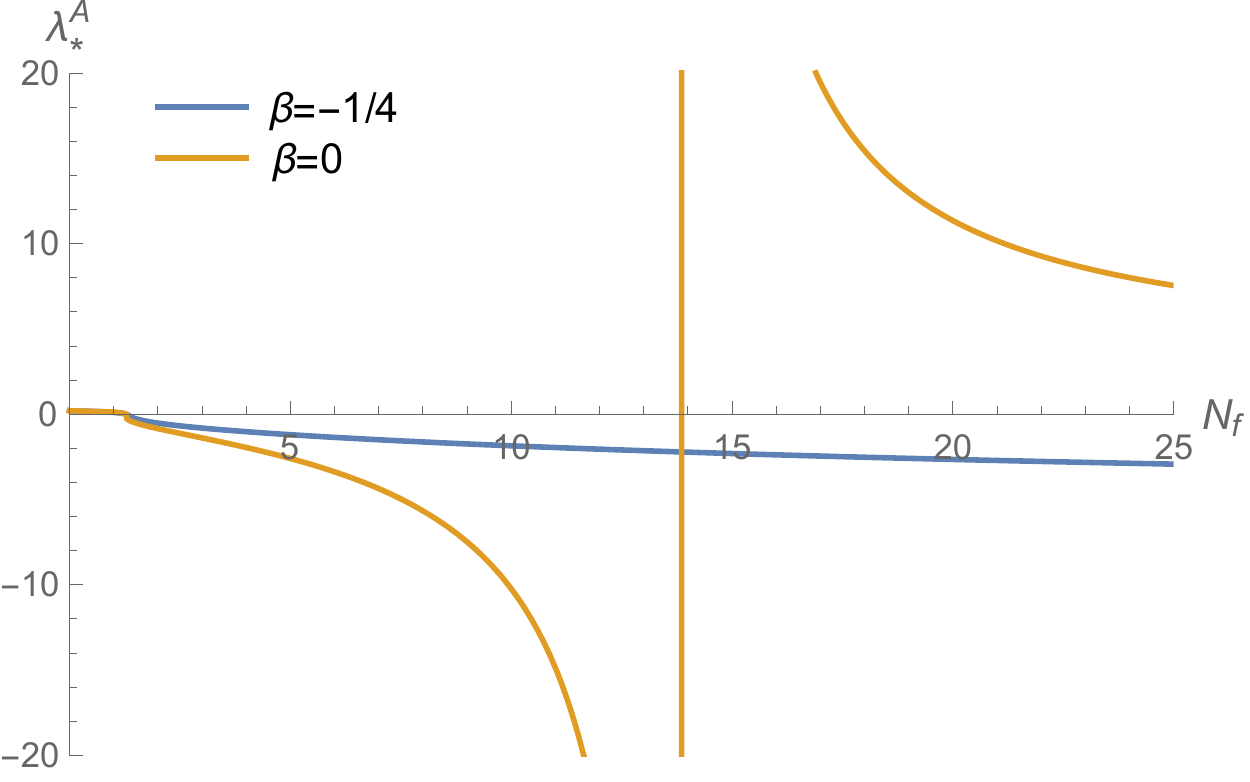}
	\includegraphics[width=0.48\textwidth]{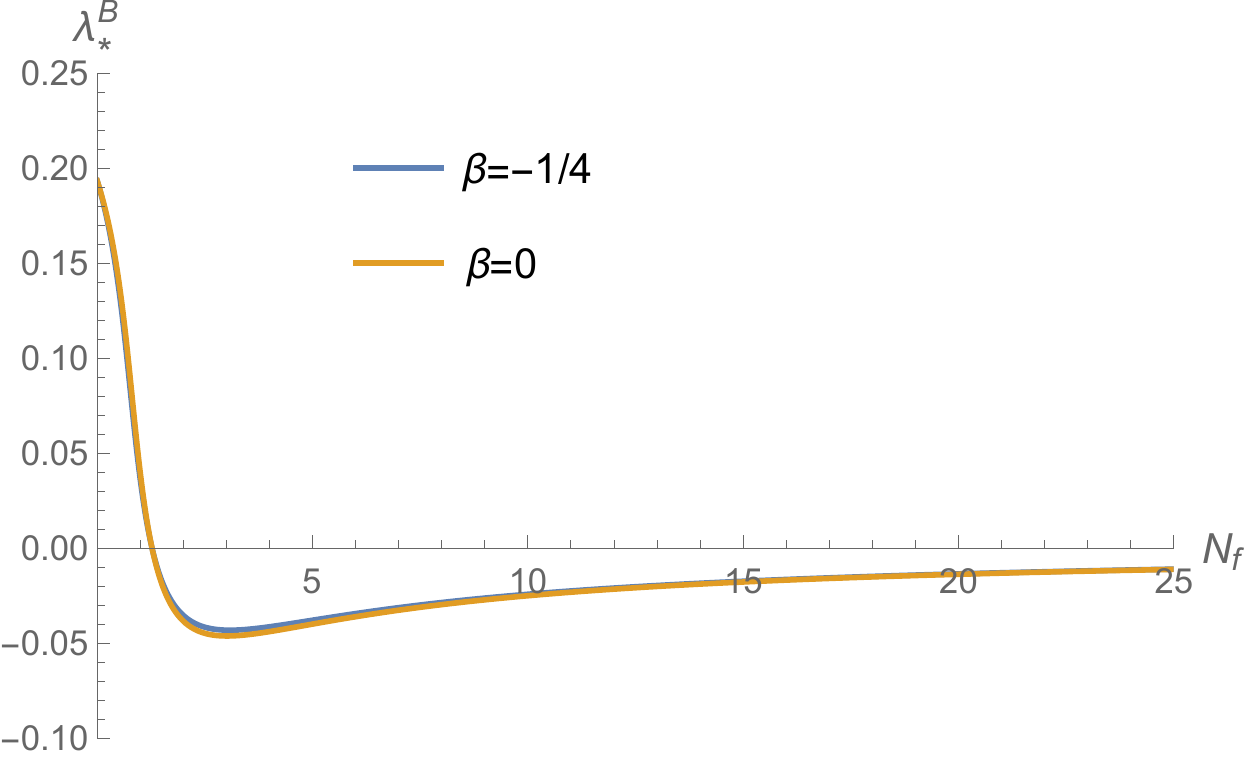}
	\includegraphics[width=0.48\textwidth]{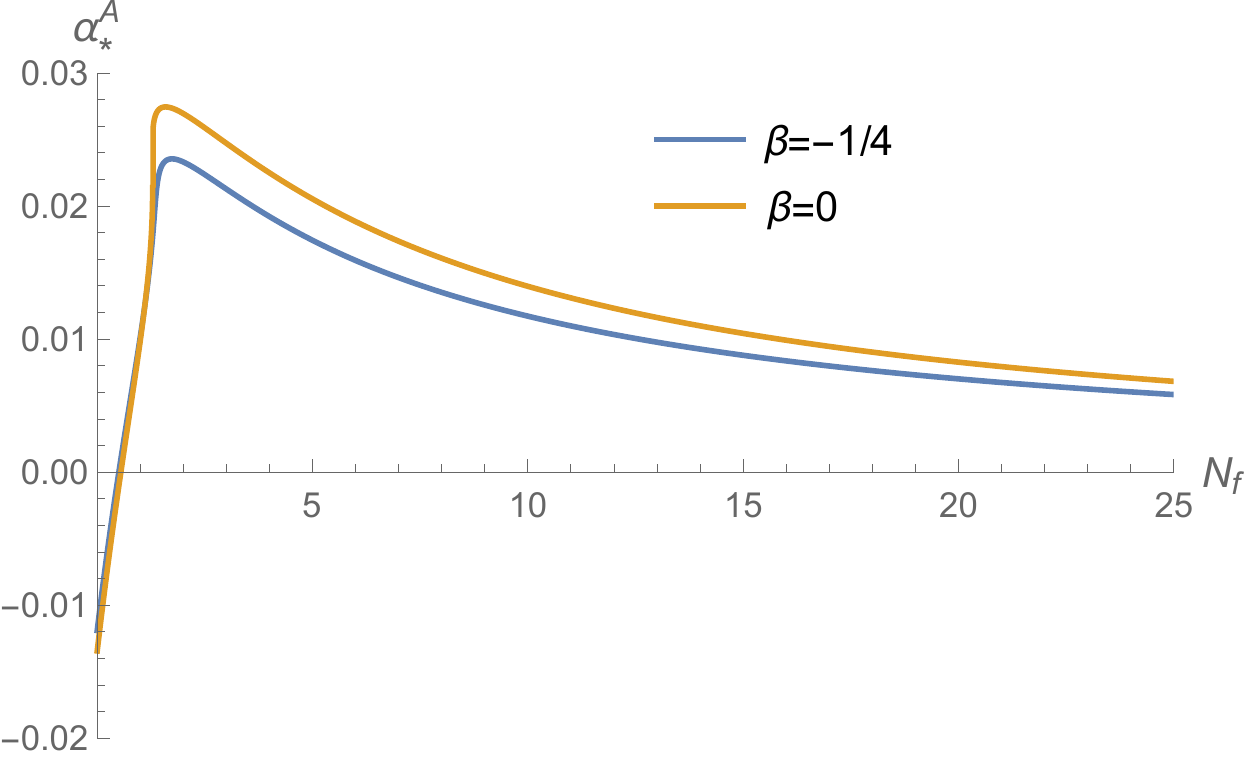}
	\includegraphics[width=0.48\textwidth]{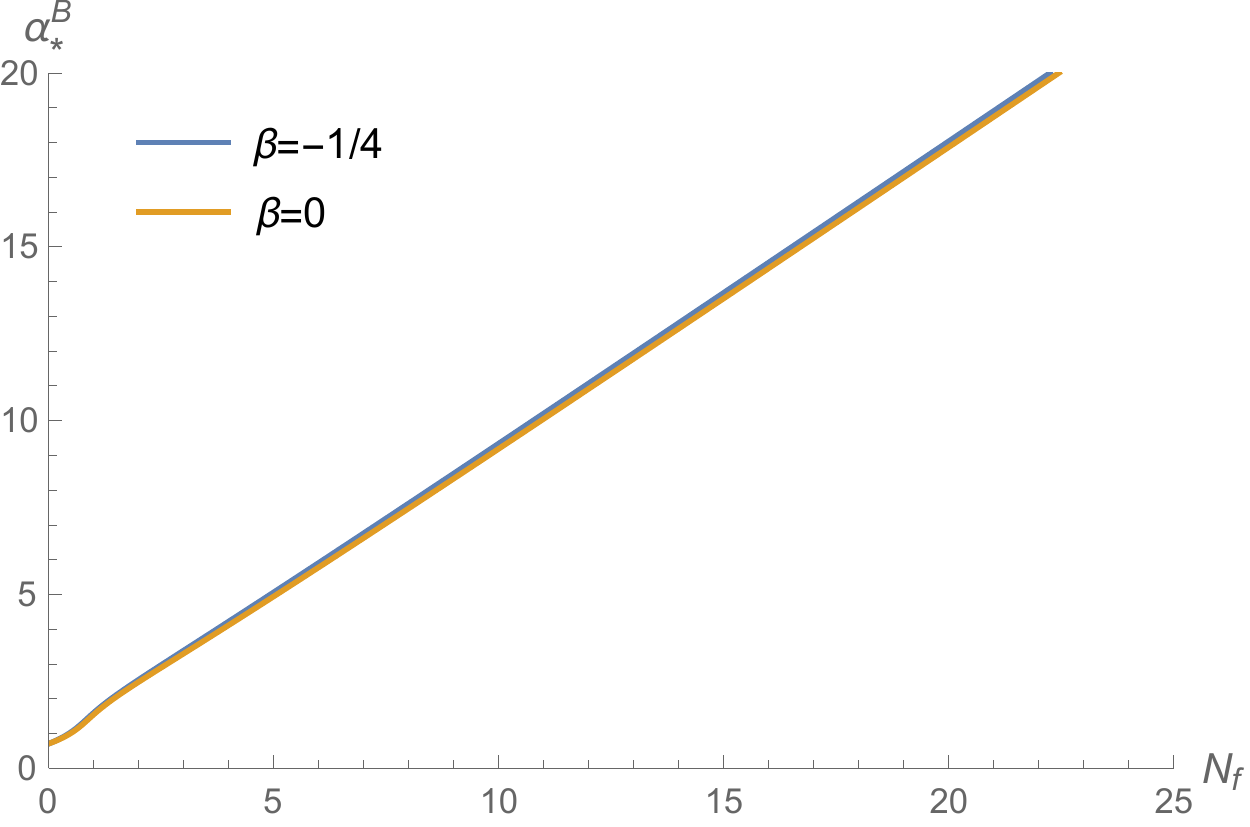}
	\caption{\label{NMCfps} The position of the fixed points as a function of $N_f$ is shown for the system \eqref{betafnctsNMC}. The blue and yellow lines indicate the difference in $\beta$, and therefore in the coarse-graining operator. Family $A$ is characterized by $\alpha_*^A \approx 0$ and structurally resembles the fixed points found at minimal coupling, cf.\ Fig.\ \ref{EHFPs}. For $\beta = 0$ there is a critical number of fermions $N_f^{\text{crit}}$ that changes the sign of $g_*^A$. Family $B$ on the other hand is characterized by $\alpha_*^B$ that grows proportional to $N_f$. This in turn makes it so that $\alpha_*^B \gg \beta$ and $g_*^B$ is always positive.} 
\end{figure}

\begin{figure}[t]
	\includegraphics[width=0.48\textwidth]{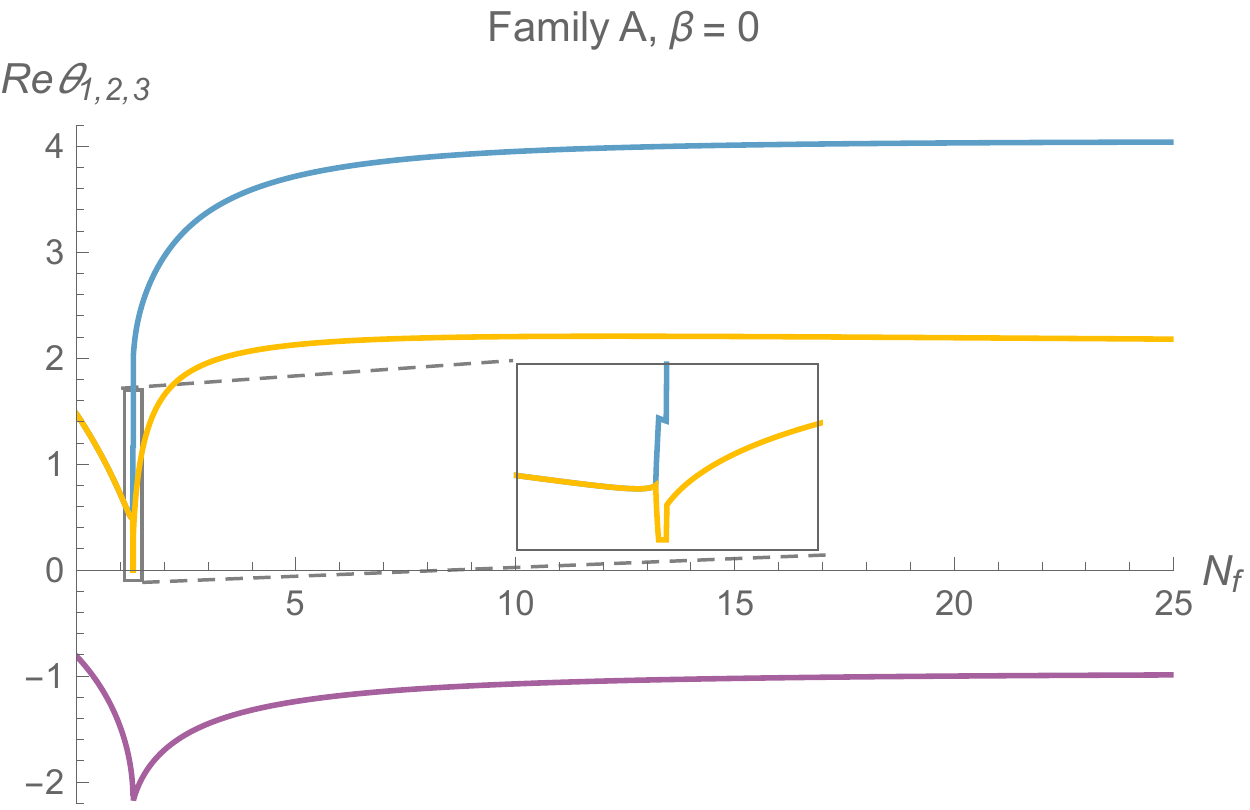}
	\includegraphics[width=0.48\textwidth]{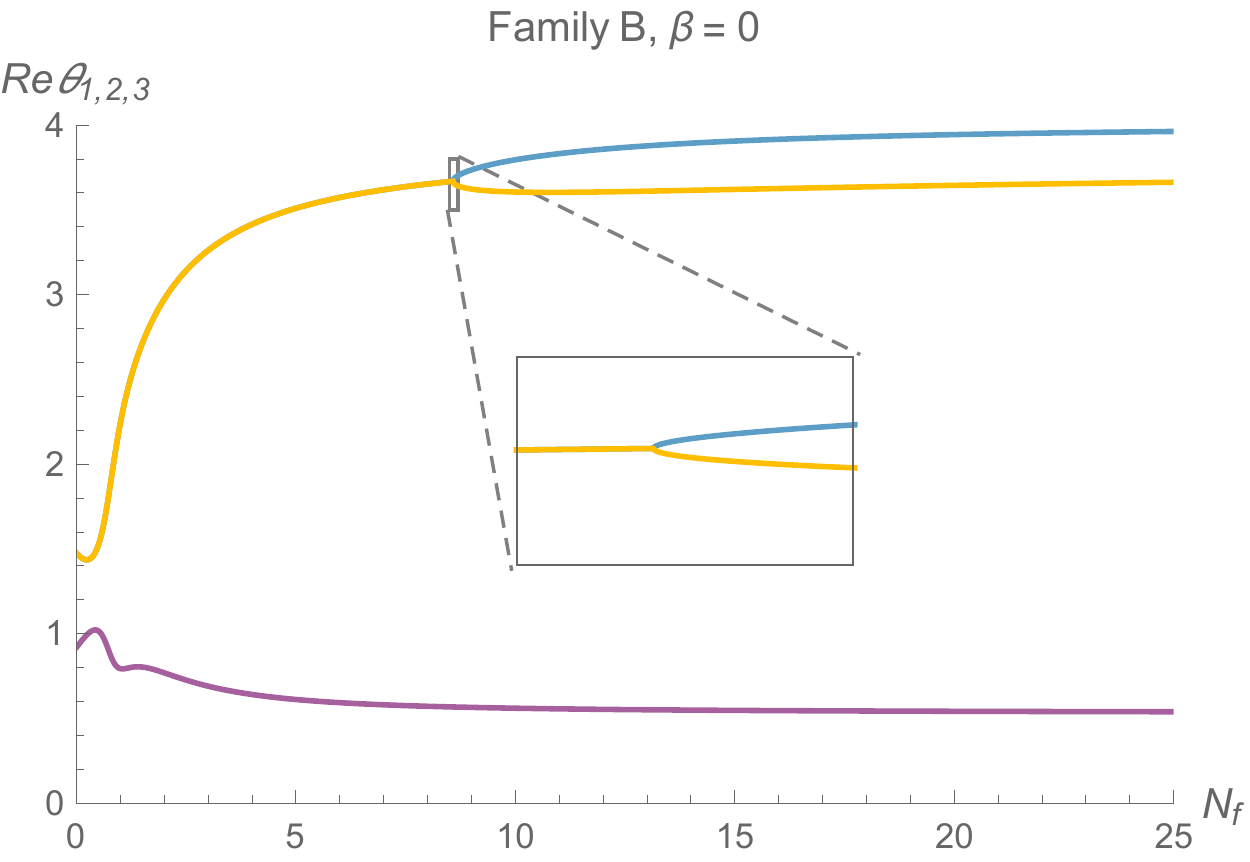}
	\includegraphics[width=0.48\textwidth]{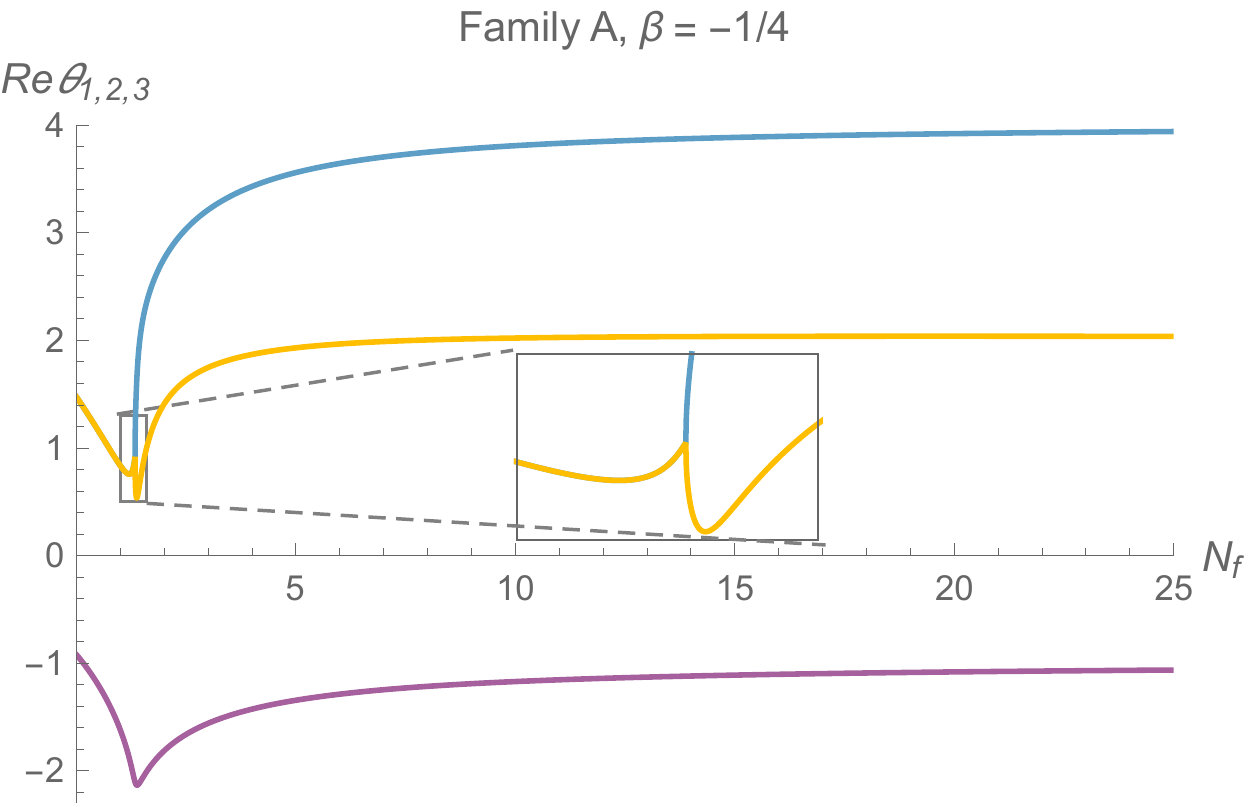}
	\includegraphics[width=0.48\textwidth]{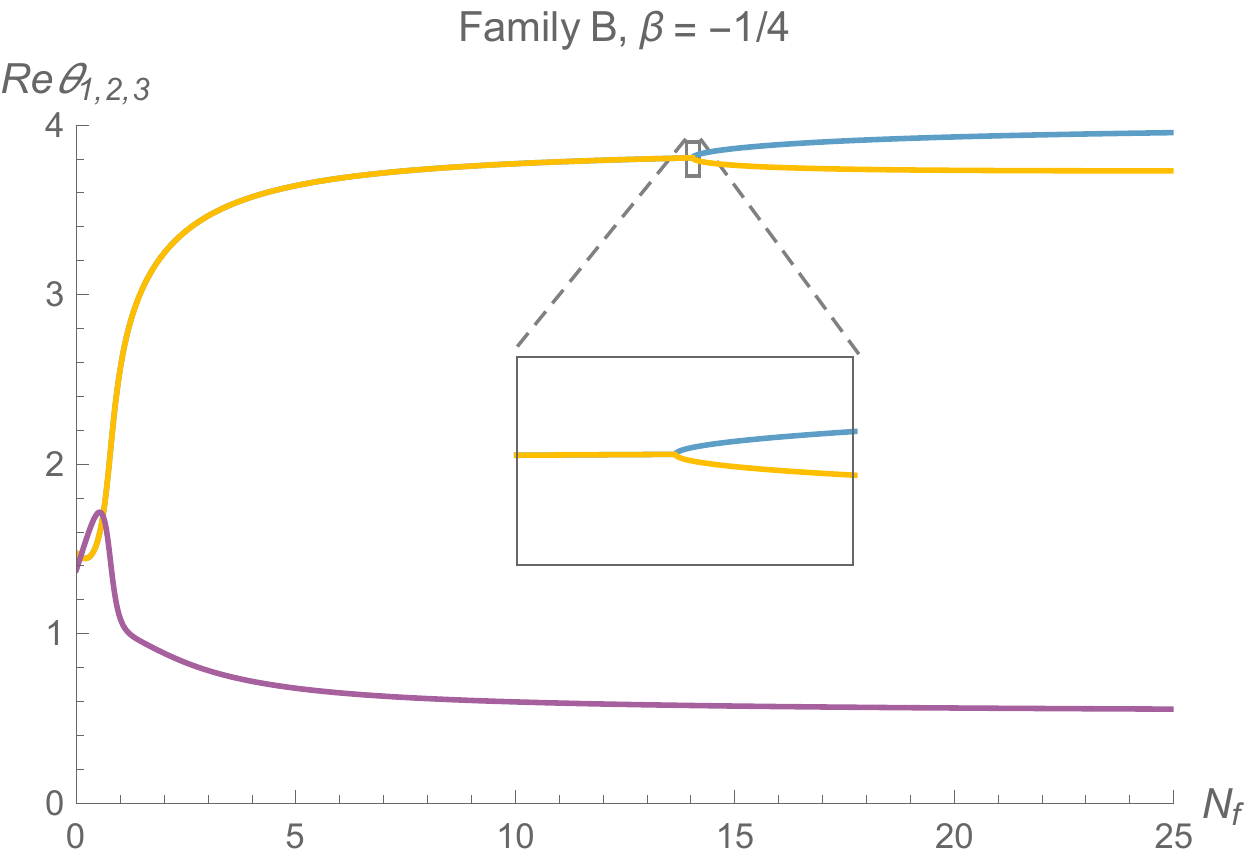}
	\caption{\label{NMCCritExps} The critical exponents associated with the NGFPs comprising Family $A$ and Family $B$ for both values of $\beta$. The families come with two and three relevant directions, respectively. This is independent of the choice of coarse-graining operator and $N_f$. Furthermore, the critical exponents become real at $N_f^A \approx 1.34$ and $N_f^B \approx 14$ in the case that $\beta = -1/4$ whereas they become real at $N_f^A \approx 1.29 $ and $N_f^B \approx 8.6$ in the case that $\beta = 0$.} 
\end{figure}

The position and stability properties associated with Family $A$ and Family $B$ are shown in Figs.\ \ref{NMCfps} and \ref{NMCCritExps}, respectively. The two families are distinguished by their value of $\alpha_*$. Family $A$ has the characteristic feature that $\alpha_*^A \ll 1$ while Family $B$ is characterized by $\alpha_*^B \propto N_f$. Family $A$ therefore corresponds to an (almost) chiral fixed point. 
Comparing the fixed point positions shown in Fig.\ \ref{EHFPs} and $g_*^A$ and $\lambda_*^A$ reveals that these exhibit a very similar behavior. Most notably, $g_*^A$ is again sensitive in the same way to the choice of $\beta$: there is a critical number $N_f^{\text{crit}}$ where $g_*^A$ switches sign. The similarity between these two cases is understood by noting that $\alpha_* = 0$ is a good approximation to the case where $\alpha_*^A \ll 1$. This suggests that the NGFPs found at minimal coupling should be identified with the fixed points comprising Family $A$ once the projection is extended to include $\alpha$.

Family $B$ is situated such that $g_*^B > 0$ for all values of $N_f$. This entails that there is no critical number of fermions $N_f^{\rm crit}$ for Family $B$. This can be understood from the fixed point condition $\eta_N^* = -2$. Considering the fact that $\alpha_*^B > 0$ continuously grows proportional to $N_f$ one sees that the contribution of $\alpha$ dominates over the contribution from the regulator, $\beta$, as $\alpha_*^B \gg \beta$. With this, one finds that $g_*^B$ is always positive. The large values for $\alpha_*^B$ indicates that Family $B$ constitutes a new class of NGFPs where chiral symmetry is broken by the coupling to gravity. Owed to the fact that this class comes with $\alpha_*^B \not = 0$ it is clear that this family may not be seen in computations which do not include the coupling $\alpha$ in the projection.

The critical exponents for both families are shown in Fig.\ \ref{NMCCritExps}. This  shows that Family $A$ and Family $B$ come with two and three relevant directions, respectively. This result is independent of the values for $N_f$ and $\beta$. Consequently, a high-energy completion based on NGFP$^A$ has the power to predict the value of $\alpha$ appearing in the effective action $\Gamma_{k=0}$. Conversely, a high-energy completion based on NGFP$^B$ has this coupling as a free parameter which needs to be fixed by experimental observations. Besides this, one can see that the critical exponents become real at $N_f^A \approx 1.34$ and $N_f^B \approx 14$ in the case that $\beta = -1/4$ whereas they become real at $N_f^A \approx 1.29 $ and $N_f^B \approx 8.6$ in the case that $\beta = 0$.

The position of the NGFPs for large $N_f$ can again be obtained analytically by performing a scaling analysis within the fixed point equations. For NGFP$^A$ the leading terms in the expansion are given by
\be\label{largeNfalpha}
\begin{split}
 \quad g_*^A & \, \simeq - \tfrac{12\pi}{(\pi-2) \, \xi \, N_f} \, , \; \,
\lambda_*^A  \, \simeq \tfrac{\pi}{(\pi-2) \xi} \, , \; \,
\alpha_*^A  \, \simeq - 
\tfrac{1008 \pi^2 + 42 \pi (\pi-2) (16 - 15 \pi) \xi - 
	3 (\pi-2 )^2 (796 - 
	273 \pi) \xi^2}{224 \, (\pi(\xi-2) - 2 \xi)^3 \, N_f}
\, ,
\end{split}
\ee
while for NGFP$^B$ 
\be\label{largeNfalphaB}
\begin{split}
	\quad g_*^B & \, \simeq  \frac{\pi(3608-945 \pi)}{560 \, N_f^2} \, , \quad
	\lambda_*^B  \, \simeq- \frac{\pi}{12} \, \frac{3608-945 \pi}{560} \, \frac{1}{N_f} \, , \quad
	\alpha_*^B  \, \simeq  \frac{560}{3608-945 \pi} N_f
	\, .
\end{split}
\ee
These asymptotic formulas again match the behavior displayed in Fig.\ \ref{NMCfps}, obtained by numerical methods. Thus both Family $A$ and Family $B$ exist for arbitrary positive values $N_f$. Moreover, the system becomes weakly coupled as $N_f \rightarrow \infty$.

\emph{In summary, the inclusion of $\alpha$ leads to new fixed points. The fixed points {\rm NGFP}$^{\rm A}$ comes with $\alpha_* \ll 1$ and corresponds to an (almost) chiral fixed point. In addition there is a family of non-chiral fixed points {\rm NGFP}$^B$. The later family is not resolved by the projections analyzed in Sect.\ \ref{ssect:4.1}, since these do not track the chiral symmetry breaking coupling $\alpha$ discriminating the two families. The large-$N_f$ expansion shows that both families admit a weak-coupling limit, $\lim_{N_f \rightarrow \infty} g_* = 0$.}
%======================================================
\subsection{The complete system}
\label{ssect:4.4}
\begin{figure}[p!]
	\includegraphics[width=0.5\textwidth]{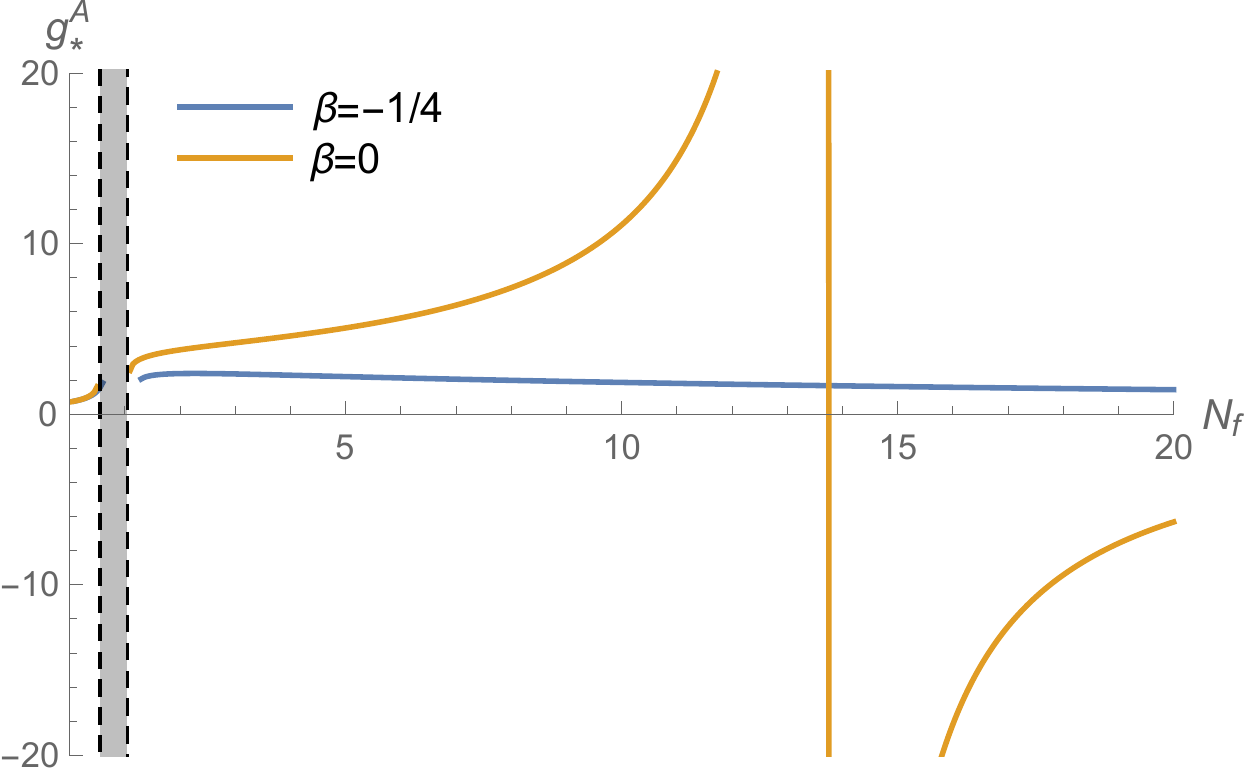}
	\includegraphics[width=0.5\textwidth]{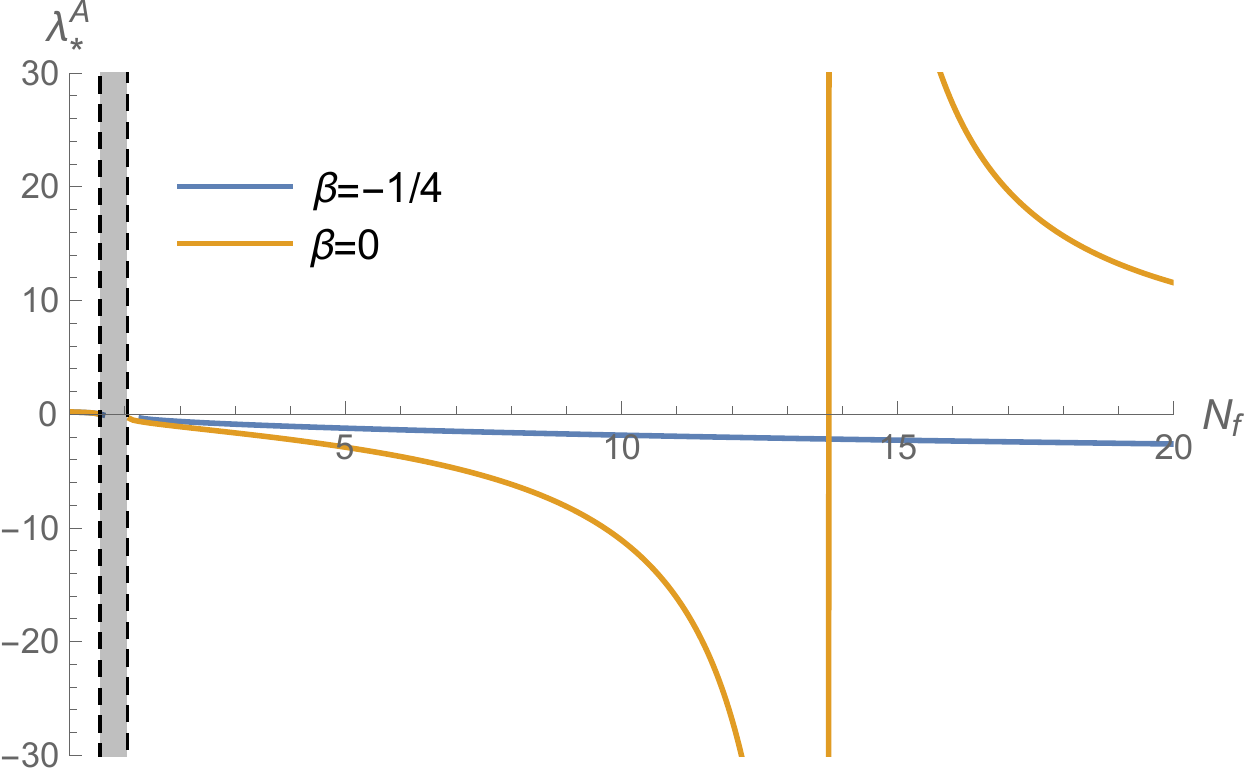} \\[3ex]
	\includegraphics[width=0.5\textwidth]{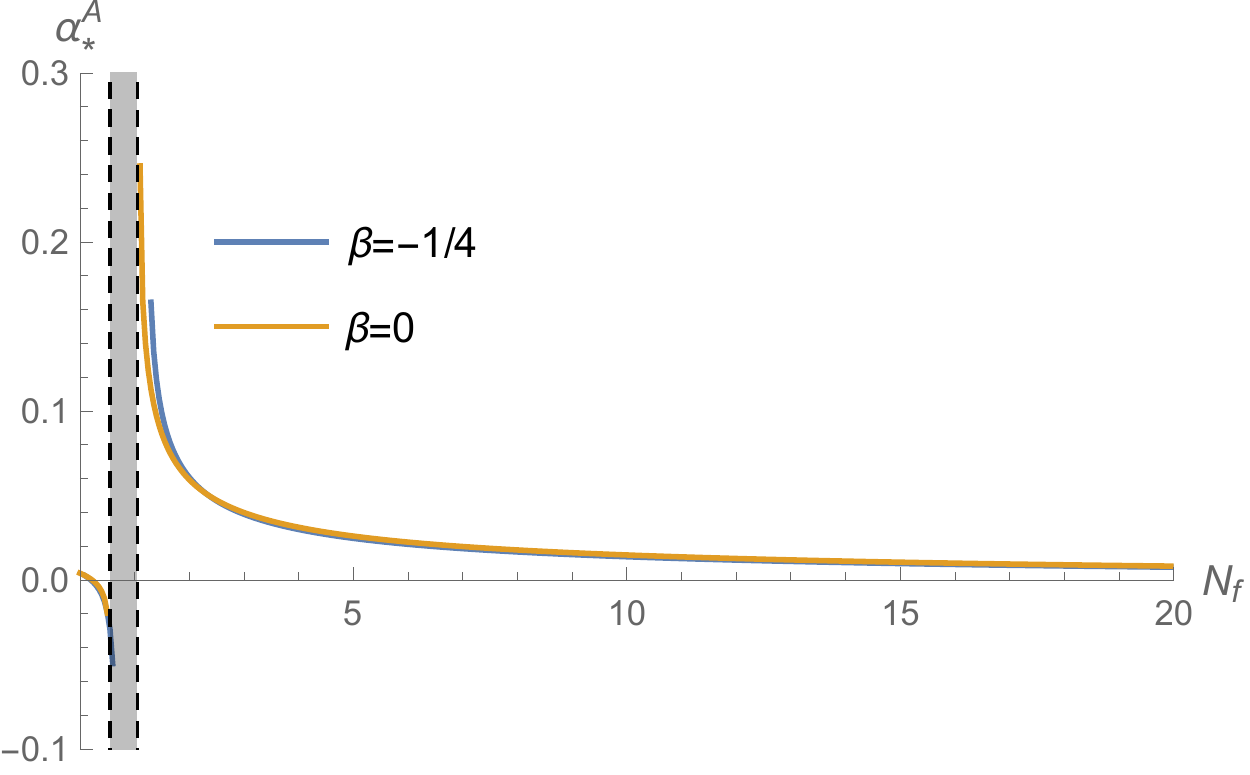}
	\includegraphics[width=0.5\textwidth]{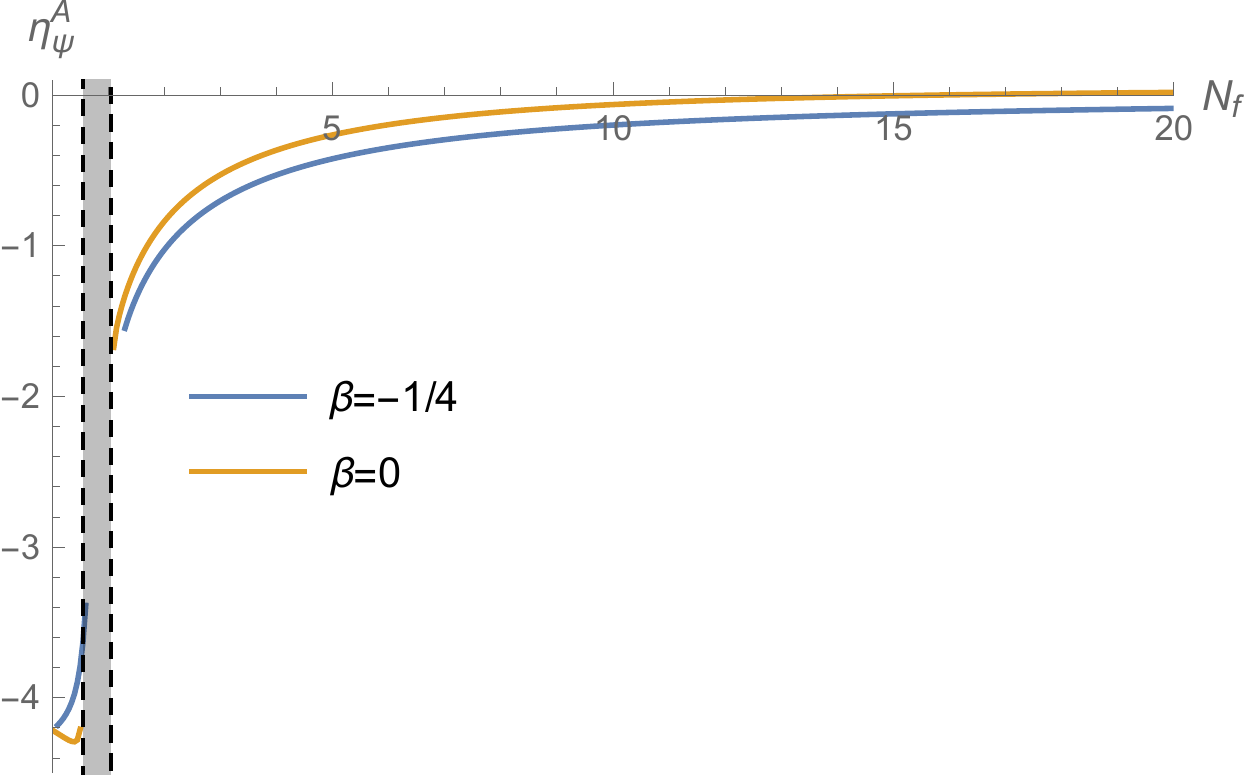} \\[3ex] 
	\includegraphics[width=0.5\textwidth]{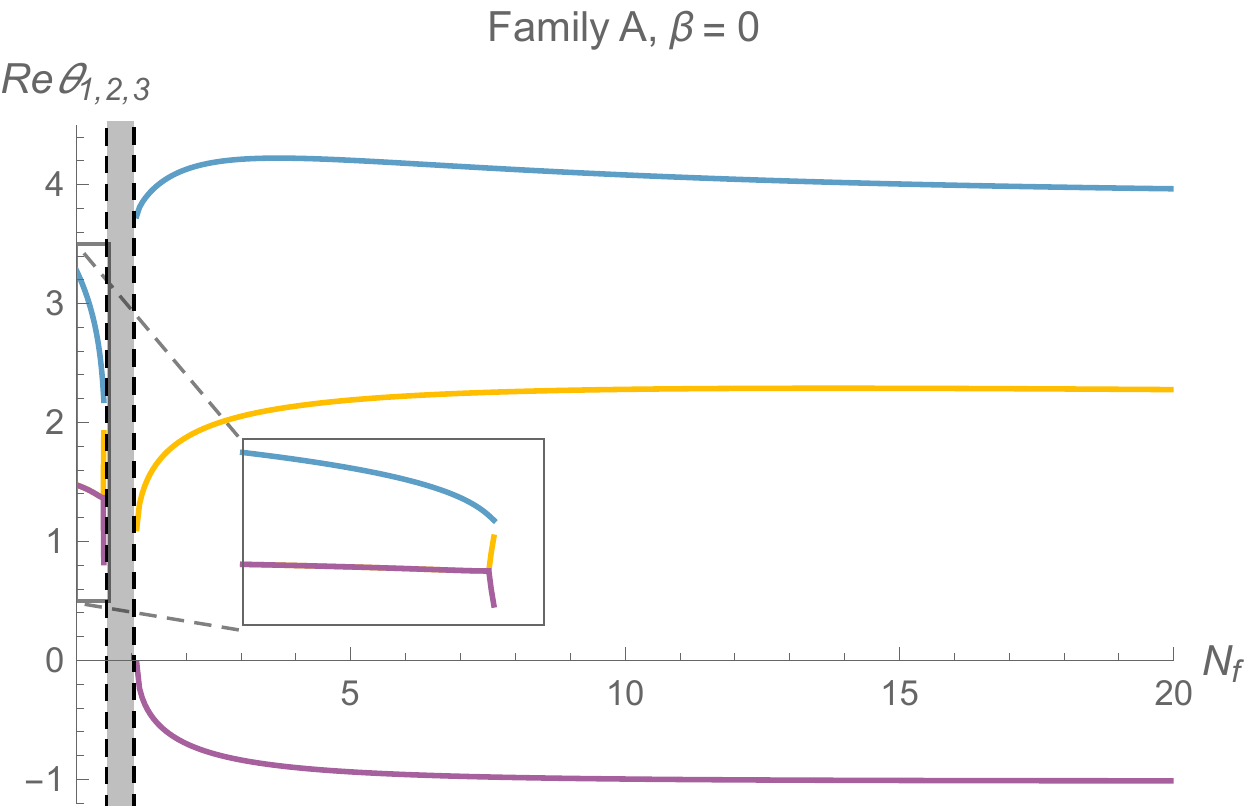}
	\includegraphics[width=0.5\textwidth]{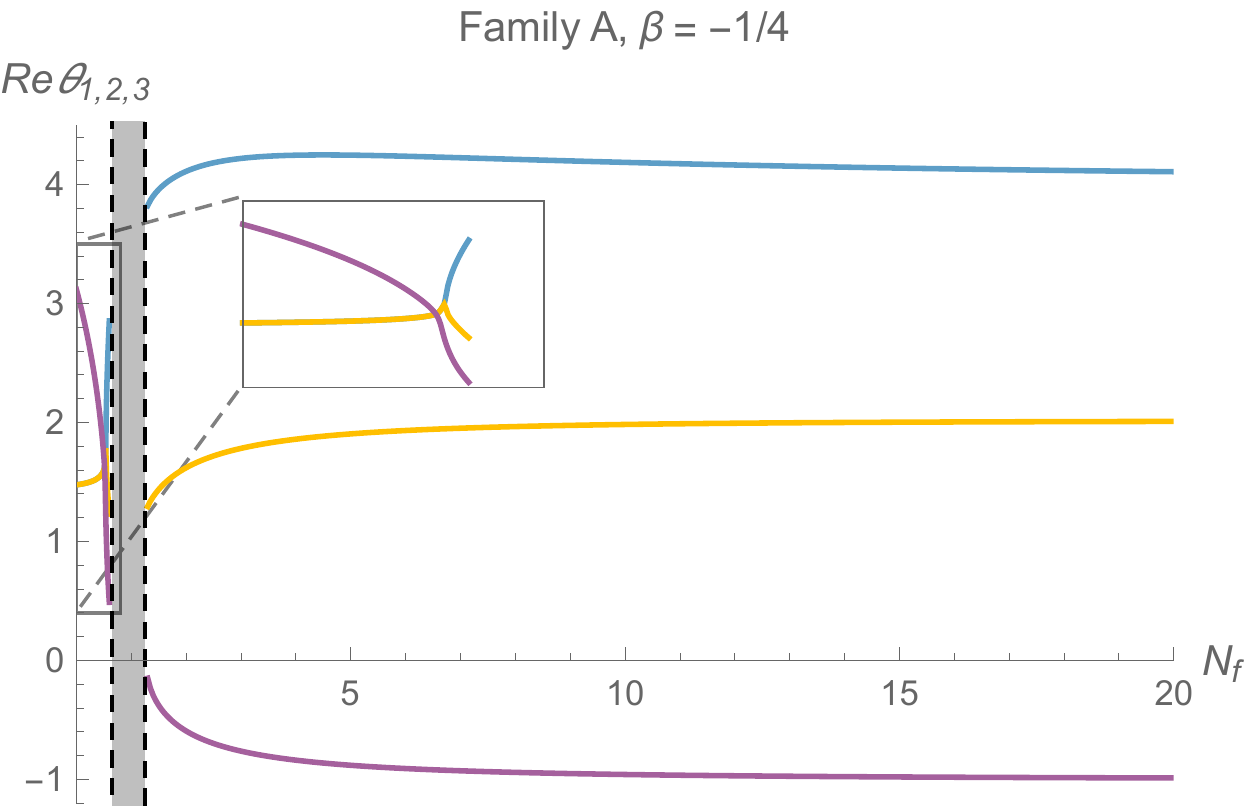} \\[2ex]
	\caption{\label{FullSysA} Results from the numerical investigation of the NGFPs belonging to Family $A$. Reading the diagrams from top-left to bottom right, the first three plots show the position of the fixed points as a function of $N_f$. The blue and yellow lines correspond to the choice in coarse-graining operator, $\beta = -1/4$ and $\beta = 0$, respectively. The fermion anomalous dimension $\eta_\psi^*$ is shown in the fourth diagram. The stability coefficients are plotted in the bottom row of the diagram indicating that the fixed points come with two UV-attractive and one UV-repulsive eigendirection once $N_f$ exceeds the upper bound of the window \eqref{complexwindow}.}
\end{figure}

\begin{figure}[p!]
	\includegraphics[width=0.5\textwidth]{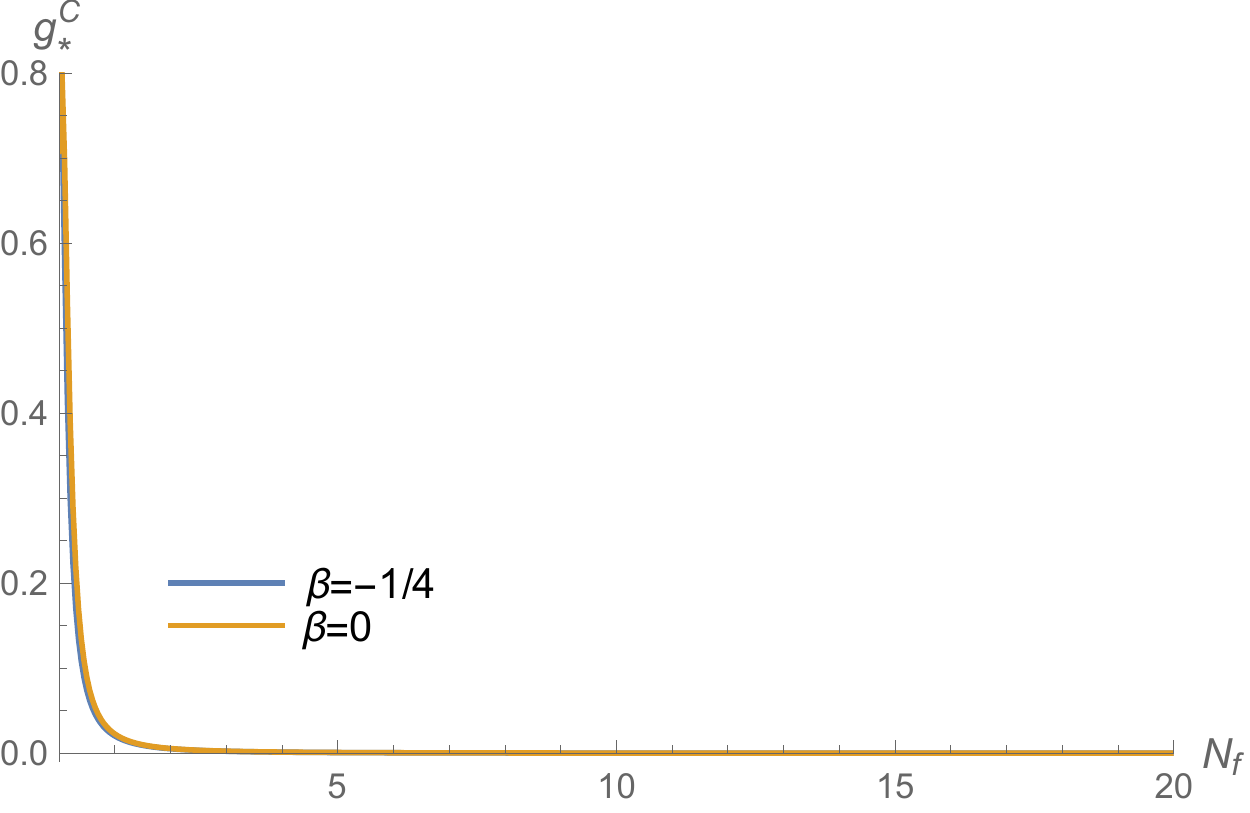}
	\includegraphics[width=0.5\textwidth]{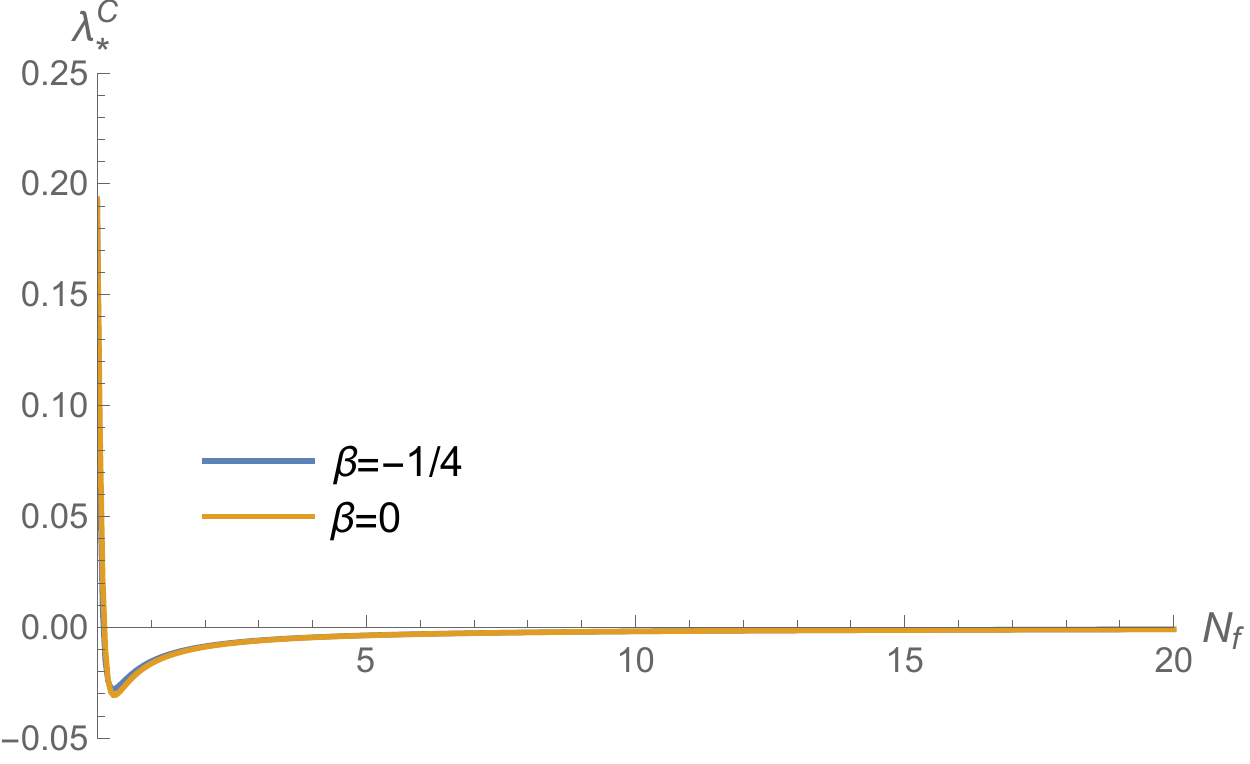} \\[3ex]
	\includegraphics[width=0.5\textwidth]{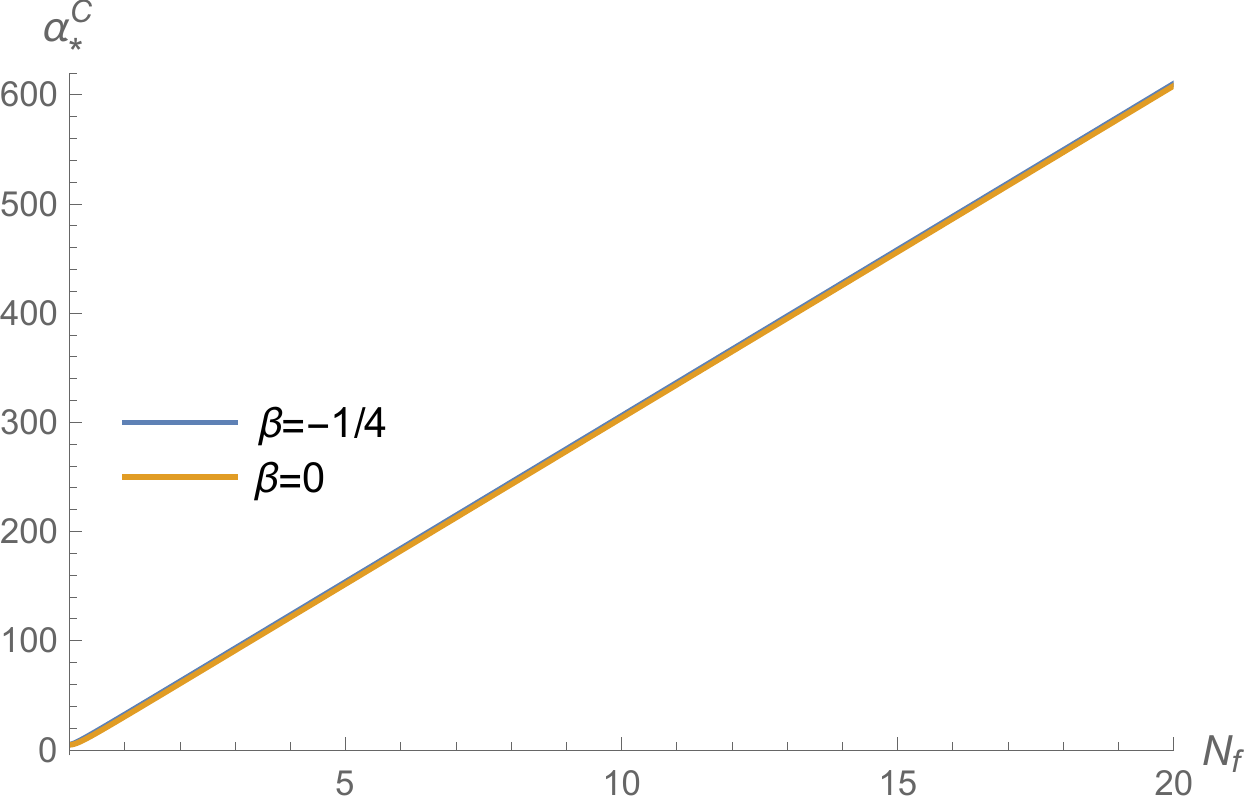}
	\includegraphics[width=0.5\textwidth]{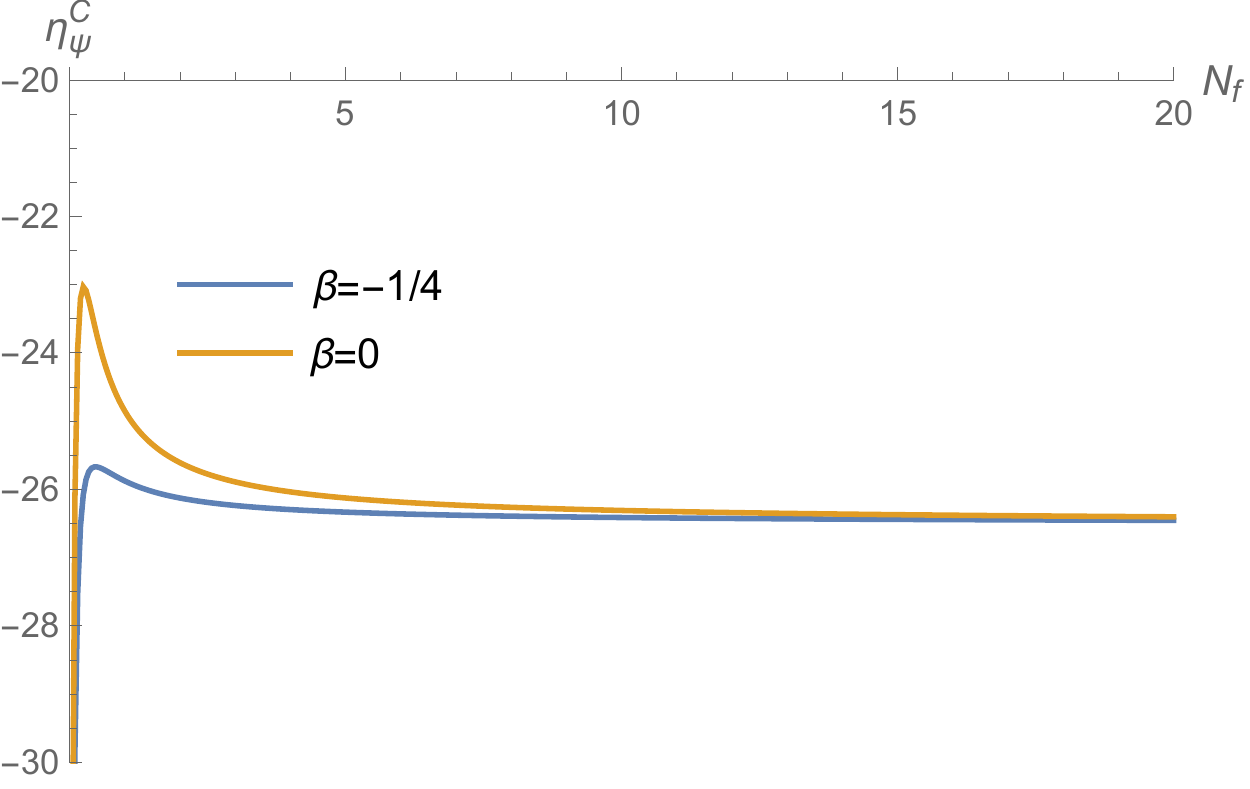} \\[3ex]
	\includegraphics[width=0.5\textwidth]{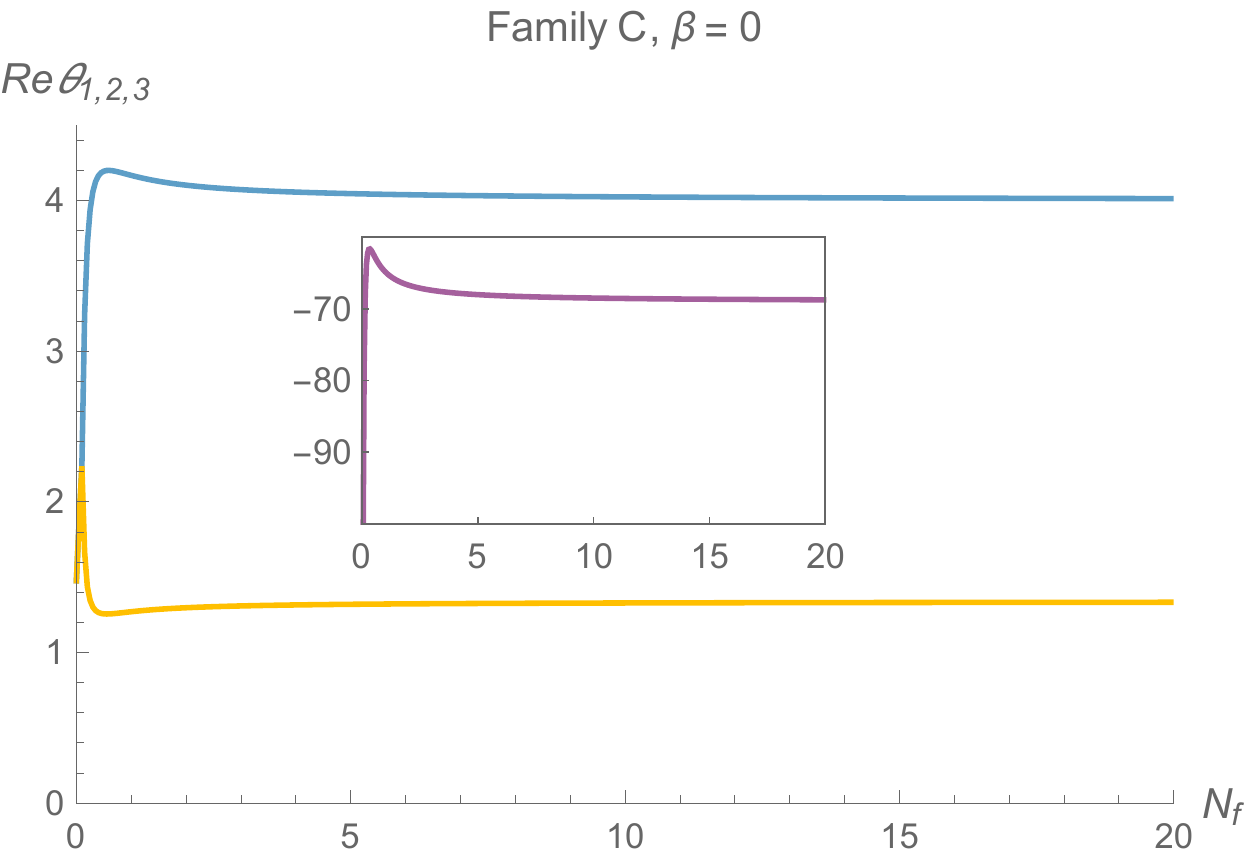}
	\includegraphics[width=0.5\textwidth]{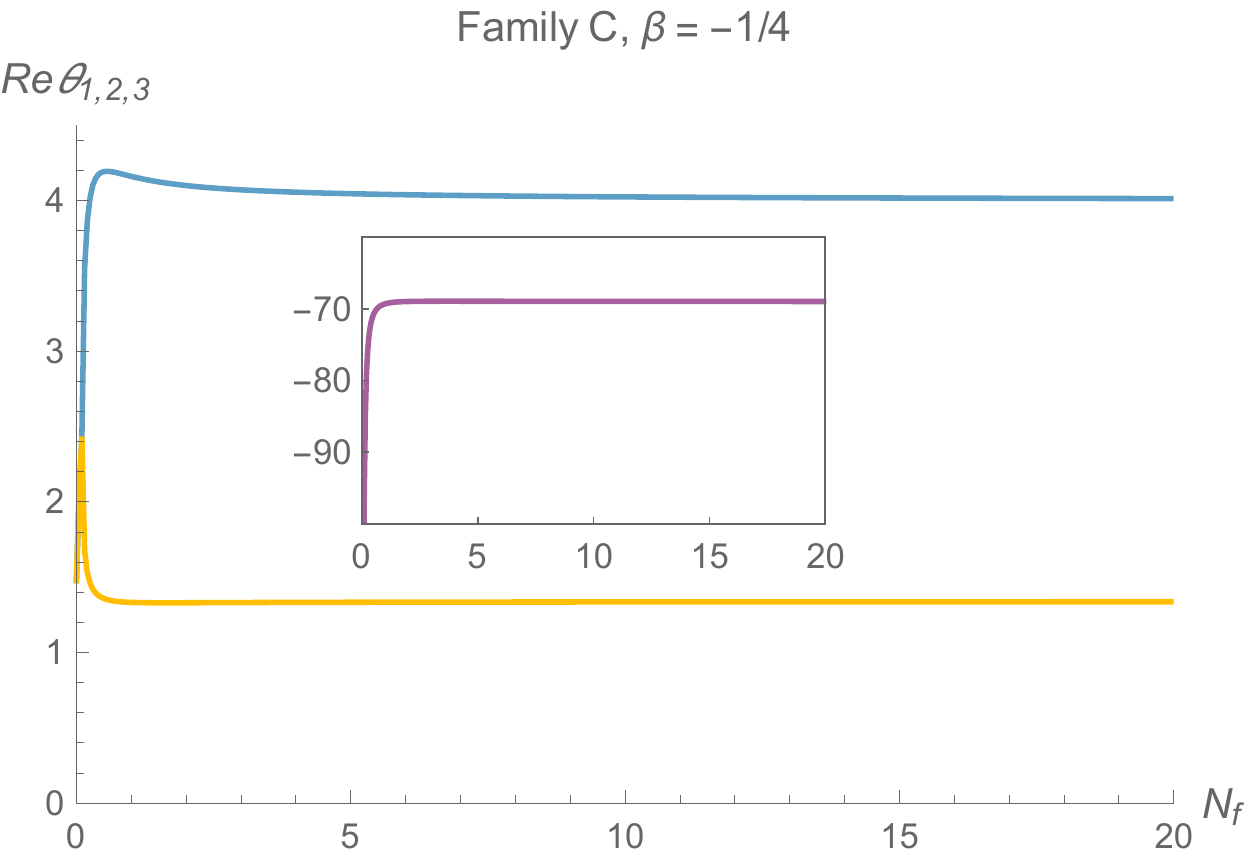} \\[2ex]
	\caption{\label{FullSysC} Characteristic properties of Family $C$ obtained from investigating the system eqs. \eqref{betagrav} and \eqref{betaalpha} numerically. The first three diagrams show the position of the fixed points as a function of $N_f$ for $\beta = -1/4$ (blue line) and $\beta = 0$ (orange line). NGFP$^{\rm C}$ exists for all values $N_f$ and is fairly insensitive to the choice of $\beta$. It is unaffected by the window \eqref{complexwindow}. The fermion anomalous dimension at the fixed points is depicted in the fourth diagram, showing that it is negative and rather large. The stability analysis identifies two UV-attractive and one UV-repulsive stability coefficient, indicating that NGFP$^{\rm C}$ is a saddle point in the $\{g,\lambda,\alpha\}$-plane.} 
\end{figure}

%======================================================
We now investigate the fixed point-structure of the complete set of beta functions \eqref{betagrav} and \eqref{betaalpha}. Our main finding is the existence
of three continuous families of NGFPs (denoted as NGFP$^A$, NGFP$^B$, and NGFP$^C$) whose properties are summarized in Figs.\ \ref{FullSysA}, \ref{FullSysB}, and \ref{FullSysC}, respectively. Families $A$ and $B$ exhibit the same qualitative features as the fixed points found in Sect.\ \ref{ssect:4.3}. Family $C$ is novel and ows its existence to the inclusion of $\eta_\psi$. While this new family of fixed points passes several non-trivial tests for its validity, the evidence supporting this class of fixed points is not on the same footing as the one for the other families which are observed in ``simpler'' truncations of the system already.

Before reporting on the results, two technical remarks are in order. First, eqs.\ \eqref{etaN} and \eqref{etapsi} are implicit equations determining the anomalous dimensions. Since they are linear in $\eta_\psi$ and $\eta_N$, it is straightforward to solve them leading to explicit expressions for $\eta_\psi$ and $\eta_N$ depending on the couplings $\{g,\lambda,\alpha\}$ and parameters ${\beta, N_f}$. The corresponding expressions are quite lengthy and little illuminating, so we refrain from reproducing them at this point. Secondly,  the inclusion of $\eta_\psi$ complicates the structure of the beta functions such that the {\tt NSolve}-algorithm provided by {\tt Mathematica} is unable to provide a complete list of fixed points for fixed values of ${\beta, N_f}$. Consequently, we adapted our search strategy as follows. In the first step, we performed a detailed search for fixed points at $N_f = 10^{-4}$ and $N_f = 20$, systematically varying the seeds of {\tt Mathematica}'s {\tt FindRoot}-routine on a three-dimensional cube spanned by $\{g,\lambda,\alpha\}$. This resulted in two lists of solutions which served as seeds in the subsequent analysis. In order to reduce the numerical complexity of the search, the list obtained for $N_f = 10^{-4}$ was reduced to the fixed points emanating from the Reuter fixed point \eqref{ReuterFP}, but coming with arbitrary values for $\alpha_*$. The seeds at $N_f = 10^{-4}$ ($N_f = 20$) were then extended towards increasing (decreasing) values of $N_f$. In this way our search algorithm covered the parameter space $N_f \in [0,20]$ twice. This allowed to eliminate outliers and spurious solutions appearing in the numerical search and resulted in the robust picture of the fixed point structure detailed below. \\

\begin{figure}[t!]
	\centering
	\includegraphics[width=0.5\textwidth]{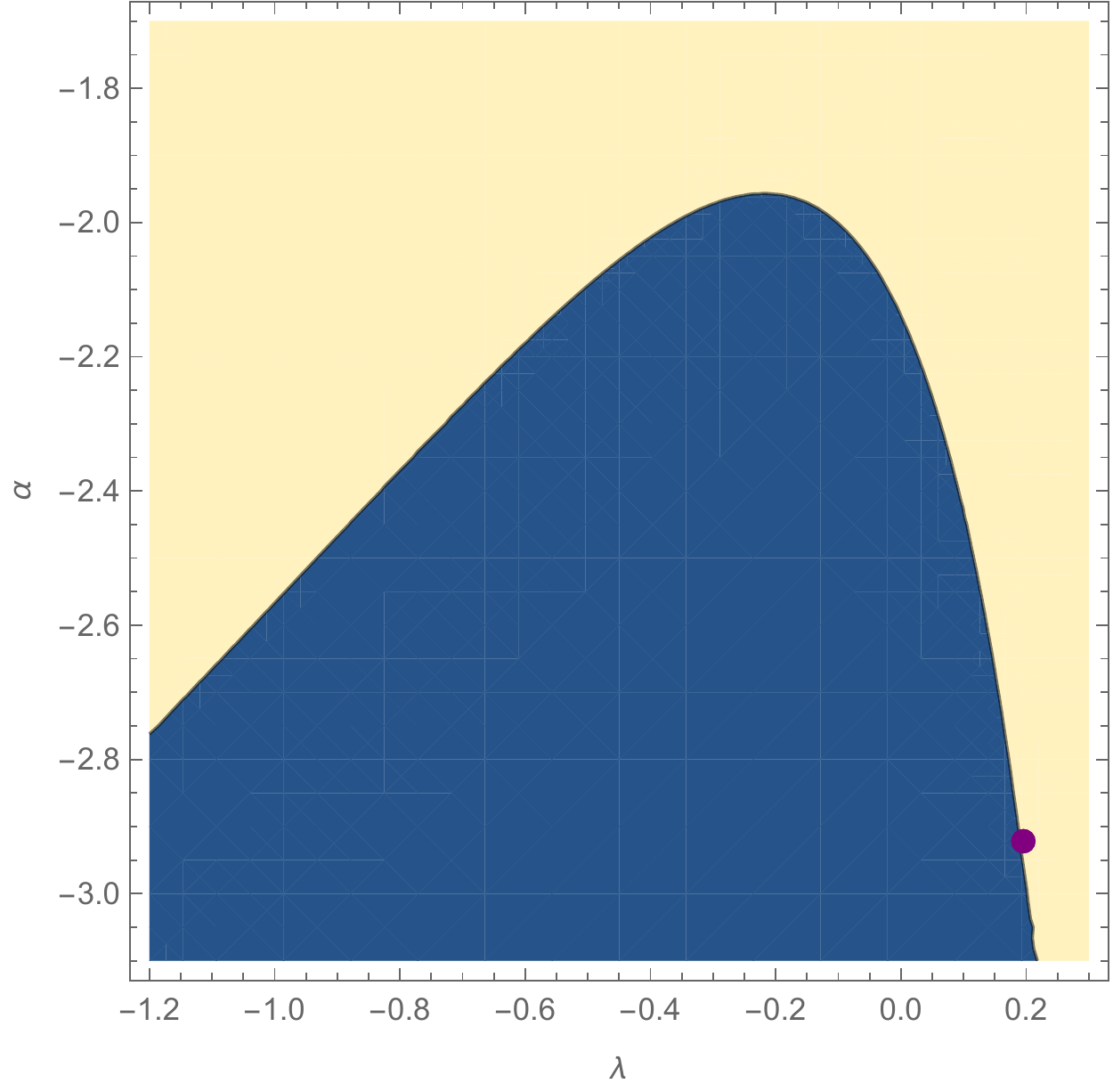}
	\caption{\label{gsqrt} In the presence of the  the fermion anomalous dimension $\eta_\psi$, the fixed point condition $\eta_N^* = -2$ constitutes a quadratic equation for $g_*$ whose solution includes a square root. The figure displays the sign of its argument as a function of $\lambda$ and $\alpha$ for fixed parameters $N_f = 0.05$ and $\beta = 0$. Negative values are marked in blue while a positive argument is obtained in the sandy region. The purple dot represents the fixed point from Family $B$. As the value of $N_f$ increases, the fixed point moves into the blue region where the argument of the square root is negative.} 
\end{figure}
\noindent
\emph{Merging fixed points - an algebraic consideration} \\
The analysis of the fixed point structure conveniently starts with the following observation. The condition $\beta_g = 0$ together with the requirement $g_* \not = 0$ requires $\eta_N^* = -2$. The latter relation can be solved explicitly for $g_*$ which then becomes a function of the remaining couplings $\{\lambda_*, \alpha_* \}$ and the parameters $\{N_f, \beta\}$. For $\eta_\psi = 0$, the condition $\eta_N^* = -2$ is linear in $g_*$ so that there is a unique solution. Including $\eta_\psi$ turns this relation into a quadratic equation for $g_*$. While this equation can still be solved algebraically, one now obtains two branches of solutions. The coordinates $\{g_*,\lambda_*\}$ obtained from the first branch connect continuously to the Reuter fixed point \eqref{ReuterFP}. It is this branch that is traced in the sequel.
The novel feature in the fixed point function $g_*(\lambda_*, \alpha_*; N_f, \beta)$ is the appearance of a square-root whose argument depends on $\{\lambda_*, \alpha_*, N_f, \beta\}$. Demanding that $g_*(\lambda_*, \alpha_*; N_f, \beta)$ is real then requires that the argument of the square-root is positive. This in turn leads to an additional constraint on the position of the NGFPs. The explicit form of this inequality is rather bulky and will not be given here. Instead we illustrate its content for a specific choice of parameters $\{\beta = 0, N_f = 0.05\}$ in Fig.\ \ref{gsqrt}. The collision of NGFP$^B$ with this region shifts the fixed points into the complex plane from which they re-emerge at larger values $N_f$.
\begin{figure}[t!]
	\centering
	\includegraphics[width=0.6\textwidth]{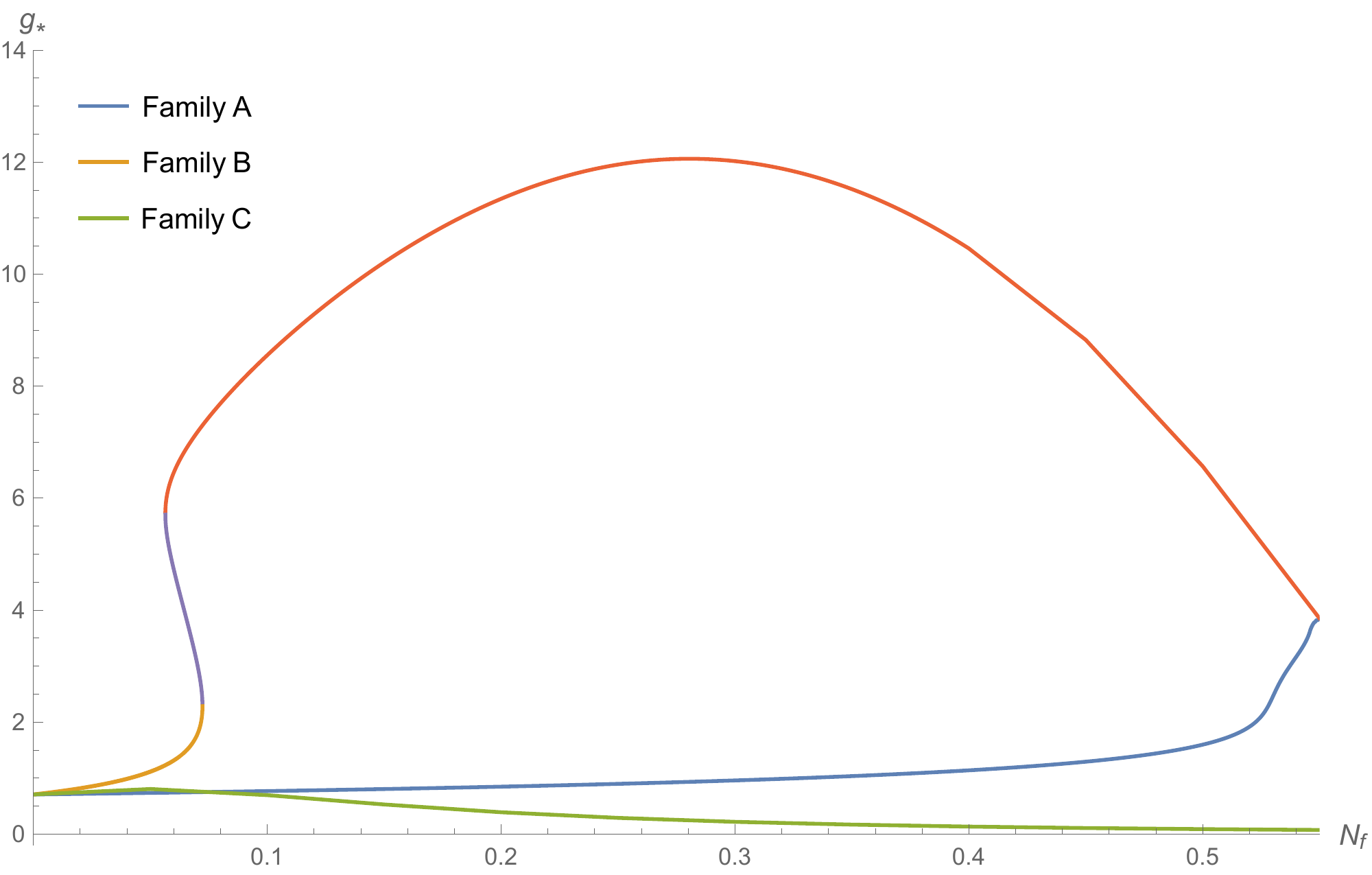}
	\caption{\label{manyFPfamilies} Illustration of the interplay between the NGFPs and the region of instability exemplified in Fig.\ \ref{gsqrt} ($\beta=0$). The figure gives the $g_*$ coordinate of various fixed point solutions for values $N_f$ below the window \eqref{complexwindow}.} 
\end{figure}

In addition, a detailed numerical investigation of the fixed point structure for $\beta = 0$ and values $N_f < 0.55$ shows an intricate pattern of fixed points moving into and out of the complex plane. This pattern is illustrated in Fig.\ \ref{manyFPfamilies}. Reading the figure from left to right, we first encounter three fixed points branching off the Reuter fixed point for $N_f \ll 1$. At $N_f \approx 0.06$ a new pair of fixed points emerges from the complex plane. At $N_f \approx 0.08$ the lower branch of this pair merges with the branch NGFP$^B$ and both fixed points vanish into the complex plane. The upper branch of the new pair merges with Family $A$ at the lower boundary given in \eqref{complexwindow}, $N_f \approx 0.55$. These fixed points eventually re-emerge at $N_f \approx 1.05$. The precise values of $N_f$ where these fixed points take a detour through the complex plane depends on the choice for $\beta$. For the two cases investigated in the present work this window is found at
	\be\label{complexwindow}
	\begin{split}
		\beta = -1/4: & \qquad 0.65 < N_f < 1.25 \, , \\
		\beta = 0: & \qquad 0.55 < N_f < 1.05 \, . 
	\end{split}
	\ee
Family $C$ is unaffected by the region of instability and exists for arbitrary values $N_f$.
The comparison with the results	obtained when setting $A_0 = 0$ (see the next section) where the window \eqref{complexwindow} is absent  suggests that it is the chiral symmetry breaking contribution $A_0$ which is responsible for creating this ``region of instability''. \\
\begin{table}[t!]
	\begin{tabular}{cccccccccc}
		%fixed point 
		&   $N_f$ \;  &   $\beta$ \;  & \; $\theta_1$  \; &  \; $\theta_2$  \; &  \; $\theta_3$  \; & \; $g_{*}$ \; & \; $\lambda_{*}$ \; & \; $\alpha_{*}$ \; & \; $\eta_{\psi}$ \; \\ \hline
		{NGFP$^{\rm A}$}
		& $3$ & $0$ & $4.21$ & $2.05$ & $- 0.84$ & $4.18$ & $-1.65$ & $0.04$ & $-0.53$ \\
		& $3$ & $-1/4$ &  $4.22$ & $1.78$ & $- 0.76$ & $2.35$ & $-0.89$ & $0.04$ & $-0.70$  \\
		& $20$ & $0$ & $3.96$ & $2.28$ & $-1.01$ & $-6.32$ & $11.6$ & $0.0081$ & $0.017$ \\
		& $20$ & $-1/4$ & $4.11$ & $2.01$ & $-0.99$ & $1.43$ & $-2.63$ & $0.0075$ & $-0.089$ \\ \hline
		{NGFP$^{\rm B}$}
		& $3$ & $0$ & $4.29$& $2.52$ & $0.68$ & $0.43$ & $-0.12$ & $1.86$ & $-0.95$ \\
		& $3$ & $-1/4$ &  $4.14$ & $2.53$ & $0.61$ & $0.44$ & $-0.12$ & $1.62$ & $-0.86$ \\
		& $20$ & $0$ & $4.40$ & $3.46$ & $0.58$ & $0.016$ & $-0.032$ & $8.95$ & $-0.65$ \\
		& $20$ & $-1/4$ & $4.38$ & $3.45$ & $0.58$ & $0.016$ & $-0.032$ & $8.74$ & $-0.62$ \\ \hline
		{NGFP$^{\rm C}$}
		& $3$ & $0$ & $4.07$ & $1.31$ & $-67.4$ & $0.0025$ & $-0.0061$ & $91.0$ & $-25.9$ \\
		& $3$ & $-1/4$ & $4.07$ & $1.33$ & $-68.9$ & $0.0024$ & $-0.0059$ & $93.0$ & $-26.2$ \\
		& $20$ & $0$ & $4.01$ & $1.33$ & $-68.7$ & $5.45 \cdot 10^{-5}$ & $-9.40 \cdot 10^{-4}$ & $608$ & $-26.4$ \\
		& $20$ & $-1/4$ & $4.01$ & $1.34$ & $-69.0$ & $5.42 \cdot 10^{-5}$ & $-9.36 \cdot 10^{-4}$ & $610$ & $-26.5$ \\ \hline
	\end{tabular}
	\caption{\label{Tab.3} Examples illustrating the generic properties associated with the three families of fixed points for selected values $N_f$ and $\beta$. The NGFP's of Family $A$ and $C$ constitute saddle points for the RG flow while NGFP$^B$ act as UV-attractors for the RG flow in the $g$-$\lambda$-$\alpha-$hyperplane. In contrast to the Reuter fixed point \eqref{ReuterFP}, all examples come with real stability coefficients.}
\end{table}
\noindent
\emph{Fixed points - numerical results} \\
We now summarize the key properties of the three continuous families of NGFPs identified by our numerical search algorithm. Explicit examples corresponding to specific choices of the parameters $N_f$ and $\beta$ are listed in Table \ref{Tab.3}. When plotting the $N_f$-dependence of the solutions, the blue and yellow lines appearing in the position plots correspond to the two choices of coarse-graining operators, $\beta = -1/4$ and $\beta = 0$, respectively. Gray bands mark the position for $\beta = 0$ of region \eqref{complexwindow} where NGFP$^{\rm A}$ and NGFP$^{\rm B}$ have been shifted to complex positions. \\

\noindent
NGFP$^{\rm A}$:
The properties of the non-Gaussian fixed points comprising Family $A$ are shown in Fig.\ \ref{FullSysA}. 
\begin{figure}[p!]
	\includegraphics[width=0.5\textwidth]{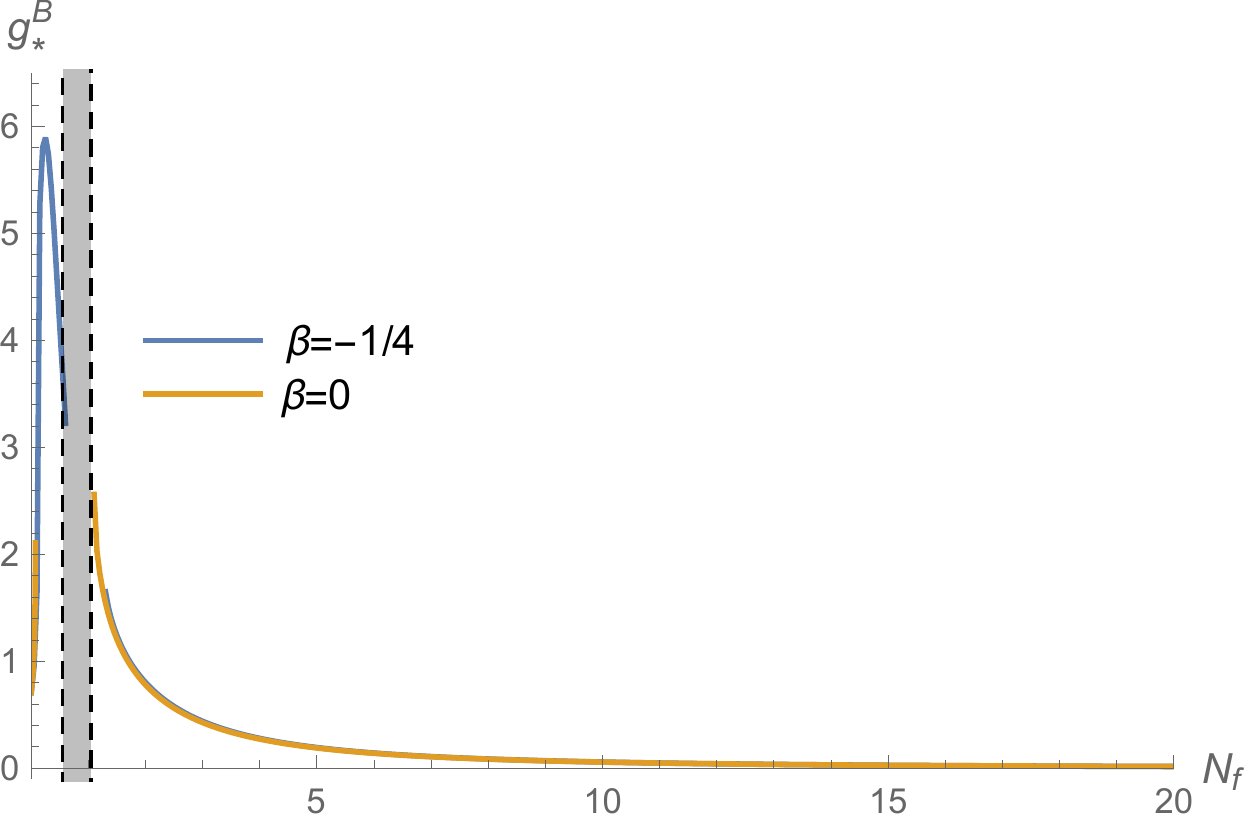}
	\includegraphics[width=0.5\textwidth]{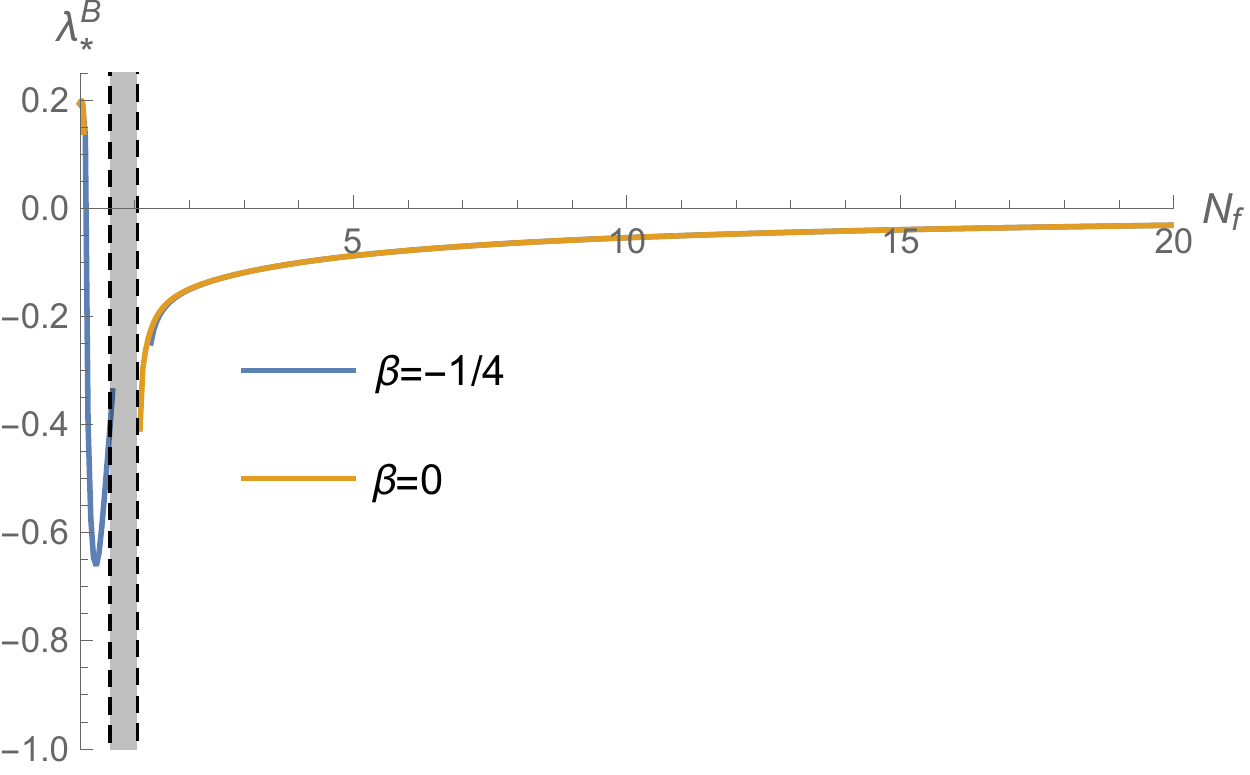} \\[3ex]
	\includegraphics[width=0.5\textwidth]{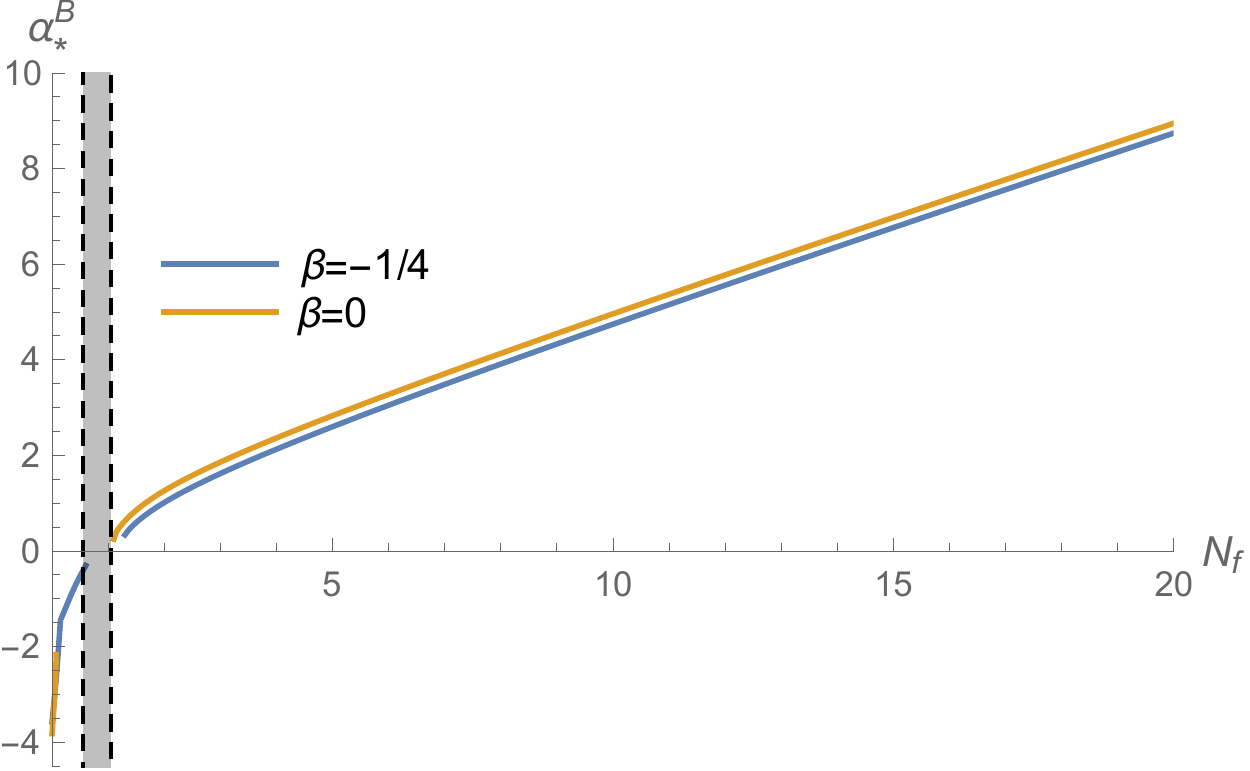}
	\includegraphics[width=0.5\textwidth]{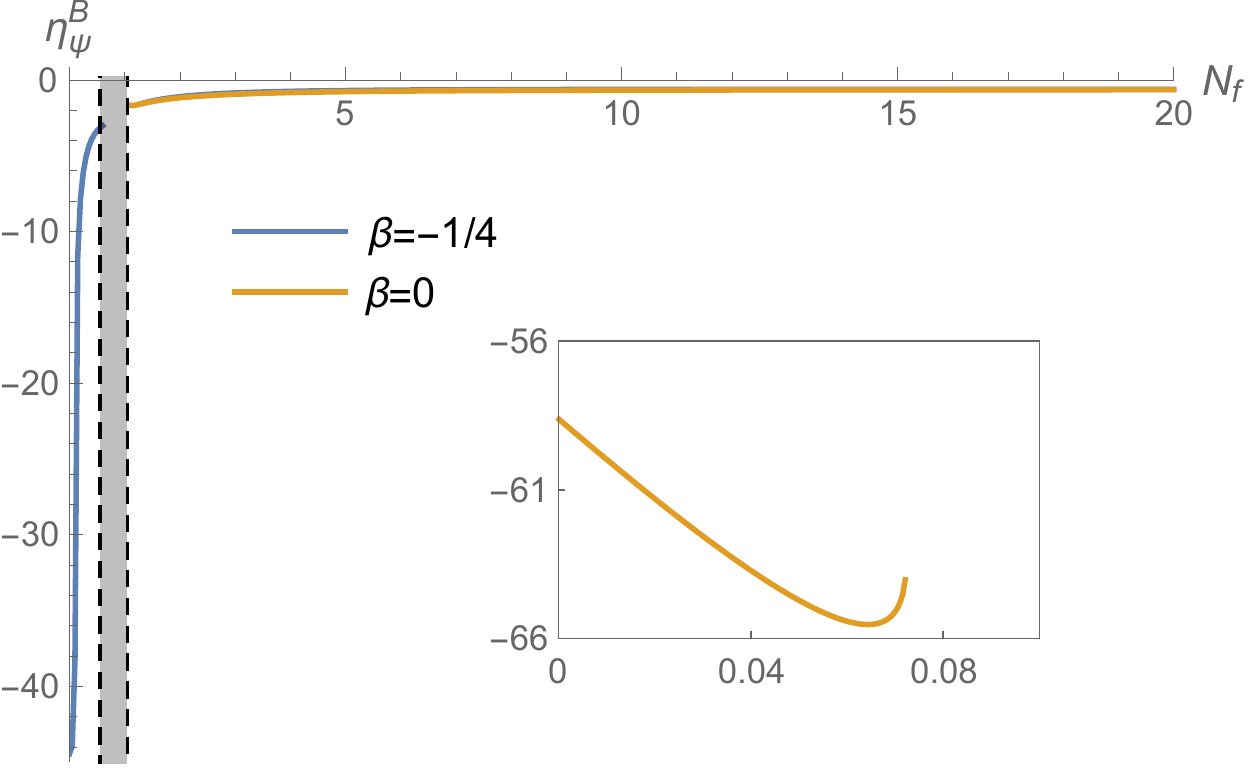} \\[3ex]
	\includegraphics[width=0.5\textwidth]{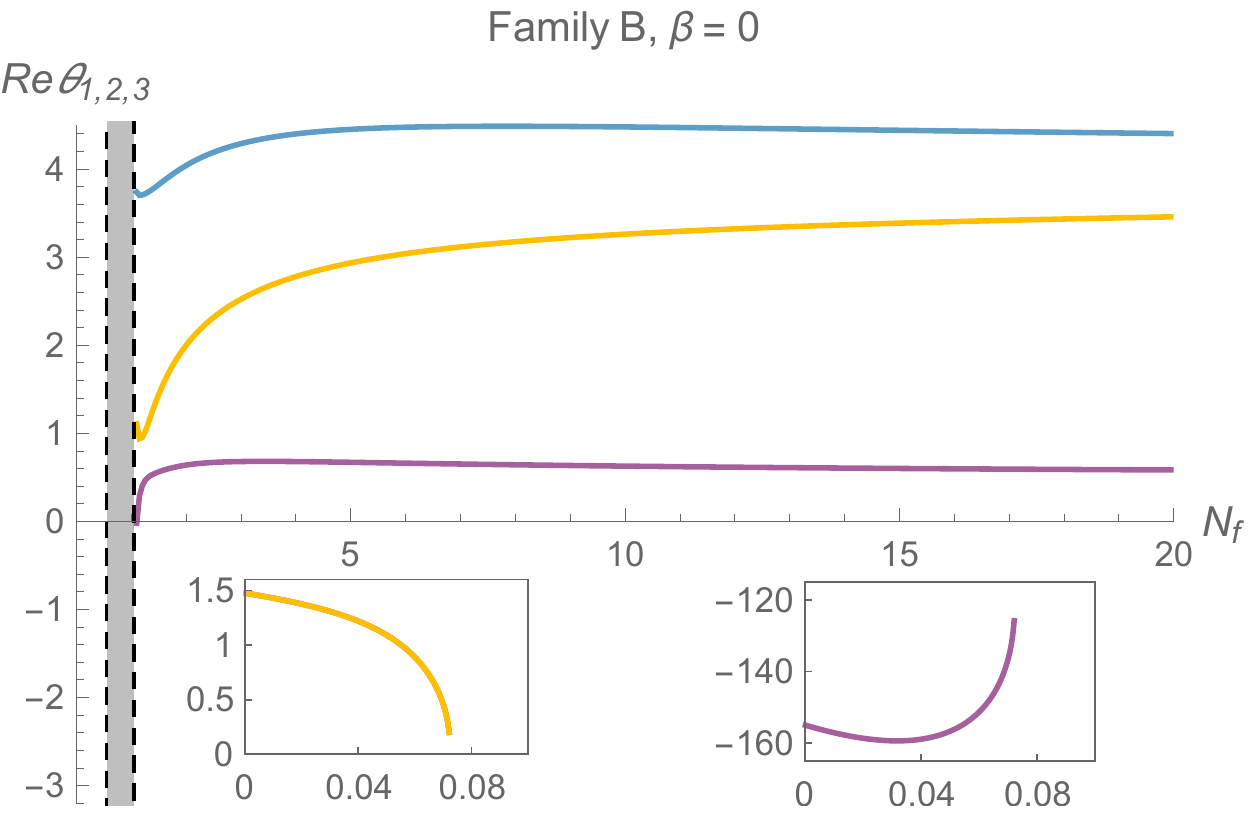}
	\includegraphics[width=0.5\textwidth]{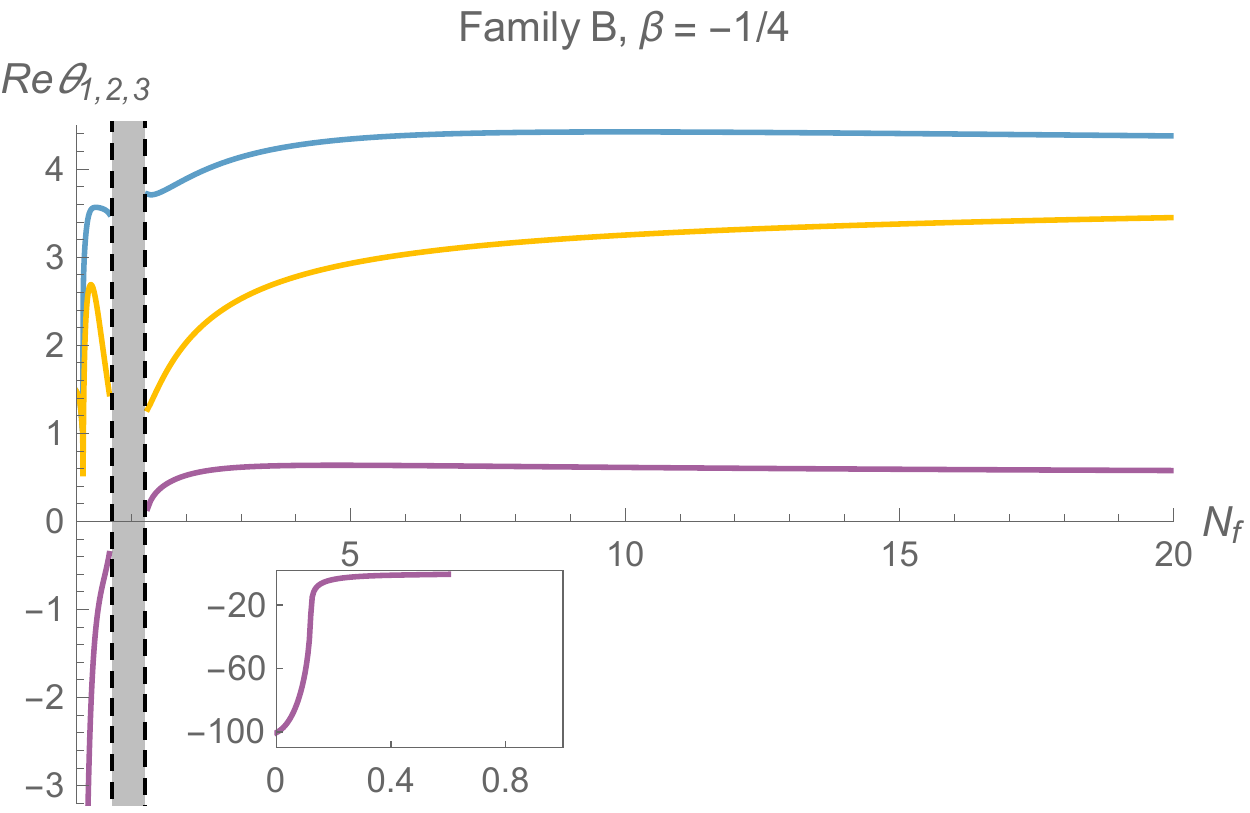} \\[2ex]
	\caption{\label{FullSysB} Results from the numerical investigation of the fixed points belonging to Family $B$. The first three diagrams show the position of the fixed points as a function of $N_f$ for $\beta = -1/4$ (blue line) and $\beta = 0$ (orange line). The fixed points exist for all values $N_f$ outside the window \eqref{complexwindow}. The fermion anomalous dimension evaluated at the fixed point is depicted in the fourth diagram, indicating that $\eta_\psi^* < 0$ in all cases. Notably, small values of $N_f$ lead to rather large absolute values $|\eta_\psi^*|$ while values $N_f$ to the right of the region of instability have $\eta_N^* \approx -1$ (cf.\ Table \ref{Tab.3}). The stability coefficients show that the NGFP$^B$ are saddle points with two UV-relevant directions (three UV-attractors) for values $N_f$ below (above) the bounds \eqref{complexwindow}. The inserts magnify the fermion anomalous dimension and the stability coefficients obtained for the NGFP$^B$ along the orange line in Fig.\ \ref{manyFPfamilies}.}
\end{figure}
The fixed points exist for all values $N_f$ outside the window \eqref{complexwindow}. The effect of the fermion anomalous dimension on the system is most pronounced for small fermion numbers $N_f \lesssim 2$ where $|\eta_\psi^*| > 1$. For values $N_f$ located to the left of the gray region, the fixed points act as UV-attractors and are characterized by a complex pair of stability coefficients. For $N_f$ larger than the bound \eqref{complexwindow}, the NGFPs are characterized by two real positive and one negative stability coefficient, indicating that they are saddle points for the RG flow. In this region the effect of including $\eta_\psi$ is rather small. As a consequence, the properties of the fixed points are in qualitative agreement with the findings in Sect.\ \ref{ssect:4.3}. In particular, the case $\beta = 0$ still exhibits a critical number of fermions $N_f^{\text{crit}}$ at which the sign of $g_*$ changes and $\alpha_* \ll 1$ so that the fixed points remain (approximately) chiral. \\

\noindent
NGFP$^{\rm B}$: 
The properties of the NGFPs comprising Family $B$ are shown in Fig.\ \ref{FullSysB}. We first focus on the region where $N_f$ is larger than the values indicated in \eqref{complexwindow}. Here the NGFPs exist for all $N_f$. The signature feature of Family $B$ in this region is the linear increase of $\alpha_*^B$ with $N_f$. This feature is accompanied by a negative fermion anomalous dimension $\eta^B_\psi \simeq -1$. Moreover, the stability analysis reveals that the three stability coefficients are real and positive, indicating that the fixed points serve as UV-attractors for all three couplings $\{\lambda,g,\alpha\}$. This picture matches the one established without considering the effect of the fermion wave function renormalization (cf.\ Figs.\ \ref{NMCfps} and \ref{NMCCritExps}). This indicates that the small negative value of $\eta_\psi^B$ does not affect the fixed point structure of Family $B$ in a qualitative way.

For values $N_f$ to the left of the gray vertical band, $\eta_\psi^B$ is negative with relatively large absolute values. As a consequence, the inclusion of $\eta_\psi^B$ affects the fixed point structure rather drastically. This includes the rather intricate pattern of fixed points being created and annihilated as a function of $N_f$ which is detailed in Fig.\ \ref{manyFPfamilies}. Moreover, $\eta_\psi^B$ turns the real part of one of the stability coefficient from positive to negative so that the NGFP$^{\rm B}$ are saddle points for low values $N_f$. \\

\noindent
NGFP$^{\rm C}$: The genuinely novel feature arising from the inclusion of the fermion anomalous dimension is a new family of NGFPs which we label Family $C$. Its properties are summarized in Fig.\ \ref{FullSysC}. Notably, Family $C$ exists for all values $N_f$: it is unaffected by the instability window \eqref{complexwindow}. We find that, again, $\alpha_*^C$ increases linearly with $N_f$. As a consequence, the family is rather robust against changes in the coarse-graining operator encoded in $\beta$. As its characteristic feature, Family $C$ comes with a negative and rather large fermion anomalous dimension, approaching $\eta_\psi^C \approx -26$ for large values $N_f$. The stability analysis shows that NGFP$^C$ is a saddle point with two (real) UV-attractive and one UV-repulsive eigendirection. Clearly, it is the interplay between $\alpha$ and the large absolute value of $\eta_\psi^C$ which gives existence to this family of fixed points.

\emph{In summary, the inclusion of $\eta_\psi$ leads to three continuous families of NGFPs: Family $A$ (saddle point), Family $B$ (UV-attractor), and Family $C$ (saddle point). Families $A$ and $B$ are qualitatively similar to the fixed points identified in the previous section while Family $C$ is novel. All families posses a weak coupling limit $\lim_{N_f \rightarrow \infty} g_* = 0$. The effect of $\eta_\psi$ is most notable for small values $N_f \lesssim 2$. In this region the fermion anomalous dimension leads to an intricate pattern of fixed points moving in and out of the complex plane, see Fig.\ \ref{manyFPfamilies}. Family $C$ is unaffected by these transitions and exists continuously for all values $N_f$.}

%======================================================
\subsection{The complete system excluding chiral symmetry breaking contributions}
\label{ssect:4.42}
%======================================================
We conclude our analysis of the fixed point structure by studying the beta functions \eqref{betagrav} and \eqref{betaalpha} with $A_0$ set to zero by hand. This approximation mimics the structure of $\beta_\alpha$ found in flat-background computations \cite{Eichhorn:2016vvy}. The main effect is  
the disappearance of the bounds \eqref{complexwindow} for Family $A$ and $B$. 
Family $C$ is structurally left unaltered. Our analysis follows the one of the previous subsection. In particular, the technical remarks made at the beginning of section \ref{ssect:4.4} again apply. \\

\noindent
\textit{Merging fixed points - algebraic considerations}\\
We start by analyzing the condition $\eta_N^* = -2$. Notably, $A_0$ does not enter into this equation. As a consequence, one again obtains a quadratic equation fixing $g_*$ as in terms of $\{\lambda_*, \alpha_*, N_f, \beta\}$. One branch is connected  continuously to the Reuter fixed point \eqref{ReuterFP}. The equation for $g_*$ again contains square roots whose arguments depend on $\{\lambda_*, \alpha_*, N_f, \beta\}$. The condition that the arguments should be positive in order to ensure that $g_*$ is real again leads to the constraints illustrated in Fig.\ \ref{gsqrt}.

\begin{figure}[t!]
	\includegraphics[width=0.6\textwidth]{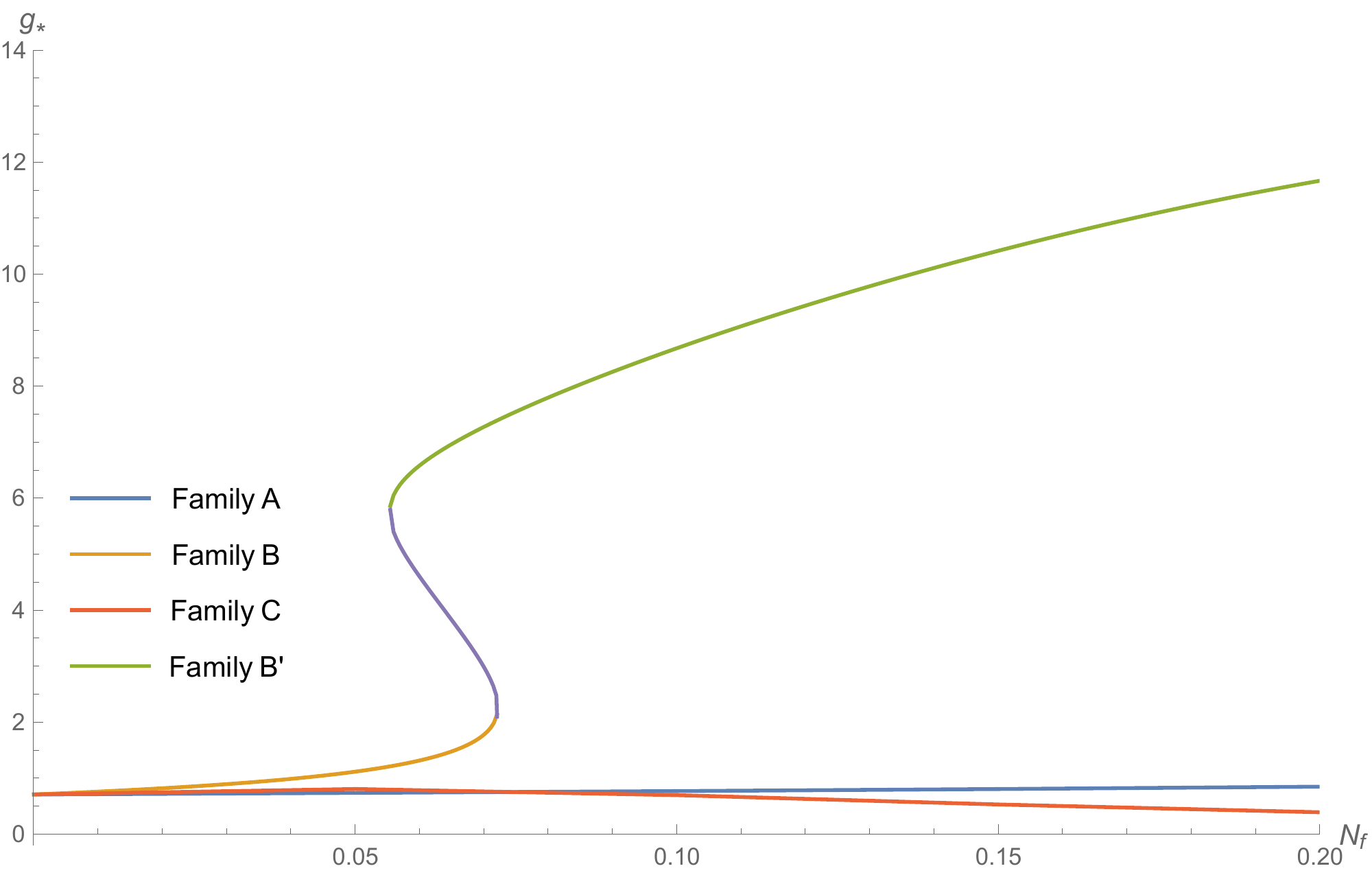}
	\centering
	\caption{\label{AllFamiliesNotFullSys} Illustration of the interplay between the fixed points for $\beta = 0$ and small values $N_f < 0.2$. Family B (orange line) crosses over into the complex plane upon colliding with a short-lived fixed point given by the purple line. The Family B' (green line) extends to arbitrary values of $N_f$. On this basis, we identify the Family B' as the ``continuation'' of Family B to arbitrary large values $N_f$.} 
\end{figure}

For $N_f < 0.2$ this property again leads to an intricate interplay of fixed points moving in and out of the complex plane. For $\beta = 0$, the resulting pattern is illustrated in Fig.\ 
\ref{AllFamiliesNotFullSys}.
 The result is very similar to the case where $A_0 \neq 0$. However, the region \eqref{complexwindow} has disappeared and Family B is unable to return from the complex plane. Instead one has a new branch, dubbed Family B', which extends to arbritrary values of $N_f$.\\

\noindent
\textit{Fixed points - numerical results}\\
We now illustrate the key properties of the three continuous families of NGFPs identified by our numerical search algorithm. When plotting the $N_f$-dependence of the solutions, the blue and yellow lines appearing in the position plots correspond to the two choices of coarse-graining operators, $\beta = -1/4$ and $\beta = 0$, respectively.\\

\begin{figure}[p!]
	\includegraphics[width=0.5\textwidth]{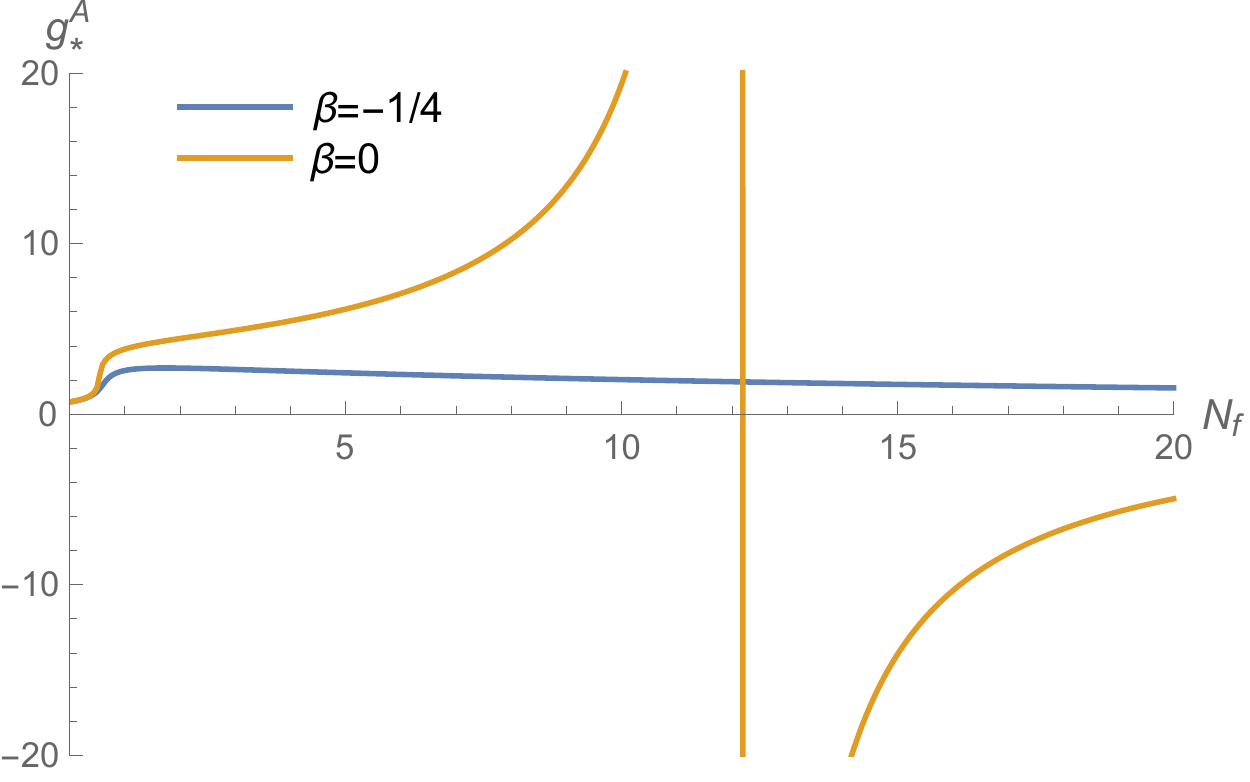}
	\includegraphics[width=0.5\textwidth]{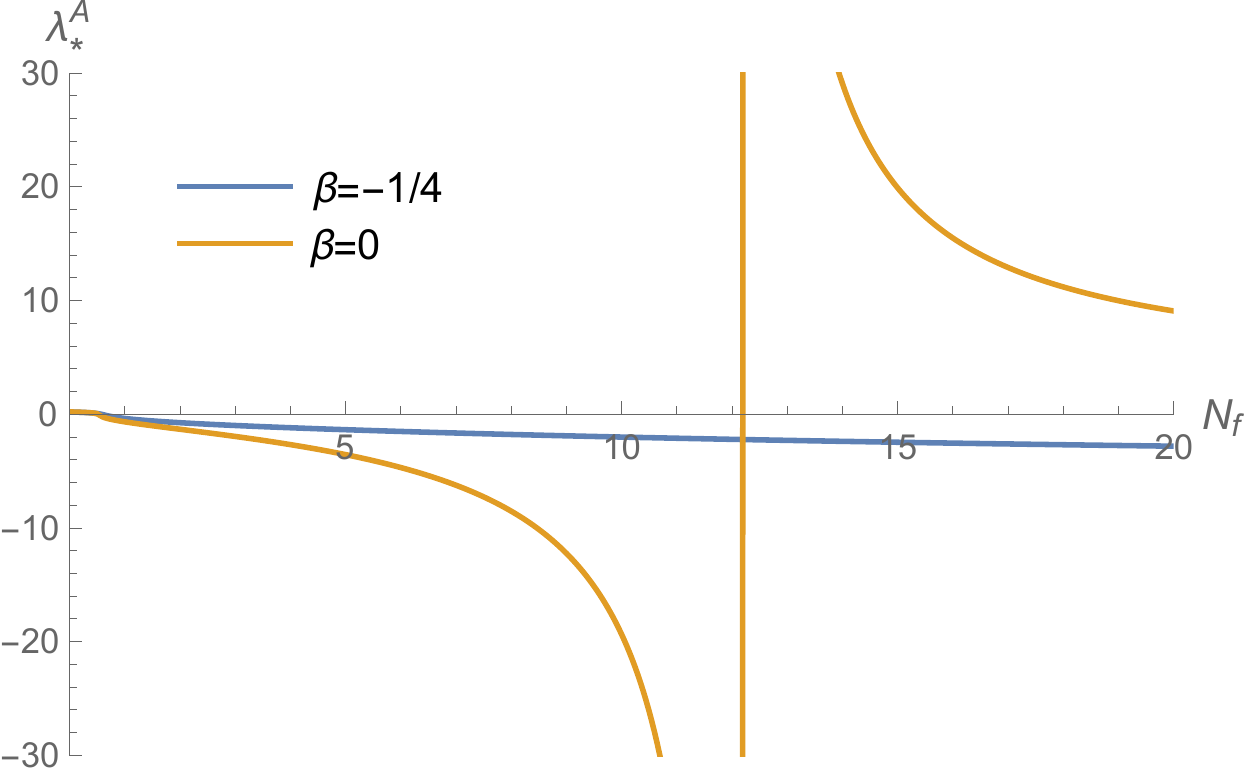} \\[3ex]
	\includegraphics[width=0.5\textwidth]{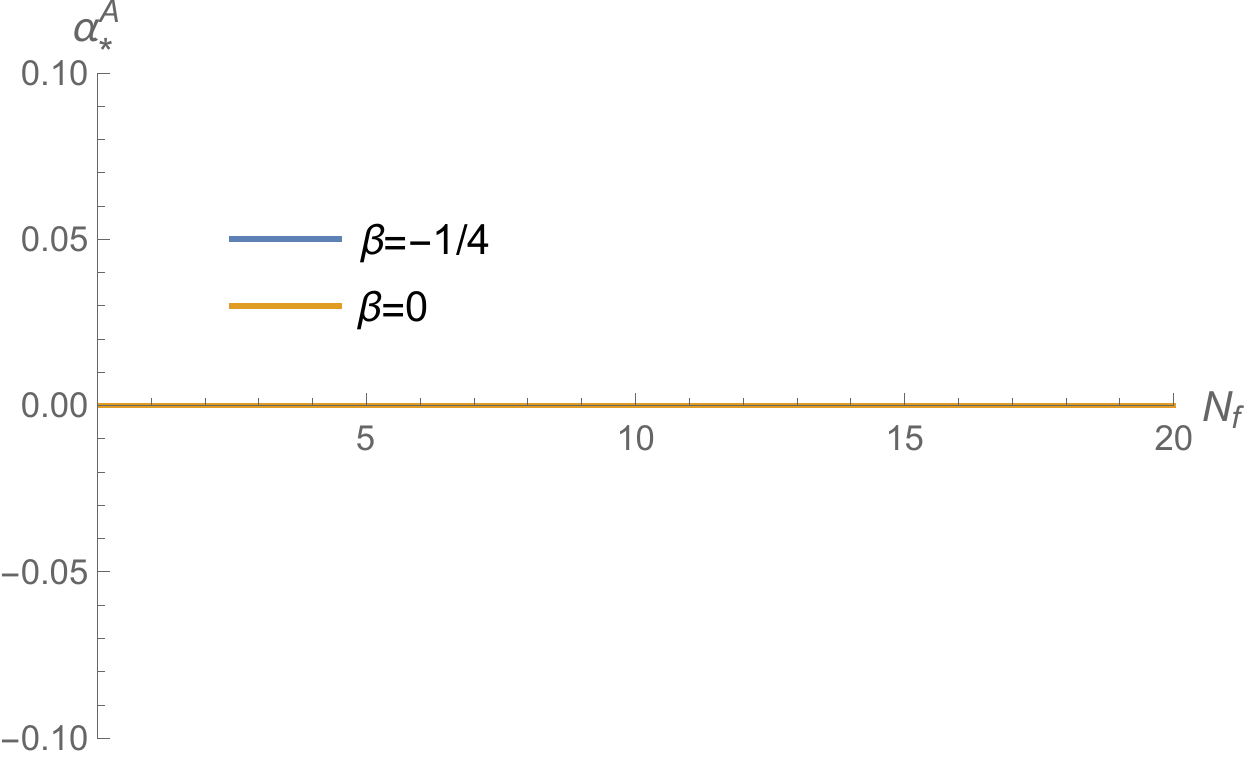}
	\includegraphics[width=0.5\textwidth]{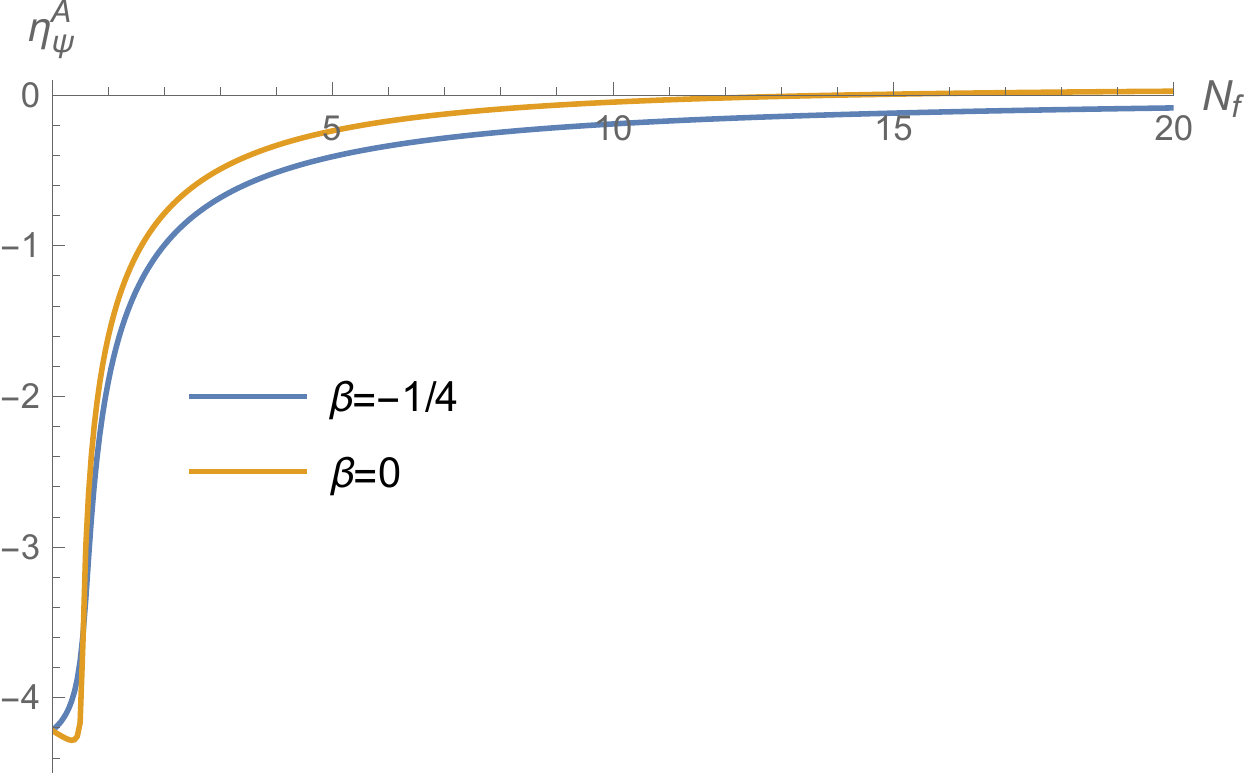} \\[3ex]
	\includegraphics[width=0.5\textwidth]{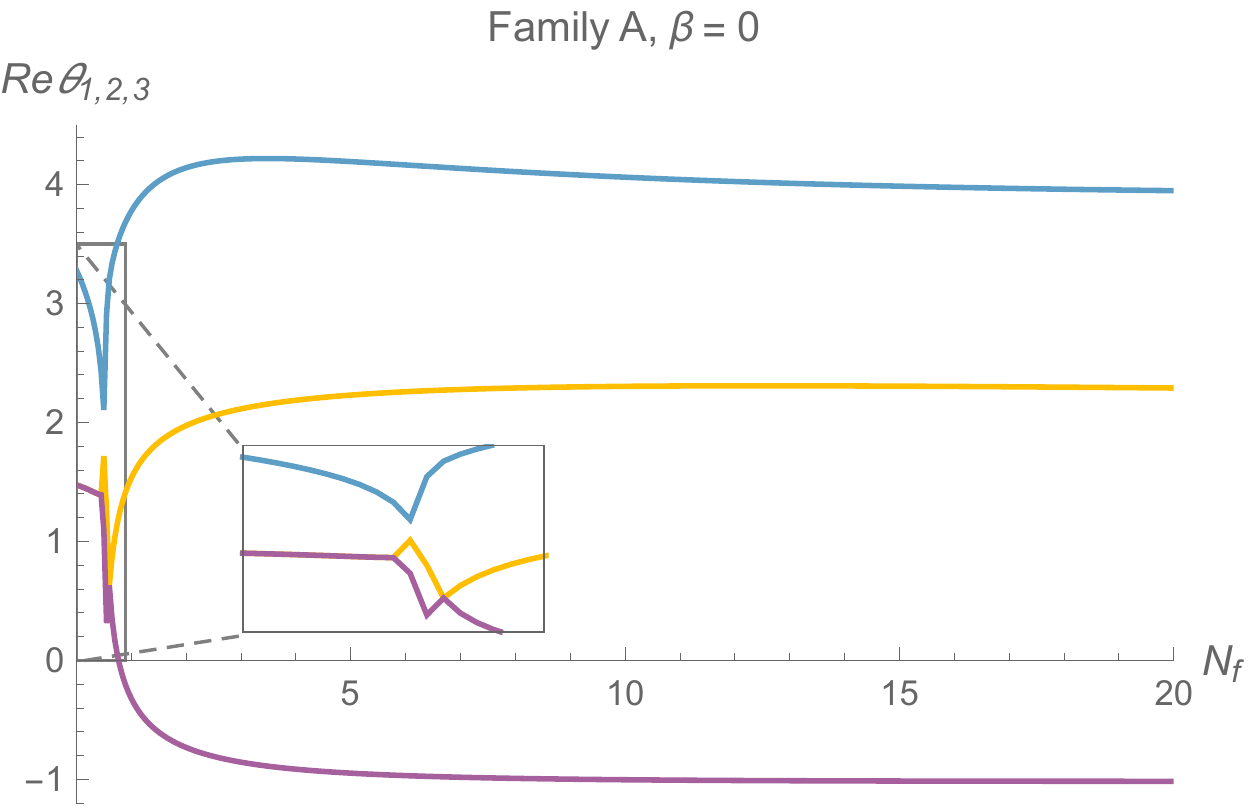}
	\includegraphics[width=0.5\textwidth]{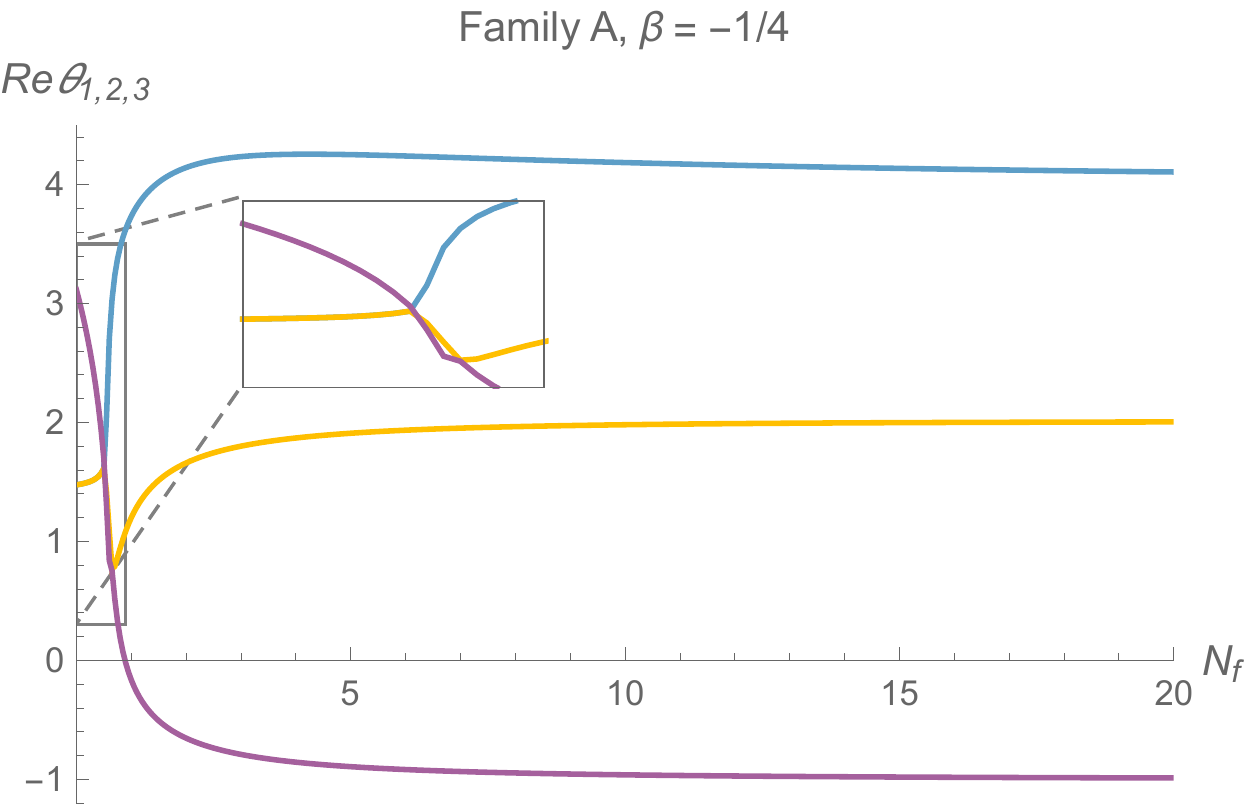} \\[2ex]
	\caption{\label{ChiralBreakExcludedA} Characteristic properties of Family $A$ obtained from investigating the system eqs. \eqref{betagrav} and \eqref{betaalpha} with $A_0 = 0$. The first three diagrams show the position of the fixed points as a function of $N_f$ for $\beta = -1/4$ (blue line) and $\beta = 0$ (orange line). NGFP$^A$ exists for all values $N_f$ and is characterized by $\alpha_*^A = 0$. The fermion anomalous dimension at the fixed points is depicted in the fourth diagram, showing no structural differences with respect to the previous section apart from the disappearence of the bounds \eqref{complexwindow}. The stability analysis identifies two UV-attractive and one UV-repulsive stability coefficient for values of $N_f$ extending beyond the window \eqref{complexwindow}, indicating that NGFP$^A$ is a saddle point in the $\{g,\lambda,\alpha\}$-plane. For smaller values of $N_f$ the stability analysis identies three UV-attractive stability coefficients.} 
\end{figure}

\noindent
NGFP$^A$: The properties of the fixed points comprising Family A are shown in Fig. \ref{ChiralBreakExcludedA}. It is then suggestive to call this family of fixed points ``quasi-chiral''. The fixed points exist for all values of $N_f$ as the window \eqref{complexwindow} is now absent. As a consequence of excluding the chiral symmetry breaking term $A_0$, Family A is now situated at $\alpha_*^A = 0$ for all values of $N_f$. The stability coefficients are structurally equal to the ones from the previous section: for low values $N_f$ the fixed points act as UV-attractors while they turn into saddle points once $N_f$ exceeds one. \\
\begin{figure}[p!]
	\includegraphics[width=0.5\textwidth]{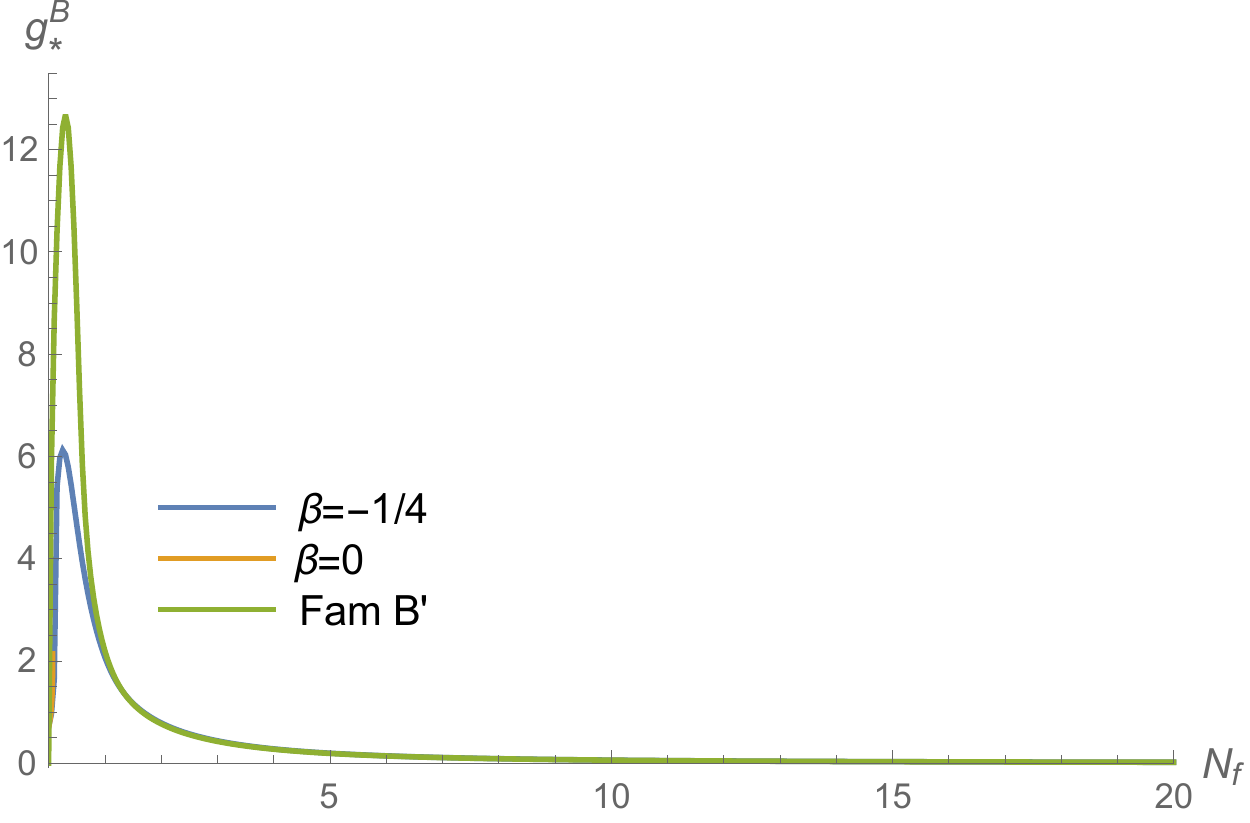}
	\includegraphics[width=0.5\textwidth]{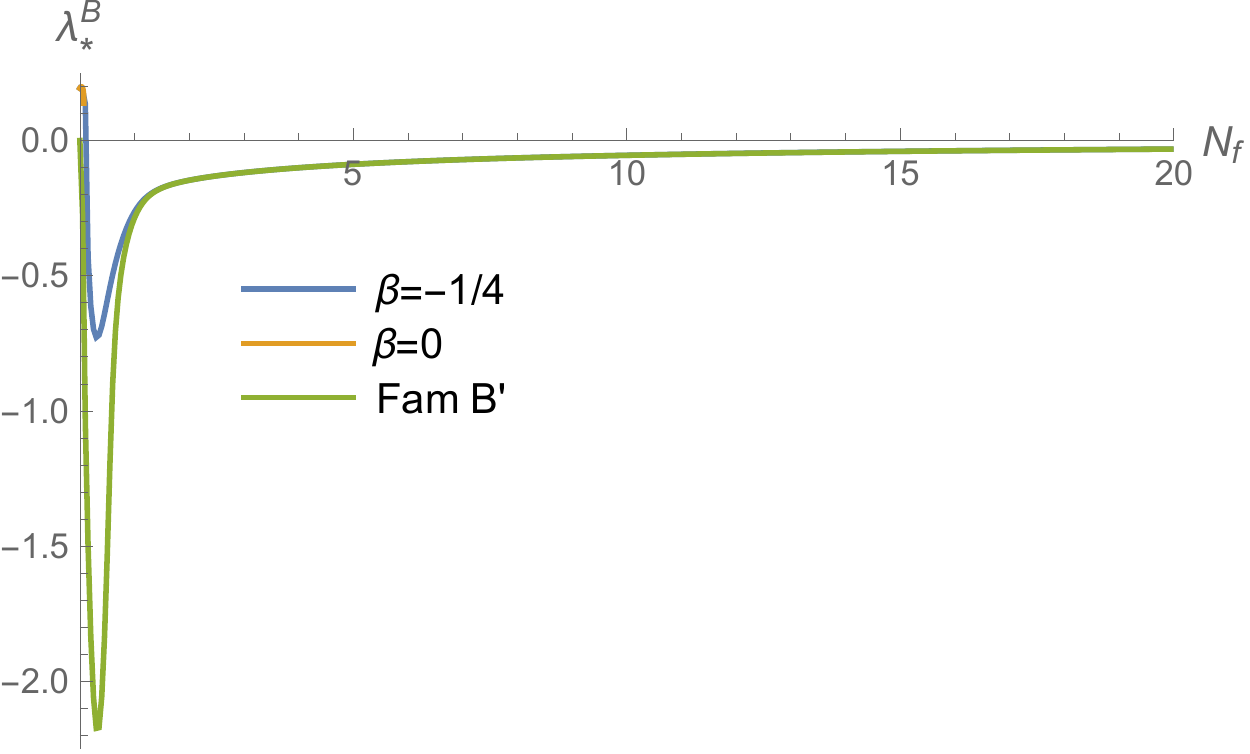} \\[3ex]
	\includegraphics[width=0.5\textwidth]{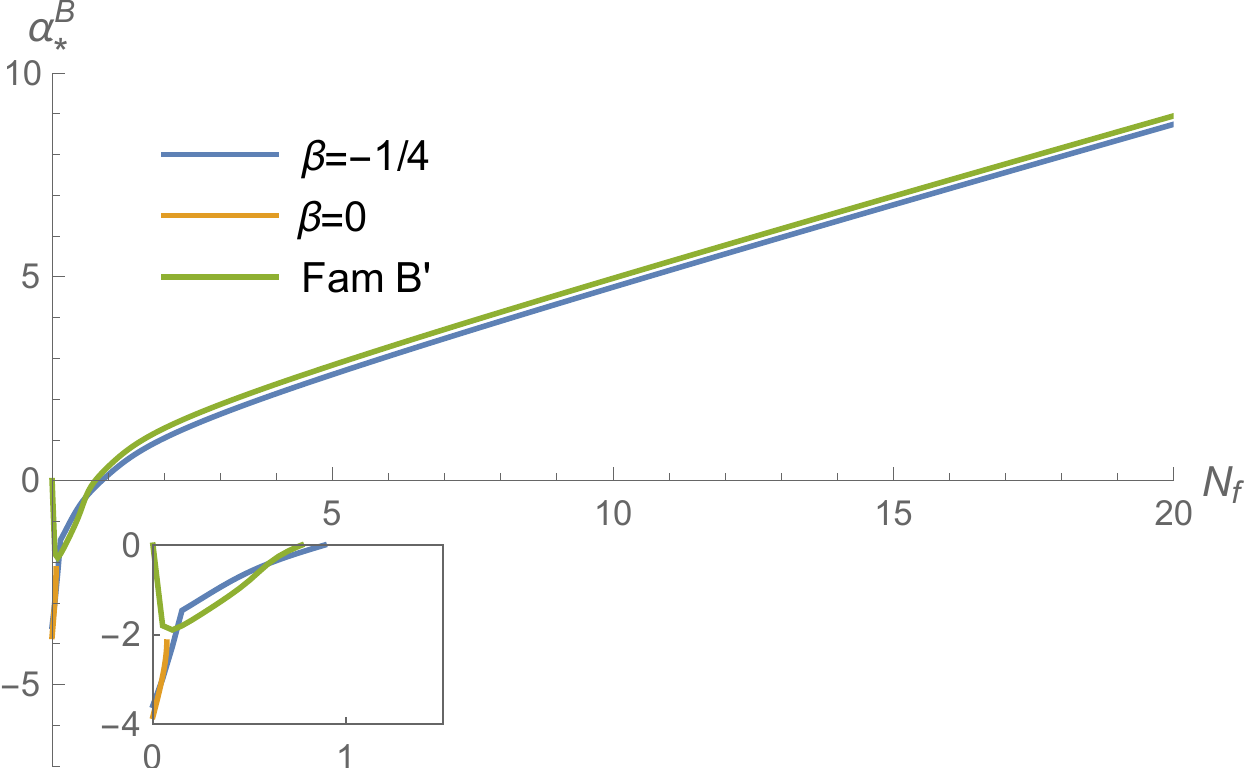}
	\includegraphics[width=0.5\textwidth]{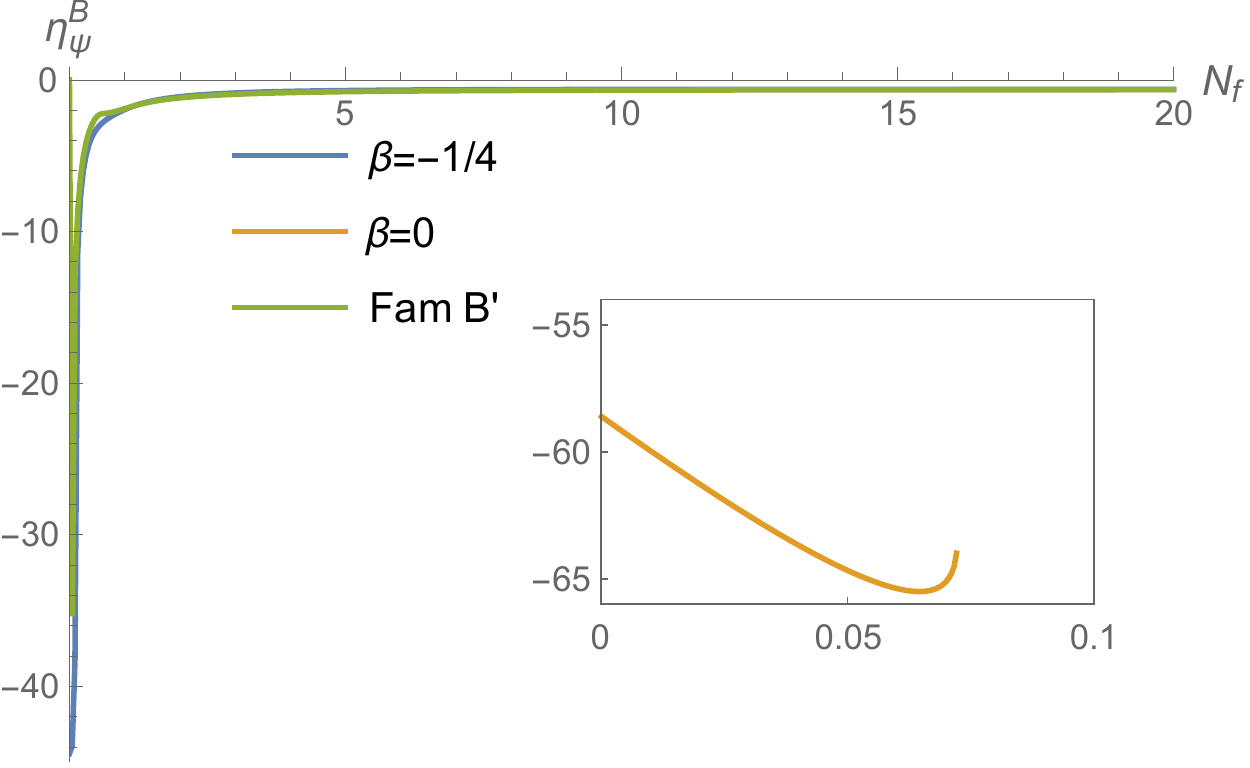} \\[3ex]
	\includegraphics[width=0.5\textwidth]{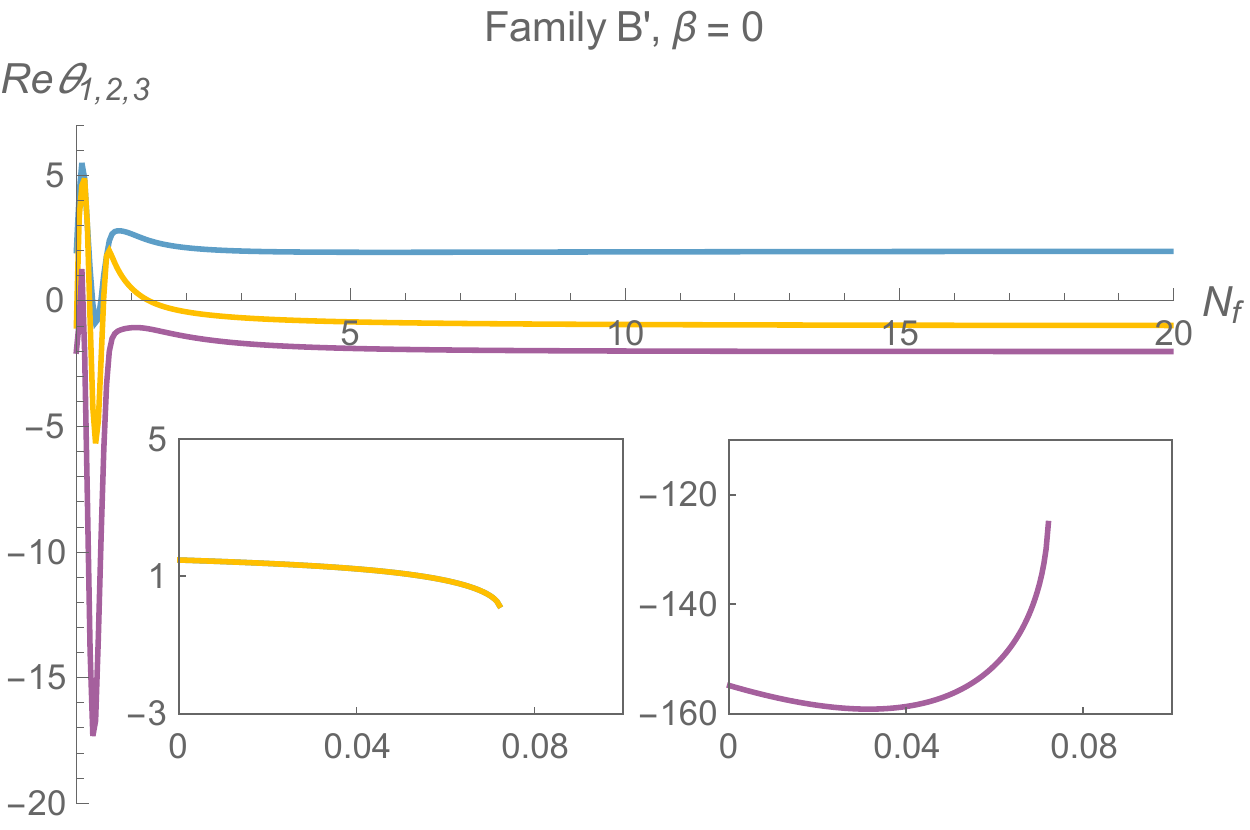}
	\includegraphics[width=0.5\textwidth]{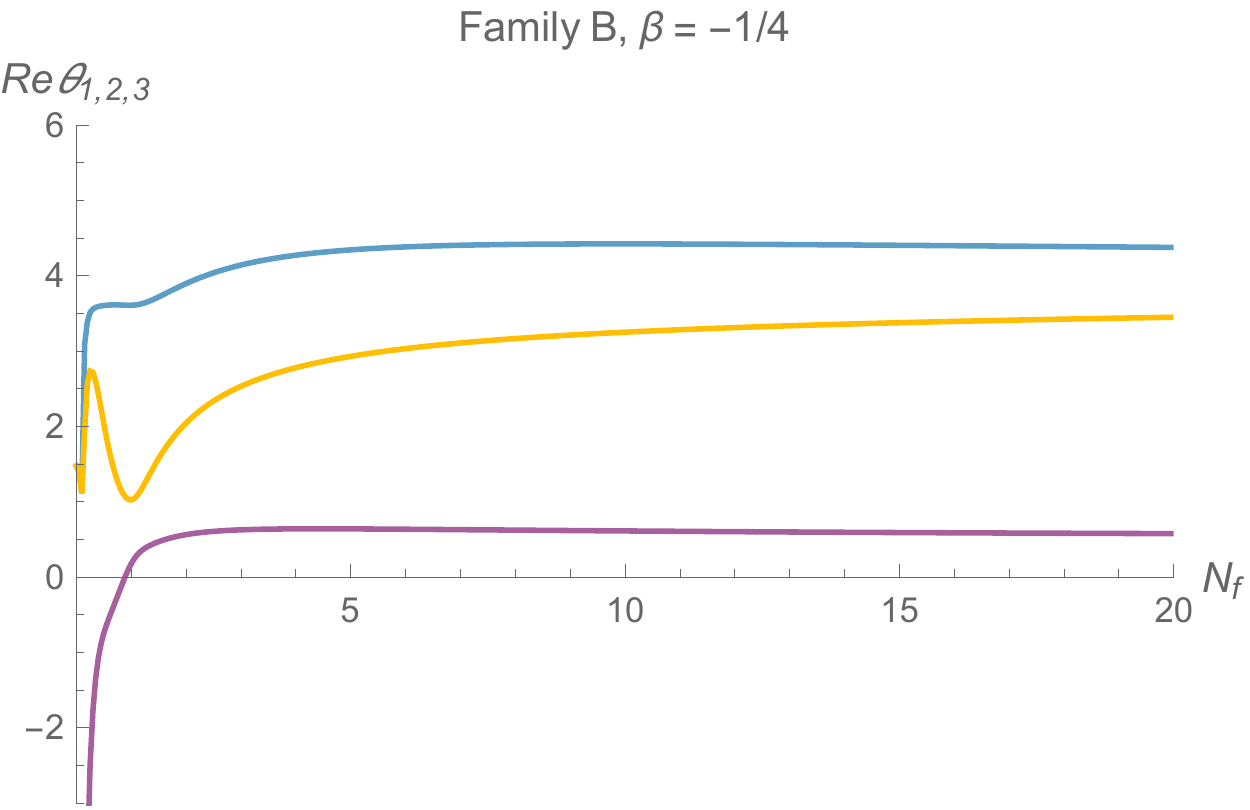} \\[2ex]
	\caption{\label{ChiralBreakExcludedB} Characteristic properties of Family $B$ obtained from investigating the system eqs. \eqref{betagrav} and \eqref{betaalpha}  with $A_0 = 0 $. The first three diagrams show the position of the fixed points as a function of $N_f$ for $\beta = -1/4$ (blue line) and $\beta = 0$ (orange line). NGFP$^B$ exists for all values $N_f$ for $\beta = -1/4$. For $\beta = 0$ family B moves into the complex plane at $N_f \approx 0.6$ and does not re-emerge due to the absence of the window \eqref{complexwindow}. Family B' continues and is shown as a green line. The fermion anomalous dimension at the fixed points is depicted in the fourth diagram, showing that it is negative and rather large. The stability analysis identifies two UV-attractive and one UV-repulsive stability coefficient for values below the outer edge of \eqref{complexwindow}, indicating that NGFP$^B$ is a saddle point in this region. For values beyond this edge, the fixed points function as UV-attractors.}
\end{figure}
\begin{figure}[p!]
	\includegraphics[width=0.5\textwidth]{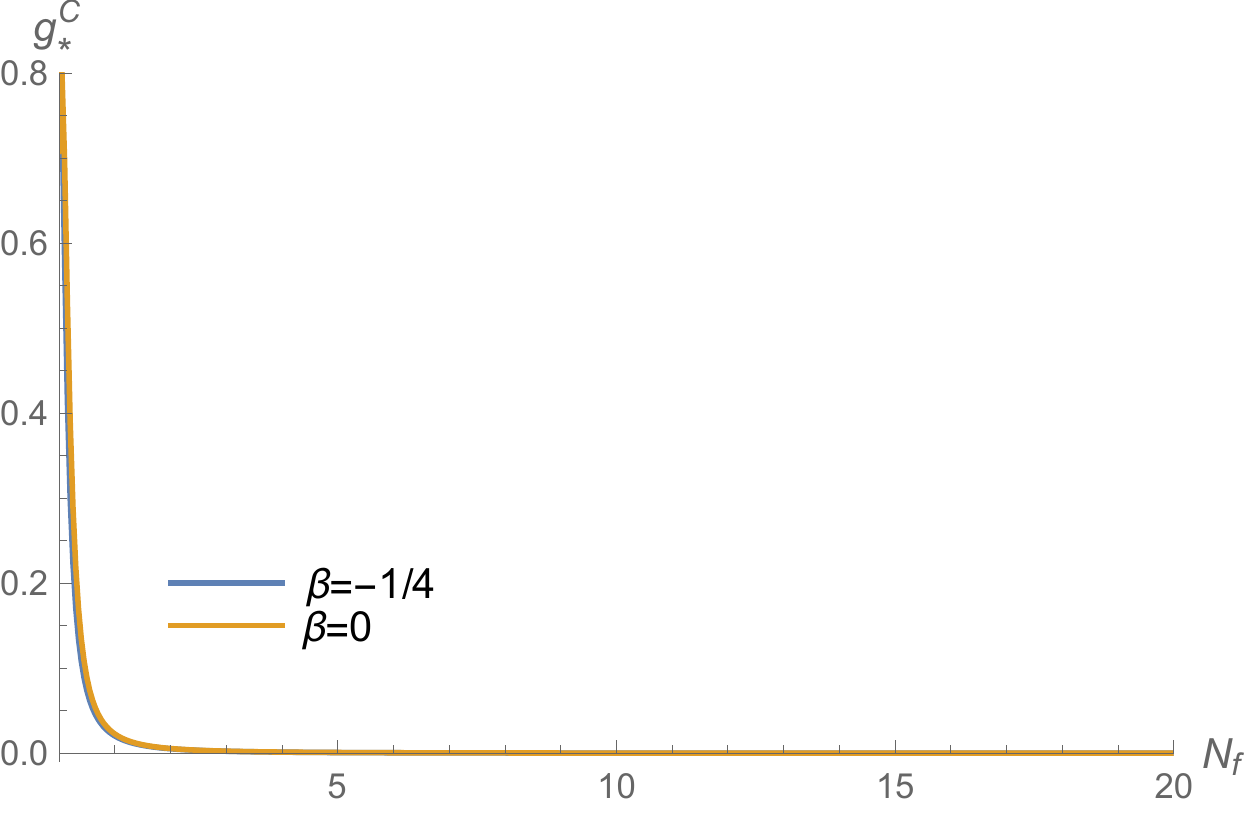}
	\includegraphics[width=0.5\textwidth]{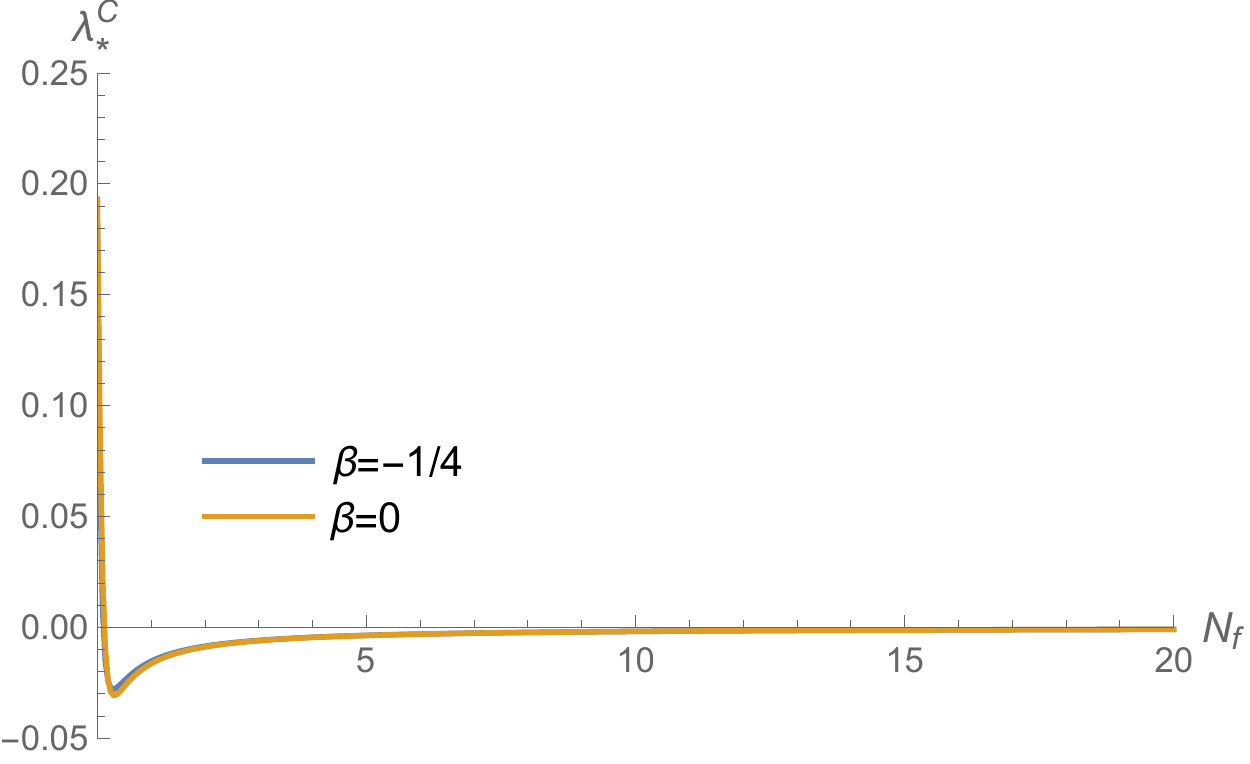} \\[3ex]
	\includegraphics[width=0.5\textwidth]{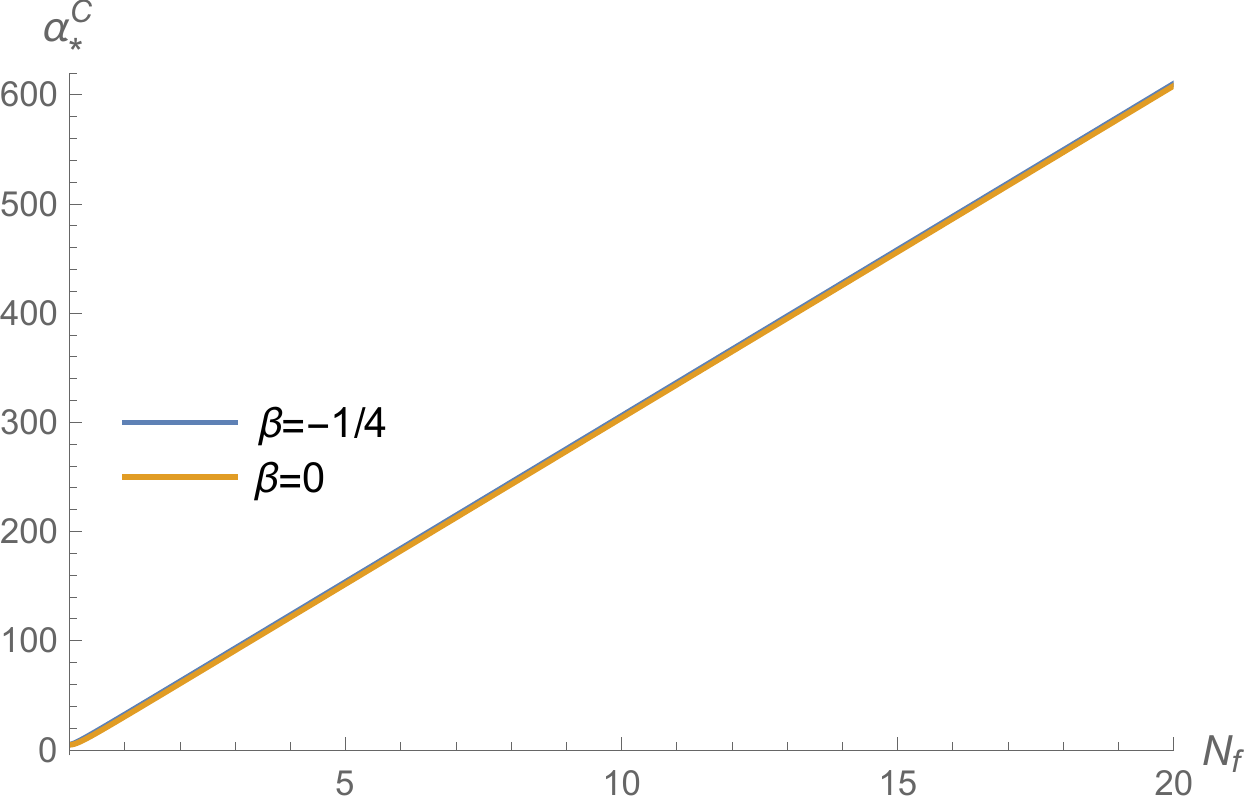}
	\includegraphics[width=0.5\textwidth]{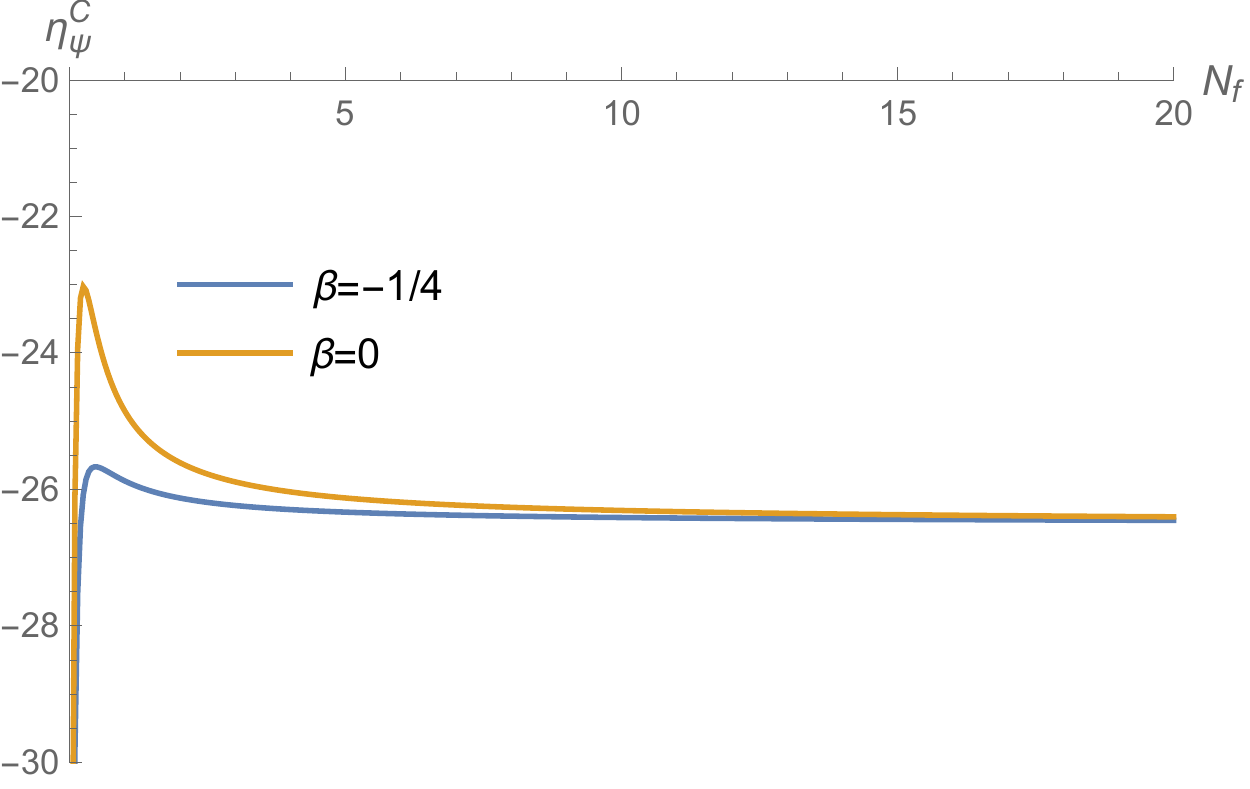} \\[3ex]
	\includegraphics[width=0.5\textwidth]{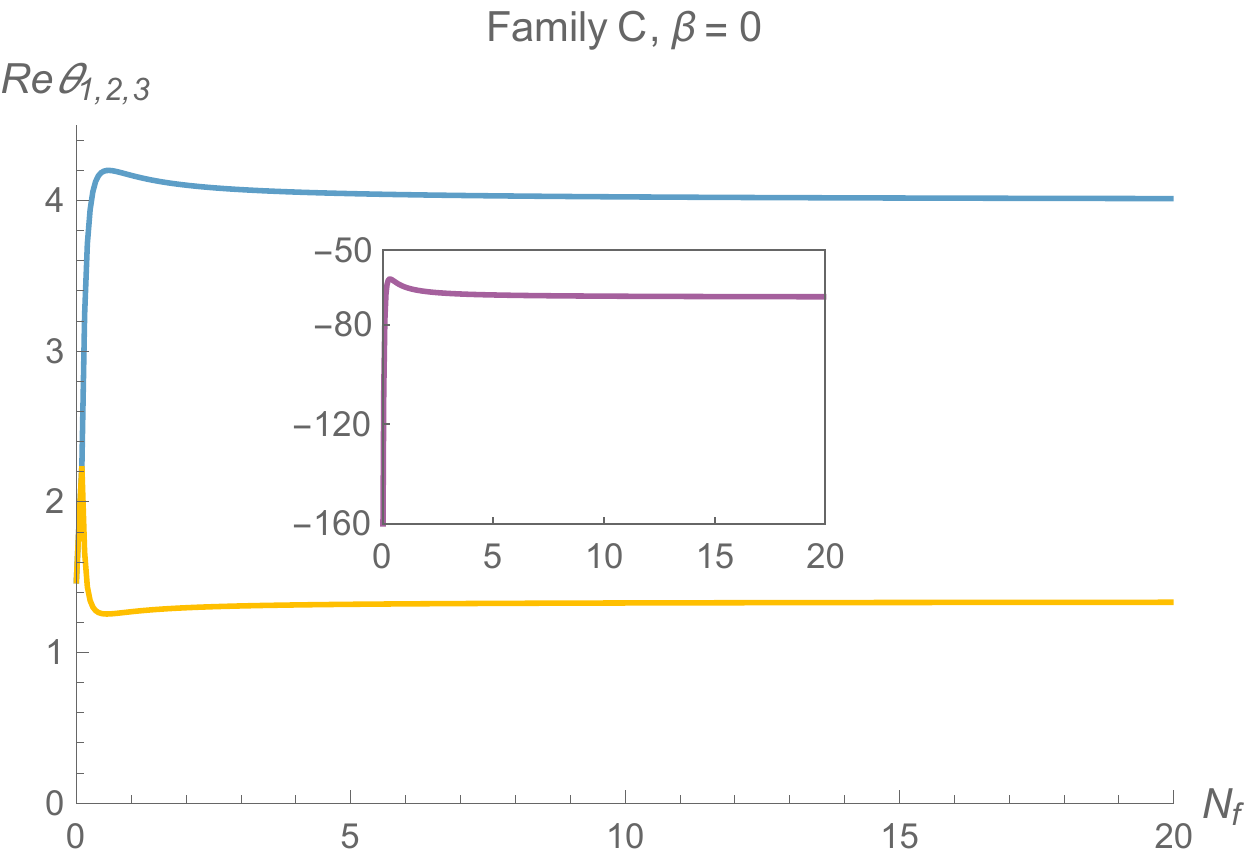}
	\includegraphics[width=0.5\textwidth]{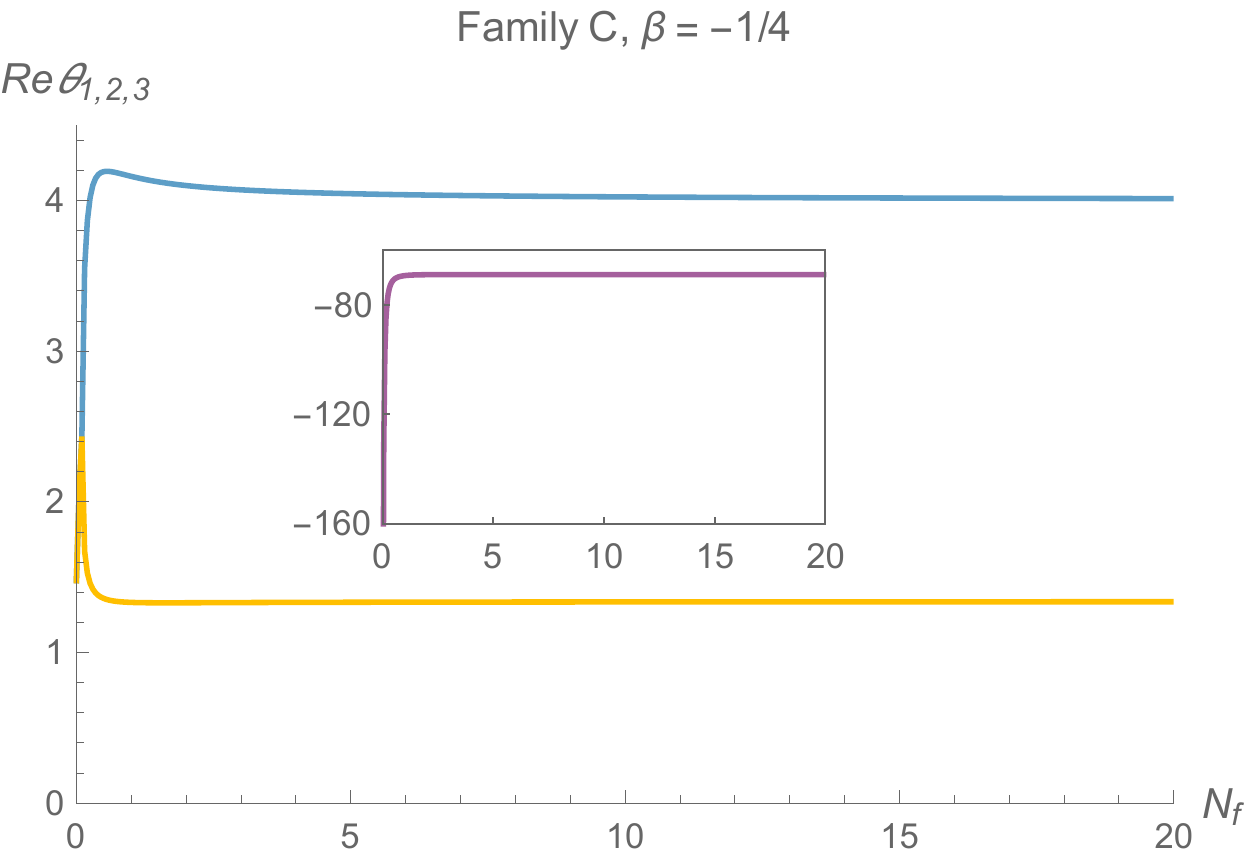} \\[2ex]
	\caption{\label{ChiralBreakExcludedC} Characteristic properties of Family $C$ obtained from investigating the system eqs. \eqref{betagrav} and \eqref{betaalpha} with $A_0 = 0 $. The first three diagrams show the position of the fixed points as a function of $N_f$ for $\beta = -1/4$ (blue line) and $\beta = 0$ (orange line). NGFP$^C$ exists for all values $N_f$ and is fairly insensitive to the choice of $\beta$. The fermion anomalous dimension at the fixed points is depicted in the fourth diagram, showing that it is negative and rather large. The stability analysis identifies two UV-attractive and one UV-repulsive stability coefficient, indicating that NGFP$^C$ is a saddle point in the $\{g,\lambda,\alpha\}$-plane.} 
\end{figure}

\noindent
NGFP$^{\rm B}$ and NGFP$^{\rm B'}$: The properties of the NGFPs comprising Families B and B' are shown in Fig. \ref{ChiralBreakExcludedB}. Their position and fermion anomalous dimension resembles the one found for NGFP$^{\rm B}$ shown in Fig.\ \ref{FullSysB}. Apart from the disappearance of the forbidden region \eqref{complexwindow} there are no structural changes with respect to the full system. Moreover, for $\beta = -1/4$ we find that Family B is UV-attractive, in agreement with the previous section.  The fixed point collisions resolved in Fig.\ \ref{AllFamiliesNotFullSys} have a drastic effect on the stability coefficients of the system though: the Family B' (located at $N_f \gtrsim 0.06$) comes with one UV-relevant and two UV-irrelevant directions. This indicates that the inclusion of $A_0$ has a significant stabilization effect on the fixed point structure. The direct comparison to the case with $A_0 \neq 0$ shows that the interplay of $A_0$ with the other terms in the beta functions leads to a fixed point structure which is less sensitive to varying $\beta$ in the sense that Family B can be recovered for both $\beta=0$ and $\beta = -1/4$. \\

\noindent
NGFP$^{\rm C}$: Structurally, this family of fixed points is left unaltered. For completeness the properties of the NGFPs comprising Family C are shown in Fig.\ \ref{ChiralBreakExcludedC}. \\

\noindent
\emph{In summary, switching of the chiral symmetry breaking contribution in $\beta_\alpha$ by hand has little effect on the solutions NGFP$^{\rm A}$ and NGFP$^{\rm C}$. NGFP$^{\rm A}$ is now situated at $\alpha_* =0$, warranting the label ``quasi-chiral'' fixed point. In this approximation, Family  B develops a significant $\beta$-dependence though. Only for $\beta =0$ one recovers the qualitative features exhibited by the full system, indicating that $A_0$ plays an intriguing role in stabilizing the fixed point structure.}

%--------------------------------------------------------------------------------
\section{Bounding chiral symmetry through asymptotic safety}
\label{sect:chiralsymmetry}
%--------------------------------------------------------------------------------
In Sect.\ \ref{ssect:4.4} we presented an in-depth analysis of the fixed points encoded in the beta functions \eqref{betagrav} and \eqref{betaalpha}. In this course, we identified four fixed points: the GFP, NGFP$^{A}$, NGFP$^{B}$ and NGFP$^C$. The goal of this section is to illustrate the RG flow resulting from the interplay of these fixed points and its consequences for the chiral symmetry breaking coupling $\alpha$. For concreteness, we set $\beta = -1/4$ and $N_f = 3$. The position and stability coefficients for this case are listed in Table \ref{Tab.3}.  

Ultimately, physics should be extracted from the effective average action $\Gamma_k$ at $k=0$ where all quantum fluctuations have been integrated out \cite{Knorr:2019atm,Bonanno:2020bil}. Asking for a (semi-)classical regime then requires that there is a cross-over in the RG flow from the NGFP providing the UV-completion and the GFP controlling the low-energy behavior of the flow \cite{Reuter:2001ag,Reuter:2004nx}. This restricts the discussion to the interplay of the GFP, the quasi-chiral fixed point NGFP$^A$ and the non-chiral fixed point NGFP$^{B}$. The NGFP$^C$ is distached from the semi-classical regime and will not play a role in the subsequent discussion.

As a starting point, we give a more detailed analysis of the stability matrix \eqref{stabmat} associated with these fixed points. For the GFP situated at $\{g^{\text{GFP}}_*, \lambda^{\text{GFP}}_*, \alpha^{\text{GFP}}_*\} = \{0,0,0\}$ the stability coefficients are fixed by canonical power counting and the corresponding eigenvectors in the $g$-$\lambda$-$\alpha$--plane are\footnote{We give all results rounded to two decimal digests. The notation $0.$ implies that the component is non-zero but rounds to zero in this representation.}
\be\label{GFPexample}
\begin{array}{ll}
\theta_1 = 2, & \qquad V_1 = \{0,1,0\} \, , \\
\theta_2 = -2,   & \qquad V_2 = \{0.98, 0.21, 0.01\} \, , \\
\theta_3 = -1,  &	\qquad V_3 = \{0,0,1\} \, . 
\end{array}
\ee
Thus, we have two IR-attractive eigendirections which are essentially aligned with the $g$- and $\alpha$-axis, while the fixed point is IR-repulsive along the $\lambda$-direction. The quasi-chiral NGFP$^A$ is located at $\{g^A_*, \lambda^A_*, \alpha^A_*\} = \{2.35,-0.89,0.04\}$ and possesses the eigensystem
\be\label{NGFPAexample}
\begin{array}{ll}
	\theta_1 = 4.22, & \qquad V_1 = \{0.33,0.941,0.01\} \, , \\
	\theta_2 = 1.78,   & \qquad V_2 = \{0.92, -0.40, 0.\} \, , \\
	\theta_3 = -0.76,  &	\qquad V_3 = \{0.90, -0.38, -0.21\} \, . 
\end{array}
\ee
Thus, we are dealing with a saddle-point whose UV-attractive directions essentially lie within the $g$-$\lambda$-plane while the $\alpha$-direction is UV-repulsive. Finally, the non-chiral NGFP$^B$ sits at $\{g^B_*, \lambda^B_*, \alpha^B_*\} = \{0.44,-0.12,1.62\}$ and acts as a UV-attractor
\be\label{NGFPBexample}
\begin{array}{ll}
	\theta_1 = 4.14, & \qquad V_1 = \{0.29,0.19,0.94\} \, , \\
	\theta_2 = 2.53,   & \qquad V_2 = \{0.61, -0.28, 0.74\} \, , \\
	\theta_3 = 0.61,  &	\qquad V_3 = \{0.31, -0.09, -0.95\} \, . 
\end{array}
\ee

\begin{figure}[t!]
	\includegraphics[width=0.48\textwidth]{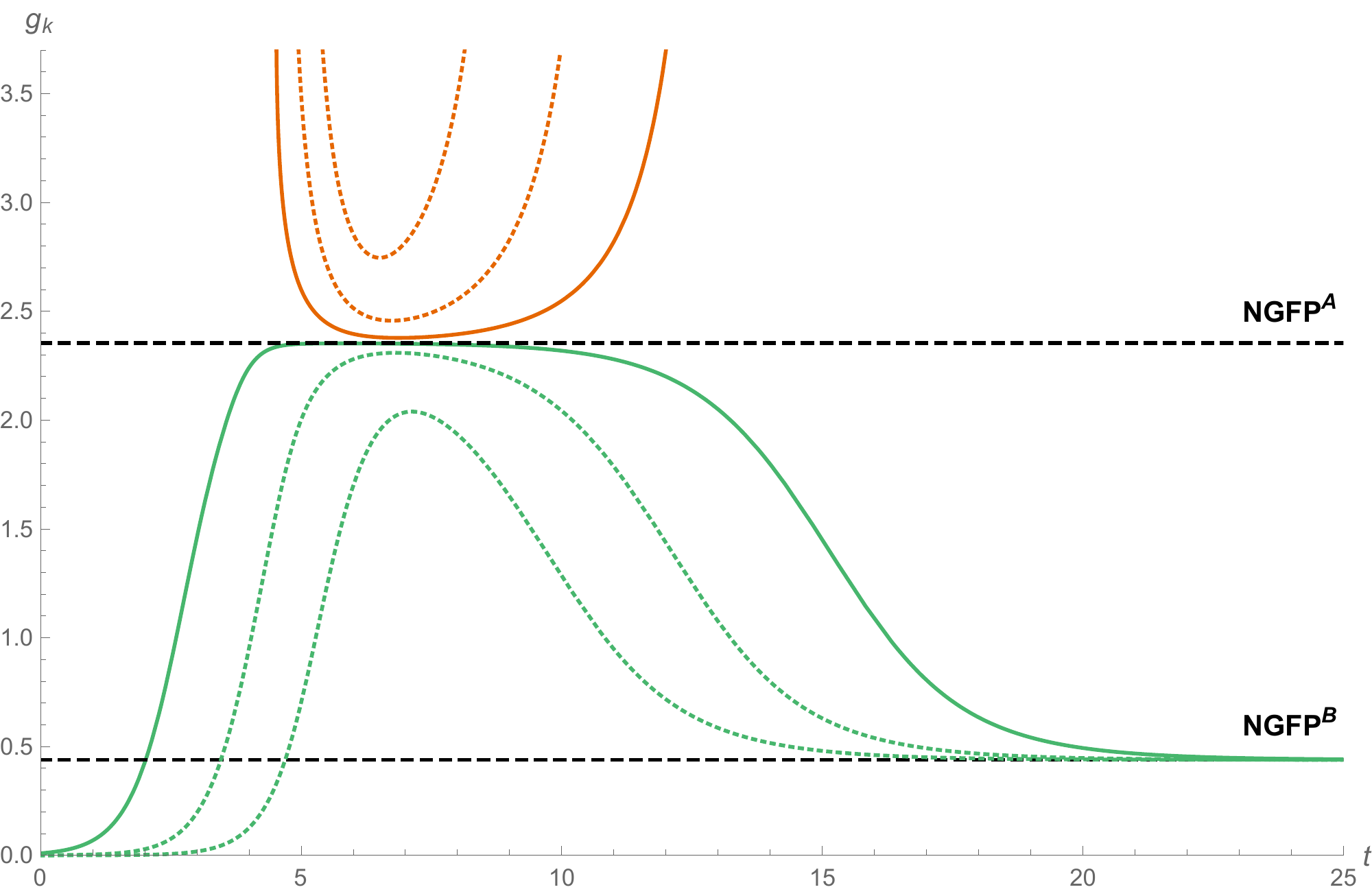} \,
	\includegraphics[width=0.48\textwidth]{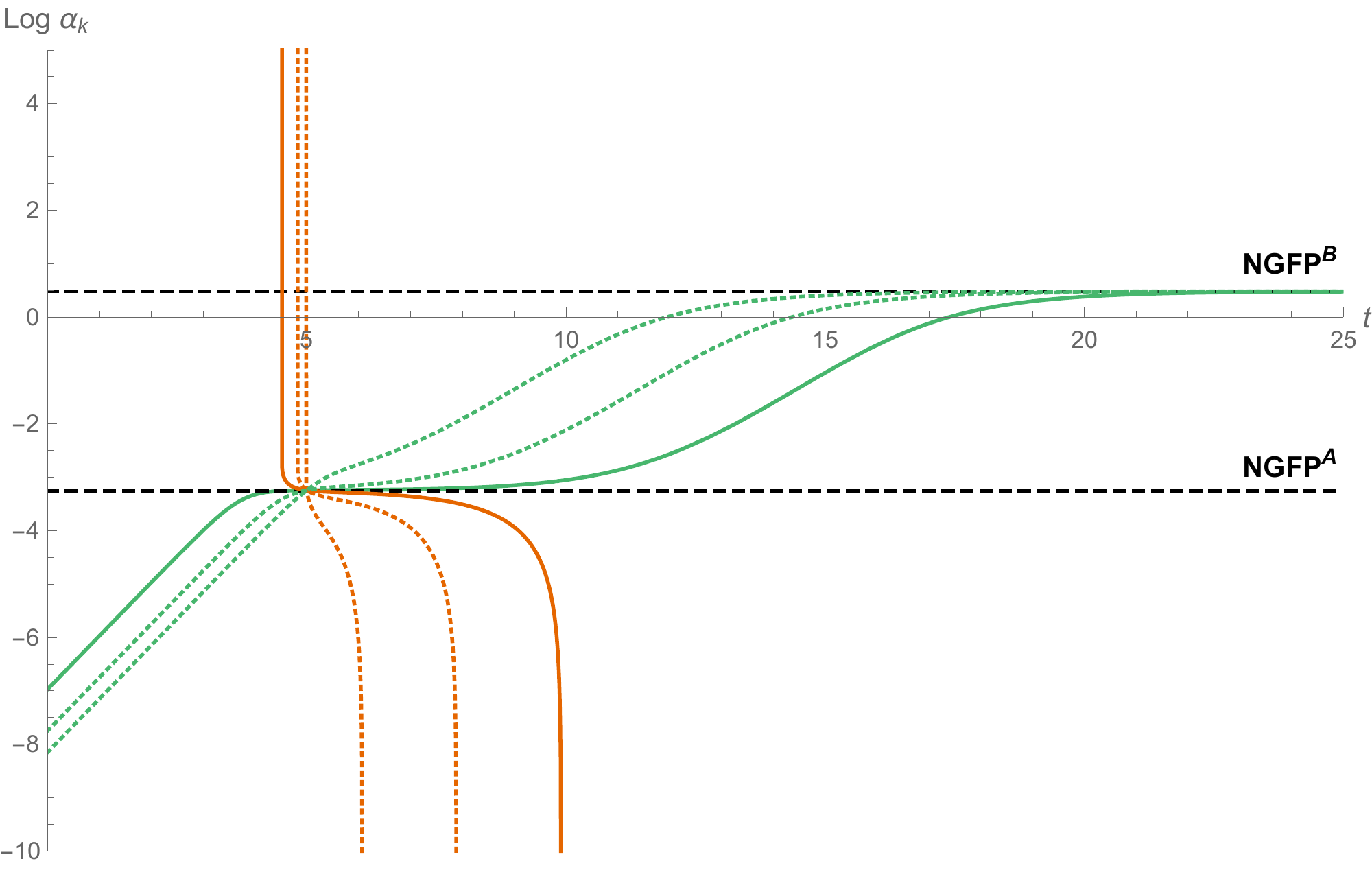}
	\centering
	\caption{
		\label{flowdiagrams} Illustration of the interplay between the three fixed points \eqref{GFPexample}, \eqref{NGFPAexample}, and \eqref{NGFPBexample} for $N_f = 3$ and $\beta = -1/4$. The left and right panel show $g_k$ and (the logarithm of)  $\alpha_k$-component obtained from solving the full system of beta functions numerically as a function of $t \equiv \log k$. 
		 The green lines correspond to the UV-safe trajectories emanating from NGFP$^B$.  The thick green line marks the crossover from NGFP$^B$ to NGFP$^A$ and constitutes the barrier between the safe and unsafe trajectories depicted in orange. For small values $t$, the flow undergoes a second cross-over from NGFP$^A$ to the GFP. The linear decrease of $\log \alpha_k$ in this regime is characteristic for the dimensionful coupling obtaining a constant ($k$-independent) IR-value.}
\end{figure}

An intriguing feature of these stability properties is that they admit RG trajectories which emanate from NGFP$^B$ in the UV, cross over to NGFP$^A$, and subsequently obtain their low-energy completion from the GFP. This process is illustrated in Fig.\ \ref{flowdiagrams}.
The interplay among the fixed points furthermore suggests that there is a window for the IR values of $\alpha_k$ provided by the asymptotic safety condition. In order to obtain a qualitative idea for the admissible values $\alpha_0$, we consider a two-parameter family of initial conditions $\{g_{\rm init}, \lambda_{\rm init}, \alpha_{\rm init}\}$ placed on a two-sphere centered on the UV-attractor NGFP$^{B}$, i.e.,
\begin{equation}
		g_{\rm init} = g_*^B + R\ \text{sin}\, \theta \ \text{cos}\, \phi, \quad \;
		\lambda_{\rm init} = \lambda_*^B + R\ \text{sin}\, \theta \ \text{sin}\, \phi, \quad \;
		\alpha_{\rm init} = \alpha_*^B + R\ \text{cos}\, \theta .
\end{equation}	
Provided that $R$ is taken sufficiently small, this construction guarantees that all trajectories constructed from these initial conditions are pulled into NGFP$^B$ as $k \rightarrow \infty$. Thus one can construct the UV-critical hypersurface of the fixed point by varying $\theta,\phi$ and integrating the RG flow towards $k \to 0$ numerically. Subsequently, one can extract the scaling of $\alpha_k$ in the vicinity of the GFP and reconstruct the value of the dimensionful coupling $\tilde{\alpha}_k$ via the relation \eqref{dimless}. Concretely, our sampling algorithm considered two different values for the radius,  $R = 0.05$ and $R=0.005$, and sampled the two-sphere with angle differences of $0.1$ and $0.3$ degrees, respectively. In the IR, this algorithm identified the following maximal and minimal values for $\tilde{\alpha}_0$: 
\be\label{boundsta}
\begin{split}
	R= 0.005: & \qquad 0.020 \leq \tilde{\alpha}_0 \leq 0.19 \, , \\
    R= 0.05: & \qquad 0.014 \leq \tilde{\alpha}_0 \leq 0.16 \, . 
\end{split}
\ee
The fact that different values for $R$ give slightly different limits results from the fact that the set of RG trajectories created by sampling the spheres at different radius is not identical: the initial conditions set at $R=0.05$ and $R=0.005$ are not positioned on the same RG trajectory and thus provide two different samplings of the UV-critical hypersurface. Our result then gives a good indication for the values of $\tilde{\alpha}_0$ which are compatible with an asymptotically safe UV completion by the non-chiral fixed point NGFP$^B$. In particular, they show that the couplings are not exponentially large, so that \eqref{boundsta} could be in agreement with the phenomenological requirements of having light chiral fermions. In a flat background where NGFP$^A$ may be exactly chiral, this cross-over may provide an elegant mechanism for starting from a non-chiral theory at high energy (NGFP$^B$) and restoring the symmetry (at least approximately) by spending a sufficient amount of RG time in the vicinity of NGFP$^A$ before flowing to the GFP.

%--------------------------------------------------------------------------------
\section{Summary and Conclusions}
\label{sect:summary}
%--------------------------------------------------------------------------------
Our work provides a detailed study of the asymptotic safety mechanism for gravity coupled to $N_f$ Dirac fermions within the background field formalism. Besides a fermion-kinetic term (and the associated fermion anomalous dimension), our study includes a non-minimal interaction coupling the fermion bilinears to the spacetime curvature. The extension beyond minimal coupling is interesting for two reasons. Firstly, the new term gives a contribution to the anomalous dimension of Newton's coupling which is proportional to the number of fermionic fields. Depending on the sign of the corresponding coupling, this leads to a novel screening/anti-screening contribution in the gravitational RG flow which works towards either stabilizing or destabilizing the asymptotic safety mechanism. Secondly, the non-minimal interaction explicitly breaks chiral symmetry, as it provides a fermion mass set by the spacetime curvature. Thus the analysis has a natural connection to interesting questions including the existence of light chiral fermions \cite{Eichhorn:2011pc} and the role of global symmetries within the asymptotic safety program \cite{Ali:2020znq}. While the effect of the non-minimal coupling has already been explored in \cite{Eichhorn:2016vvy}, our work (and its companion \cite{Daas:2020dyo}) is the first where the role of the explicit symmetry breaking terms is studied \emph{in a non-flat background}. Together with the flat-space results, our findings establish that the topology of spacetime has a crucial effect on symmetries supported by the matter sector (also see \cite{Gies:2018jnv,Hamada:2020mug} for related discussions).

The renormalization group flow of the system projected onto the subspace spanned by the ansatz \eqref{Imono} comes with an intricate fixed point structure. At the same time the ansatz is sufficiently sophisticated so that it admits several non-trivial subsystems. Thus it allows to test the robustness of the fixed point structure under the inclusion of further interaction monomials. In the absence of the non-minimal coupling, the systems exhibits a one-parameter family of interacting fixed points, called NGFP$^{A}$, existing for all values $N_f$. Their structure resembles the pure gravity fixed point. In particular, they come with two UV-relevant directions associated with Newton's coupling and the cosmological constant while the fermion anomalous dimension is negative and of order unity. Notably, this family admits a large-$N_f$-expansion showing that the theory becomes weakly coupled in the large $N_f$-limit.

Extending the projection subspace by the non-minimal coupling has two significant effects. Firstly, the ``chiral'' fixed points NGFP$^{A}$ persist but are shifted towards a non-zero value of the non-minimal coupling. The coupling behaves as a ``shifted'' Gaussian matter coupling, i.e., the cubic beta function \eqref{betaalpha} for the new coupling has a non-vanishing constant term. One source for this shift is the contribution of the background fermion fields, eq.\ \eqref{thetabackground}, which violates chiral symmetry.\footnote{This is reminiscent of the discussion of gravitational catalysis occuring in fermionic systems in hyperbolic spacetimes \cite{Ebert:2008pc}, where it is also the spectrum of the Dirac operator which is responsible for chiral symmetry breaking.} The new direction of the renormalization group flow is irrelevant, so that the symmetry breaking dies off towards low energy and does not provide an obstruction to the existence of light (standard-model like) chiral fermions. Secondly, the inclusion of the non-minimal coupling reveals the existence of a family of ``non-chiral'' fixed points, called NGFP$^{B}$, characterized by the new coupling taking values $\alpha_*^{B} \ge 1$. These come with three UV-relevant directions. The specific values for the stability coefficients and anomalous dimensions suggest that this class of fixed points also behaves ``almost gaussian'' in the sense that quantum corrections do not overrule the canonical power counting. Supplementing the full system with the fermion anomalous dimension induces a ``region of instability'' in the $N_f$-dependence of the fixed point structure. This region is situated at small values $N_f \sim 2$ and characterized by fixed points shifting into and emerging from the complex plane (see Fig.\ \ref{manyFPfamilies}). In addition, the anomalous dimension creates one more branch of fixed point solutions, NGFP$^{C}$. In this case the fermion anomalous dimension and eigenvalues deviate significantly from canonical power counting. 

These results fit well with the fixed point structure found on flat backgrounds. The properties of NGFP$^{A}$ are in qualitative agreement with the chiral fixed point investigated in \cite{Dona:2013qba,Dona:2014pla,Meibohm:2015twa} while non-chiral fixed points similar to the class NGFP$^{B}$ have been reported in \cite{Eichhorn:2016vvy}. In particular the critical exponents and fermion anomalous dimensions are in qualitative agreement when the values of $N_f$ allow for such a comparison. Our work then adds two important new insights to our understanding of the asymptotic safety mechanism of gravity coupled to fermionic matter: firstly, there is an intriguing interplay between the topology of the background and the possibility of realizing global symmetries at an interacting fixed point. Our analysis provides an explicit example where the choice of background structure induces a non-zero value for a coupling responsible for the explicit breaking of a global symmetry. Secondly, we were able to trace the gravity-matter fixed points to large values $N_f$. This feature is in agreement with the fixed point structure observed in fluctuation computations \cite{Meibohm:2016mkp}. Intriguingly, for large values of $N_f$, the fixed points become weakly coupled in the background Newton coupling. This opens the exciting possibility that this regime may actually be accessible via perturbative methods. 

Conceptually, it would be interesting to enhance the projection subspace of our work by promoting the fermionic wave-function renormalization to a (matrix-valued) momentum-dependent function $Z_k^\psi \rightarrow Z_k^\psi(\slashed{\nabla})$. When expanding $Z_k^\psi(x)$ in a power series at $x=0$, the lowest order terms correspond to a mass-term, the fermion kinetic term, and the invariant $\int d^4x\sqrt{g}\,\bar{\psi}\ds^2\psi$.\footnote{Based on a flat-background study, partial results on the fixed point structure associated with this coupling have been reported in \cite{Eichhorn:2016vvy}.} This class of interactions is singled out by the observation that, besides the structures included in \eqref{Imono}, they are the only invariants which may still contribute to the anomalous dimension of Newton's coupling at the background level. With the present choice of background fermions, it is not possible to disentangle this coupling from an $R\bar{\psi}\psi$ term though. While it is clear that the methods developed in this work can be used to analyze this coupling in an approximation excluding the $R\bpsi\psi$-term, we leave this investigation to future work.

\appendix
%---------------------------------------------------------------
\section{Fermions on curved backgrounds}
\label{App.A}
%---------------------------------------------------------------
Our evaluation of the Wetterich equation \eqref{FRGE} in the fermionic sector utilizes the spin-base formalism developed in \cite{Gies:2013noa,Gies:2015cka,Lippoldt:2015cea} and summarized in \cite{Lippoldt:2016ayw}. The most important formulas resulting from this approach together with our conventions for the spinors are summarized in this appendix.

%---------------------------------------------------------------
\subsection{Spinor conventions}
\label{App.A1}
%---------------------------------------------------------------
Let us start by considering a flat spacetime $\mathbb{R}^4$ with Euclidean metric $\delta_{ab}$, $a,b=1,2,3,4$. The spacetime admits a set of complex $4 \times 4$-matrices $\bar{\gamma}^a$, the  Dirac matrices, satisfying the Clifford algebra 
\be
\left\{\bar{\gamma}^a, \bar{\gamma}^b\right\} = 2 \, \delta^{ab} \, \unit \, , 
\ee
where $\unit$ is the $4\times4$-dimensional unit matrix. In addition there is the fifth $\gamma$-matrix , $\gamma^5 \equiv - \bar{\gamma}^1\bar{\gamma}^2\bar{\gamma}^3\bar{\gamma}^4$, satisfying $\bar{\gamma}^a\,\gamma^5 = -\gamma^5\,\bar{\gamma}^a$ and $(\gamma^5)^2 = \unit$. The $\gamma$-matrices can be used to build a convenient basis for Dirac space. Explicitly, the elements are
\begin{equation}
	\big\{\Gamma^n\big\}_{n = 1,...,16} =\big\{\unit,\bar{\gamma}^a,\frac{i}{2}\big[\bar{\gamma}^a, \bar{\gamma}^b\big], \gamma^5, i \, \bar{\gamma}^a\gamma^5\big\},
	\label{basis Dirac space}
\end{equation}
and satisfy 
\begin{equation}
	{\rm tr}(\Gamma^n\Gamma^m) = 4\,\delta^{nm}.
\end{equation}
 One can then project the structure $\bar{\psi}A\psi$, with $A$ an arbitrary matrix in Dirac space onto $\bar{\psi}\Gamma^a\psi$ using
\begin{equation}
    \bar{\psi}A\psi = \frac{1}{4} \sum_{n} \tr\big(A\Gamma^n)\,\bar{\psi}\Gamma^n\psi \, . 
    \label{projection identity}
\end{equation}
By making use of the explicit basis \eqref{basis Dirac space} basis one can derive the Fierz reordering formula
\be
\big(\bar{u}_1\Gamma^m u_2\big)\big(\bar{u}_3\Gamma^nu_4\big) = \,- \frac{1}{16}\sum_{o,p}{\rm tr}(\Gamma^m\Gamma^n\Gamma^o\Gamma^p)\big(\bar{u}_1\Gamma^ou_4\big)\big(\bar{u}_3\Gamma^p u_2\big),
\label{Fierz}
\ee
where the $u_i$ are arbitrary Dirac spinors. The minus sign is incorporated since we work with Grassmann-valued spinors. 

We will now discuss the generalization to Euclidean curved spacetimes with metric $g_{\mu\nu}$, admitting a suitable spin-structure. Here it is convenient to introduce vierbeins $e^a_{\,\mu}$ satisfying
\begin{equation}
    g_{\mu\nu} = e^a_{\,\mu}e^b_{\,\nu} \, \delta_{ab}.
\end{equation}
Notably, these vierbeins are not unique. They are fixed up to local SO$(4)$-transformations only. Requiring invariance under these transformations serves as an important guiding principle for the construction of various objects. The Dirac matrices $\gamma^\mu$ satisfying the generalized Clifford algebra 
\be\label{clifford-curved}
\big\{\gamma^\mu, \gamma^\nu\big\} = 2 \, g^{\mu\nu} \, \unit ,
\ee
can be constructed from the flat space gamma matrices $\bar{\gamma}^a$ as
\begin{equation}
    \gamma^{\mu} \equiv e^\mu_{\,a} \, \bar{\gamma}^a.
\end{equation}
Note that in general the Dirac matrices $\gamma^{\mu}$ will now depend on the spacetime coordinate.

In order to construct fermion bilinears and in particular the fermion kinetic term two additional objects are required, the spin metric $h$ and the spin connection $\Gamma_\mu$ providing the connection piece in the spin covariant derivative $\nabla_\mu$. The spin metric is used to construct the Dirac-adjoint
\begin{equation}
    \bar{\psi} \equiv \psi^\dagger h \, . 
\end{equation}
The spin metric has to satisfy $|\det (h)| = 1$ in order to ensure no new scale is introduced. In addition the reality of the spinor bilinear forces $h^\dagger = -h$ and $\gamma_\mu^\dagger = - h \gamma_\mu h^{-1}$.
Starting from some basic assumptions for both $h$ and $\nabla_\mu$ (e.g. linearity, product rule, metric compatibility and covariance) one can show that for an arbitrary spinor $\psi$ containing components of arbitrary spacetime rank we have
\begin{equation}
    \nabla_\mu\psi = D_\mu\psi + \Gamma_\mu\psi, \quad \nabla_\mu\bar{\psi} = D_\mu\bar{\psi} - \bar{\psi}\Gamma_\mu.
\end{equation}
Here $D_\mu$ is the Levi-Civita connection and $\Gamma_\mu$ is a spin connection piece. Throughout this work we assume a torsion free connection: $\tr(\Gamma_\mu) = 0$.  Through covariance of the vierbeins it can be shown that
\begin{equation}
	\nabla_\mu\gamma^\nu = 0 \, . 
\end{equation}
 From the properties of $\Gamma_\mu$ one can also derive the important commutator identity
\begin{equation}
    \big[\nabla_\mu,\nabla_\nu\big] = \frac{1}{8}R_{\mu\nu\alpha\beta}\big[\gamma^\alpha,\gamma^\beta\big].
    \label{commutator spin connection}
\end{equation}
This relation allows to derive the well-known Lichnerowicz formula 
\begin{equation}
    -\slashed{\nabla}^2\psi = \big(\Delta_\psi + \frac{R}{4} - \frac{1}{2}\,\gamma^\mu\gamma^\nu\,\big[D_\mu, D_\nu\big]\big)\psi,
    \label{Lichnerowicz formula}
\end{equation}
where $\Delta_\psi \coloneqq -\nabla_\mu\nabla^\mu$.
At this stage an action giving rise to the dispersion relation $(\Delta_\psi + m^2)\psi = 0$ can be constructed as
\begin{equation}
    S^{\rm ferm} = \int d^4x\sqrt{g}\,\bar{\psi}\big(i\slashed{\nabla} + m\gamma^5\big)\psi.
\end{equation}
The reality condition $h^\dagger = -h$ guarantees that the action is real.

We close this section by showing that, on a background given by a four-sphere, an eigenmode of the Dirac operator satisfying \eqref{thetabackground} must necessarily be the lowest eigenmode. Let the spinor $\psi$ satisfy the relation $\nabla_\mu\psi = ic\,\gamma_\mu\psi$, for some real number $c$. Then we have
\begin{equation}
    \big[\nabla_\mu, \nabla_\nu\big]\psi = -c^2\big[\gamma_\nu, \gamma_\mu\big]\psi.
\end{equation}
By rewriting the right-hand side via \eqref{commutator spin connection} and exploiting \eqref{S4curvature} one then concludes that $c$ must take the value
\begin{equation}\label{Only possible c value}
    c = \sqrt{\frac{R}{48}} \, . 
\end{equation}
This corresponds to the lowest eigenvalue of the Dirac operator \cite{Camporesi:1995fb}. Hence $\psi$ must the lowest momentum mode of the Dirac operator. Notably, this does not mean that the lowest momentum mode necessarily satisfies relation \eqref{thetabackground}. Working in an explicit basis however (see for instance \cite{Camporesi:1995fb}), it can be verified that it does. 
%---------------------------------------------------------------	
\subsection{Variations with respect to the metric field}
\label{App.B}
%---------------------------------------------------------------
The computation of the flow in the fermion sector requires the explicit form of the Hessian $\Gamma_k^{(2)}$ entering \eqref{FRGE}. For this one needs the expansion of the EAA in terms of the fluctuation fields at quadratic order. Only the fermion fields $\psi$ can give rise to the fermion-fluctuating fields $\chi$ whereas all other objects present in the EAA produce terms containing the fluctuating metric $h_{\mu\nu}$. The variations of the gamma matrices can be determined by equating the Clifford algebra \eqref{clifford-curved} order by order in the metric perturbation. For our purpose, the first two orders in this expansion are sufficient
\begin{equation}
    \delta\gamma^\mu = -\frac{1}{2}h^{\mu\nu}\gamma_\nu, \qquad \delta^2\gamma^\mu = \frac{3}{8}\gamma^\alpha h_{\beta\alpha}h^{\mu\beta}.
\end{equation}
Here indices are raised/lowered by the background metric $\bar{g}$. The matrix $\gamma^5$ can be constructed such that the variations with respect to the metric vanish. 
 For the variations of $\nabla_\mu$ we have \cite{Lippoldt:2016ayw}
\begin{equation}
    \begin{split}
        &\delta\nabla_\mu = \frac{1}{8}\big[\gamma^\alpha, \gamma^\beta\big]D_\beta h_{\mu\alpha}, \\
        &\delta^2\nabla_\mu = \frac{1}{8}\big[\gamma^\alpha, \gamma^\beta\big]\big(h^\lambda_{\,\alpha}D_\beta h_{\mu\lambda} + h^\lambda_{\, \beta} D_\lambda h_{\mu\alpha} + \frac{1}{2}h^\lambda_\alpha D_\mu h_{\beta\lambda}\big).
    \end{split}
\end{equation}
%---------------------------------------------------------------	
\subsection{Vertices and Propagators}
\label{App.B2}
%---------------------------------------------------------------
We close the discussion by giving the explicit form of the vertices entering into $\Gamma_k^{(2)}$. Given the linear split \eqref{Fermion linear split} we adopt
\begin{equation}
    \frac{\delta\psi^i(x)}{\delta\chi^j(y)} = \delta^{i}_{\,j}\,\delta(x-y), \qquad \frac{\delta\bar{\psi}_i(x)}{\delta\bar{\chi}_j(y)} = \delta^{\,j}_{i}\,\delta(x-y) \, . 
\end{equation}
All the other variations are taken to be zero, which corresponds to treating $\psi$ and $\bar{\psi}$ as independent fields. Introducing a multiplet $\Upsilon = \big\{\hat{h}_{\mu\nu}, h, \chi^a, \bar{\chi}_a\big\}$ the Hessian $\Gamma_k^{(2)}$ is defined as
\begin{equation}
    \Gamma_k^{(2)} = {\frac{\overrightarrow{\delta}}{\delta\Upsilon}}\,\Gamma_k\,\frac{\overleftarrow{\delta}}{\delta\Upsilon^T}.
\end{equation}
The variations from the right are taken with respect to the transpose in spinor space. Denoting the entries of this matrix in field space by $\Gamma_{f_if_j} \equiv \frac{\overrightarrow{\delta}}{\delta f_i}\,\Gamma_k\,\overleftarrow{\frac{\overleftarrow{\delta}}{\delta f_j^T}}$, and employing the identity $\big(\bar{\psi}A\psi\big)^T = - \psi^T A^T \psi^\dagger$, the off-diagonal entries are given by
\begin{equation}
    \begin{split}
        \Gamma_{\hat{h}\chi} / Z_k^\psi &= \frac{i}{4}\,\overleftarrow{D}^\mu\,\bar{\theta}\gamma^\nu + \bar{\alpha}_k\overleftarrow{D}^\mu\overleftarrow{D}^\nu \,\bar{\theta}\gamma^5\\
        \Gamma_{h\chi} / Z_k^\psi &= -\frac{3i}{16}\,\overleftarrow{D}^\mu\,\bar{\theta}\gamma_\mu + \frac{\bar{\alpha}_k}{4}\,(3\Delta + R)\,\bar{\theta}\gamma^5 - \frac{3}{2}\,c\,\bar{\theta}\\
        \Gamma_{\hat{h}\bar{\chi}} / Z_k^\psi &= \frac{i}{4}\,\overleftarrow{D}^\mu\,\gamma^\nu\theta - \bar{\alpha}_k\overleftarrow{D}^\mu\overleftarrow{D}^\nu\,\gamma^5\theta\\
        \Gamma_{h\bar{\chi}} / Z_k^\psi &= -\frac{3i}{16}\,\overleftarrow{D}^\mu\,\gamma_\mu\theta -\frac{\bar{\alpha}_k}{4}\,(3\Delta + R)\,\gamma^5\theta + \frac{3}{2}\,c\,\theta\\[20pt]
        \Gamma_{\chi\hat{h}} / Z_k^\psi &= -\frac{i}{4}\,\overleftarrow{D}^\mu\,\bar{\theta}\gamma^\nu - \bar{\alpha}_k\overleftarrow{D}^\mu\overleftarrow{D}^\nu\,\bar{\theta}\gamma^5\\
        \Gamma_{\chi h} / Z_k^\psi &= \frac{3i}{16}\,\bar{\theta}\gamma^\mu D_\mu -\frac{\bar{\alpha}_k}{4}\,\bar{\theta}\gamma^5\,(3\Delta + R) + \frac{3}{2}\,c\,\bar{\theta}\\
        \Gamma_{\bar{\chi}\hat{h}} / Z_k^\psi &= -\frac{i}{4}\,\gamma^\nu\theta D^\mu + \bar{\alpha}_k\,\gamma^5\theta D^\mu D^\nu \\
        \Gamma_{\bar{\chi} h} / Z_k^\psi &= \frac{3i}{16}\,\gamma^\mu \theta D_\mu + \frac{\bar{\alpha}_k}{4}\,\gamma^5\theta\, (3\Delta + R) -\frac{3}{2}\,c\,\theta.
    \end{split}
    \label{variations}
\end{equation}
Following \eqref{Only possible c value}, we have $c = \sqrt{\frac{R}{48}}$ and all objects are constructed from the background metric only. The left arrow over the derivatives arrives from partial integration and is there to signal the derivative acts on the gravitational degrees of freedom. The boson-boson variations, ignoring the contributions of the kinetic term, are given by
\begin{equation}\label{Ghhvert}
\begin{split}
    \Gamma_{\hat{h}\hat{h}} / Z_k^\psi &= \frac{5}{2}c\,\bar{\theta}\theta\, -\bar{\alpha}_k\,\bar{\theta}\gamma^5\theta\,\,(\frac{1}{2}\Delta + \frac{1}{3}R)\big(\mathbb{1}-P\big)^{\mu\nu}_{\alpha\beta}\\
    & \qquad \qquad -D_\sigma D^\rho\big(\mathbb{1}-P\big)^{\mu\nu}_{\lambda\rho}\big(\mathbb{1}-P\big)^{\lambda\sigma}_{\alpha\beta} + \mathcal{O}(D_\mu)\\
    \Gamma_{hh} / Z_k^\psi &= -\frac{3}{8}c\,\bar{\theta}\theta + \frac{3}{16}\bar{\alpha}_k\,\bar{\theta}\gamma^5\theta\,\Delta + \mathcal{O}(D_\mu) \, . 
   \end{split}
\end{equation}
Here $\mathbb{1}$ denotes the identity on the space of symmetric two-tensors and $P$ projects out the trace, meaning
\begin{equation}
    \big(\mathbb{1}-P\big)^{\mu\nu}_{\alpha\beta} = \frac{1}{2} (\delta^\mu_\alpha \delta^\nu_\beta + \delta^\nu_\alpha \delta^\mu_\beta  ) - \frac{1}{4} g_{\alpha\beta} g^{\mu\nu} \,.
\end{equation}
The explicit form of the $\mathcal{O}(D_\mu)$-terms is irrelevant since their contribution vanishes after they are traced over. Finally, the fermion-fermion variations result in
\begin{equation}\label{eq:Gppb}
\begin{split}
    \Gamma_{\bar{\psi}\psi} / Z_k^\psi &= i\ds + \bar{\alpha}_kR\gamma^5\\
    \Gamma_{\psi\bar{\psi}} / Z_k^\psi &= -\big(i\ds + \bar{\alpha}_kR\gamma^5\big)^T.
\end{split}
\end{equation}
These are all the variations that enter into the calculation for the beta function encoding the running of couplings in the fermionic sector. Combining \eqref{eq:Gppb} with the regulator \eqref{Rkpsi} and inverting the resulting expression gives the explicit form of the fermionic propagator
\begin{equation}\label{eq:fermprop}
	G_\psi^{-1} = i\ds\,\widetilde{W}_\chi(\ds^2) + \gamma^5\,W_\chi(\ds^2) \, .
\end{equation}
The functions $\widetilde{W}_\chi(\ds^2)$ and $W_\chi(\ds^2)$ are defined as
\begin{equation}
	\begin{split}
		&W_\chi(\ds^2) \equiv \frac{R_k^\psi + \bar{\alpha}R}{-\ds^2 + (R_k^\psi)^2 + 2\bar{\alpha}R\,R_k^\psi} \, , \qquad 
		\widetilde{W}_\chi(\ds^2) \equiv \frac{1}{-\ds^2 + (R_k^\psi)^2 + 2\bar{\alpha}R\,R_k^\psi} \, . 
	\end{split}
\end{equation}
Here the terms $-\ds^2$ are short-hand for the right-hand side of \eqref{eq:nabla2}. The graviton propagator resulting from the Einstein-Hilbert action in harmonic gauge is readily obtained from \cite{Reuter:1996cp,Reuter:2019byg} and reads
\be\label{gravprop}
\begin{split}
G_h^{-1} = & \, -128 \pi G_k \, (\Delta - 2 \Lambda_k)^{-1} \, , \\
G_{\hat{h}}^{-1} = & \,  32 \pi G_k \, (\Delta + \tfrac{2}{3} R - 2 \Lambda_k)^{-1}\,\big(\mathbb{1}-P\big)^{\mu\nu}_{\alpha\beta}\, . 
\end{split}
\ee

%---------------------------------------------------------------
\section{Functional traces involving fermionic background fields}
\label{App.C}
\begin{figure}[t!]
	\begin{center}
		\includegraphics[width = 0.3\textwidth]{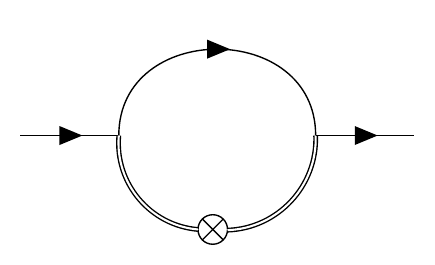} \; 
		\includegraphics[width = 0.32\textwidth]{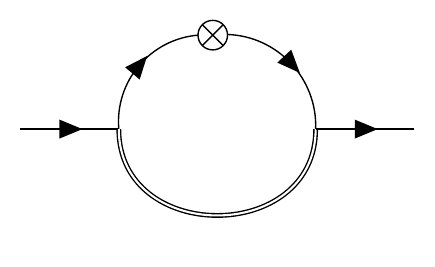} \; 
		\includegraphics[width = 0.3\textwidth]{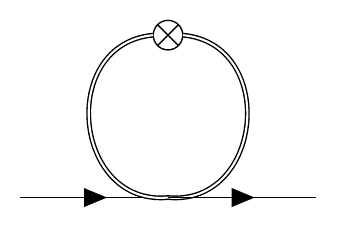}
	\end{center}
	\caption{\label{fig.feyn} Feynman diagrams encoding the contributions \eqref{diagrams prop and vert} to the RG flow. Solid and double internal lines correspond to the fermion and gravity propagators while the crossed circle marks the insertion of the corresponding regulator. The external straight lines denote background fermionic fields. The explicit expressions for the vertices are collected in section \ref{App.B}.}
\end{figure}

%---------------------------------------------------------------
In this appendix, we illustrate the techniques used for evaluating operator traces containing background spinor fields. In practise these computations amount to evaluating the Feynman diagrams shown in Fig.\ \ref{fig.feyn} for the Feynman rules obtained from \eqref{Gans} on a background four-sphere.
 For the trace-mode $h$ of the graviton fluctuations, the explicit expressions corresponding to these diagrams (from left to right) are
\begin{equation}
	\begin{split}
		D_3^{hh\chi} &= \Tr_0\big[G_h^{-1}\,\Gamma_{h\chi}\,G_\psi^{-1}\,\Gamma_{\bar{\chi}h}\,G_h^{-1}\partial_t\mathcal{R}_k^h\Big]\\
		D_3^{\chi\chi h} &= -\Tr_\psi\Big[G_\psi^{-1}\,\Gamma_{\bar{\chi}h}\,G_h^{-1}\,\Gamma_{h\chi}\,G_\psi^{-1}\partial_t\mathcal{R}_k^\psi\Big] \\
			D^h_{\rm Tad} &= -\frac{1}{2}\Tr_0\Big[G_h^{-1}\,\Gamma_{hh}\,G_h^{-1}\,\partial_t\mathcal{R}_k^h\Big] \, . 
		\label{diagrams prop and vert}
	\end{split}
\end{equation}
The explicit form of the vertices is given in eqs.\ \eqref{variations} and \eqref{Ghhvert} while the graviton and fermion propagators takes the form \eqref{gravprop} and \eqref{eq:fermprop}, respectively. In addition, there is an identical set of expressions where $h$ is replaced by the traceless fluctuations $\widehat{h}$ and Tr$_0$ substituted by the trace over traceless, symmetric matrices. 

%---------------------------------------------------------------
\subsection{Useful identities for background spinor computations}
\label{App.C2}
%---------------------------------------------------------------
Our goal is to project \eqref{diagrams prop and vert} onto the invariants $I_3$ and $I_4$ listed in eq.\ \eqref{Imono}. For the specific choice of background spinor \eqref{thetabackground}, this corresponds to extracting the contributions proportional to
\be\label{backgroundstructure}
\int d^4x \sqrt{\gb} \, \sqrt{\Rb} \,  \btheta \theta \, , \quad 
 \int d^4x \sqrt{\gb} \Rb \, \btheta \gamma_5 \theta \, . 
\ee
These structures carry the information about the fermion anomalous dimension and the scale-dependence of $\alpha_k$, respectively. The projection then entails that we can ignore all terms of order $\Rb^{3/2}$ and higher. Moreover, the trace arguments in \eqref{diagrams prop and vert} typically contain products of $\gamma$-matrices which need to be mapped to the corresponding basis elements $\Gamma^n$ appearing in the fermion bilinears. This mapping is carried out using the formula \eqref{projection identity}, for $\Gamma^n = \unit$ and $\Gamma^n = \gamma^5$, respectively. A corrollary of this projection is that all terms containing an odd number of gamma matrices between the background spinors will not contribute to the set \eqref{projection identity}.

In order to keep track of the order of the background curvature, all derivatives acting on the background spinors are eliminated. To accomplish this, we will list three identities. 
The first identity shows how to commute an arbitrary function of the squared Dirac operator through a gamma matrix and reads
\begin{equation}
    \ds^2\gamma^\alpha\psi_{\alpha ..} = \gamma^\alpha\ds^2\psi_{\alpha ..} + 2\gamma^\nu\big[D_\nu, D_\alpha\big]\psi^\alpha_{\,\,..}+\frac{R}{4}\gamma^\alpha\psi_{\alpha ..}.
    \label{Identity I}
\end{equation}
In addition to the index $\alpha$, the (generalized) spinor $\psi$ may carry an arbitrary set of spacetime indices (which is indicated by the dots). Secondly, we give an identity allowing us to commute a function of $\ds^2$ and a background spinor \eqref{thetabackground}
\begin{equation}
\begin{split}
    f(\ds^2)\,\theta\,T =& \,\theta\,f(-\Delta)\,T - 16c^2\,\theta\,f^\prime(-\Delta)\,T + \frac{1}{2}\gamma^\mu\gamma^\nu\theta\,f^\prime(-\Delta)\,\big[D_\mu,D_\nu\big]\,T\\
    &+2c^2\,\theta\,\Delta f^{\prime\prime}(-\Delta)\,T + 2ic\,\gamma^\mu\theta\,D_\mu\,f^\prime(-\Delta)\,T.
\end{split}
    \label{Identity II}
\end{equation}
Here $T$ is a tensor of arbitrary rank with the indices suppressed. This formula can be derived by repeatedly acting with $\ds^2$ onto the structure $\theta\,T$ and is valid up to terms of order $\Rb^{3/2}$. The final identity allows to pull out a spinor bilinear $\bar{\theta}\,A\,\theta$, with $A$ being an arbitrary (not necessarily constant) matrix in Dirac space out of the functional trace
\begin{equation}
    \Tr\Big[(\Delta\phi)\,g(\Delta)\Big] = 0,
    \label{Identity III}
\end{equation}
where $\phi$ is a scalar function. The reason this identity holds is because the Laplacian $\Delta$ projects out the zero mode of the spherical harmonics. Off-diagonal heat kernel methods (see \cite{Benedetti:2010nr}) then show that the functional trace vanishes.  We now have all ingredients to evaluate the traces \eqref{diagrams prop and vert}. 

For reference, we further note the explicit commutation relations appearing in explicit applications of \eqref{Identity I} and \eqref{Identity II}. On a background four-sphere and up to terms of order $\mathcal{O}(R^2)$ we have
\begin{equation}
	\big[D_\mu, W(\Delta)\big]\,\phi_{\alpha_1...\alpha_n} = W^\prime(\Delta)\big[D_\mu, \Delta\big]\,\phi_{\alpha_1...\alpha_n} \, . 
\end{equation}
For tensors of rank one and symmetric traceless tensors $\hat{\phi}^{\mu\nu}$ the commutators evaluate to the explicit expressions
\begin{equation}
	\begin{split}
		&\big[D_\mu,\Delta\big]\,\phi^\mu = -\frac{1}{4}\,R\,D_\mu\,\phi^\mu \, , \qquad 
		\big[D_\mu,\Delta\big]\,\hat{\phi}^{\mu\nu} = -\frac{5}{12}\,R\,D_\mu\,\hat{\phi}^{\mu\nu} \,.
	\end{split}
	\label{commutation relations on a sphere}
\end{equation}

%--------------------------------------------------------------
\subsection{Results for selected traces}
%--------------------------------------------------------------
We exemplify our computational strategy by evaluating $D_3^{hh\chi}$ explicitly, before giving some concluding remarks on the evaluation of the other trace expressions. In order to ease the notation, we focus on extracting the contributions proportional to $\Rb \, \btheta \gamma_5 \theta$ only. We also set $Z_k = 1$ for simplicity. The terms proportional to $\sqrt{\Rb}$ are obtained along the same lines. Inserting the explicit expressions for the propagators and vertices one has
\begin{equation}
	\begin{split}
		D_3^{hh\chi} = \Tr_0\Big[&W_h(\Delta)\,\big(\frac{3}{16}\,i\,D_\mu\bar{\theta}\gamma^\mu + \frac{\bar{\alpha}}{4}\,(3\Delta + R)\bar{\theta}\gamma^5-\frac{3}{2}ic\,\bar{\theta}\big)\\
		&\frac{i\ds + \gamma^5R_k^{\psi} + \gamma^5\,\bar{\alpha}R}{-\ds^2 + (R_k^\psi)^2 + 2\bar{\alpha}R\,R_k^\psi}
		\big(\frac{3}{16}\,i\,\gamma^\nu\theta\,D_\nu + \gamma^5\,\frac{\bar{\alpha}}{4}\theta\,(3\Delta + R) - \frac{3}{2}ic\,\theta\big)\Big],
		\label{Diagram hhc}
	\end{split}
\end{equation}
where the function $W_h(\Delta)$ is given by
\begin{equation}
	W_h(\Delta) \coloneqq G_h^{-2}(\Delta)\,\partial_t\mathcal{R}_k^h(\Delta).
\end{equation}
The strategy for evaluating the right-hand side of \eqref{Diagram hhc} is then the following
\begin{enumerate}
    \item Factor out the vertex terms in order to obtain traces with a specific combination of $\gamma$-matrices sandwiched between the background fermions.
    \item For each of these combinations, use the representation \eqref{eq:fermprop} to identify the part of the fermion propagator which will give the non-vanishing contribution to the trace (recalling that traces with odd numbers of $\gamma$-functions will vanish under the projection and keeping terms up to order $\Rb$).  
    \item Commute the gamma matrix with the function of $\ds^2$ with identity \eqref{Identity I}.
    \item Use identity \eqref{Identity II} to deal with the derivatives acting on the background spinor.
    \item Take the spinorial scalar out of the functional trace (allowed by identity \eqref{Identity III}) and project onto correct spinor structure using \eqref{projection identity}. 
    \item Compute all the commutators and contract the indices.
    \item Add all the terms together.
\end{enumerate}
 This algorithm results in the following set of scalar traces
\begin{equation}
\begin{split}
    D_3^{hh\chi}\Big|_{\bar{\theta}\gamma^5\theta} =& -\frac{9}{256}\Tr_0\big[W_h(\Delta)\,\Delta\,W_\chi(-\Delta)\big]
    %\\     &
    +\frac{3}{64}\Tr_0\big[W_h(\Delta)\,R\,W_\chi(-\Delta)\big]
    \\     &
    +\frac{3}{1024}\Tr_0\big[W_h(\Delta)\,\Delta\,R\,W^\prime_\chi(-\Delta)\big]
    -\frac{3}{2048}\Tr_0\big[W_h(\Delta)\,\Delta^2\,R\,W^{\prime\prime}_\chi(-\Delta)\big]\\
    &+\frac{9}{32}\bar{\alpha}\,\Tr_0\big[W_h(\Delta)\,\Delta^2\,\widetilde{W}_\chi(-\Delta)\big]
    +\frac{3}{32}\bar{\alpha}\,\Tr_0\big[W_h(\Delta)\,\Delta\,R\,\widetilde{W}_\chi(-\Delta)\big]\\
    &-\frac{9}{128}\bar{\alpha}\,\Tr_0\big[W_h(\Delta)\,\Delta^2\,R\,\widetilde{W}^\prime_\chi(-\Delta)\big]
    +\frac{3}{256}\bar{\alpha}\,\Tr_0\big[W_h(\Delta)\,\Delta^3\,R\,\widetilde{W}^{\prime\prime}_\chi(-\Delta)\big]\\
    &+\frac{9}{16}\bar{\alpha}^2\,\Tr_0\big[W_h(\Delta)\,\Delta^2\,W_\chi(-\Delta)\big]
    +\frac{3}{8}\bar{\alpha}^2\,\Tr_0\big[W_h(\Delta)\,\Delta\,R\,W_\chi(-\Delta)\big]\\
    &-\frac{3}{16}\bar{\alpha}^2\,\Tr_0\big[W_h(\Delta)\,\Delta^2\,R\,W^\prime_\chi(-\Delta)\big]
    +\frac{3}{128}\bar{\alpha}^2\,\Tr_0\big[W_h(\Delta)\,\Delta^3\,R\,W^{\prime\prime}_\chi(-\Delta)\big] \, ,
    \label{trace rep scalar bosonic loop diagram}
\end{split}
\end{equation}
where the subscript on the right-hand side indicates that the diagram has been projected onto the fermion-bilinear ${\bar{\theta}\gamma^5\theta}$. At this stage the argument of all scalar traces has been reduced to functions of the Laplacian $\Delta$. These expressions are then readily evaluated using the early-time expansion of the heat-kernel \cite{Reuter:1996cp,Codello:2008vh,Reuter:2019byg}, in $d=4$ dimensions and up to terms quadratic in the curvature 
\be\label{eq:heat}
{\rm Tr}_s[W(\Delta)] = \frac{1}{(4 \pi)^2} \, {\rm tr}_s[\unit] \, \int d^4x \sqrt{g} \, \left( Q_2[W] + \frac{1}{6} R \, Q_1[W] \right) + \cO(R^2) \, , 
\ee
with
\be\label{eq:Qfcts}
Q_n[W] \equiv \frac{1}{\Gamma(n)} \int_0^\infty dz z^{n-1} W(z) \, , \qquad n > 0 \, . 
\ee
Applying these formulas to \eqref{trace rep scalar bosonic loop diagram} gives the final expression for $ D_3^{hh\chi}\Big|_{R\,\bar{\theta}\gamma^5\theta}$
\be\label{D3hhx-final}
\begin{split}
 D_3^{hh\chi}\Big|_{R\,\bar{\theta}\gamma^5\theta} = &\frac{G\,k}{(1-2\lambda)^2\pi}\Bigg\{-\frac{29}{256} + (-\frac{89}{1280}+\frac{33}{1024}\pi)\eta_N + (-\frac{9}{16}+\frac{9}{64}\pi)\beta + (\frac{9}{128}-\frac{9}{512}\pi)\beta\,\eta_N \\
 &+\alpha\,\Big[\frac{3}{16}-\frac{27}{128}\pi + (-\frac{13}{128} + \frac{15}{256}\pi)\eta_N -\frac{3}{8}\beta + (-\frac{117}{80} + \frac{63}{128}\pi)\beta \eta_N\Big]\\
 &+\alpha^2\Big[\frac{371}{80}-\frac{21}{16}\pi + (-\frac{5}{14}+\frac{3}{32}\pi)\eta_N + (\frac{27}{4}-\frac{27}{16}\pi)\beta + (\frac{183}{40}+\frac{45}{32}\pi)\beta\,\eta_N\Big]\\
 &+\alpha^3\Big[-\frac{51}{20}+(\frac{237}{56}-\frac{81}{64}\pi)\eta_N\Big]\Bigg\}.
\end{split}
\ee

We now discuss the contribution of $D_3^{\chi\chi h}$. A priori, $D_3^{\chi\chi h}$ constitutes a trace in spinor-space,
\be\label{D3cch}
 D_3^{\chi\chi h} = -\sum_n \langle\bar{\psi}_n|\Gamma_{\bar{\chi}h}\,G_h^{-1}\Gamma_{h\chi}G_\psi^{-1}\partial_t\mathcal{R}_k^\psi\,G_\psi^{-1}|\psi_n\rangle \, , 
\ee
where the sum is over a complete basis of spinors on the four-sphere. One can then exploit that the vertices $\Gamma_{h\chi}$ act as ``intertwiners'' converting from a spinorial to a scalar expression. This allows to rewrite \eqref{D3cch} as a scalar trace by inserting a complete basis of scalar functions satisfying $\sum_m |\phi_m \rangle \langle \phi_m| = \unit$
\be\label{D3cch1}
\begin{split}
D_3^{\chi\chi h}	&=  -\sum_{n,m}\langle\bar{\psi}_n|\Gamma_{\bar{\chi}h}\,G_h^{-1}\,|\phi_m\rangle\langle\phi_m|\,\Gamma_{h\chi}G_\psi^{-1}\partial_t\mathcal{R}_k^\psi\,G_\psi^{-1}|\psi_n\rangle\\
&=-\sum_{n,m}\langle\phi_m|\,\Gamma_{h\chi}G_\psi^{-1}\partial_t\mathcal{R}_k^\psi\,G_\psi^{-1}|\psi_n\rangle\langle\bar{\psi}_n|\Gamma_{\bar{\chi}h}\,G_h^{-1}\,|\phi_m\rangle \\
& =  -\sum_{m}\langle\phi_m|\,G_h^{-1}\,\Gamma_{h\chi}G_\psi^{-1}\partial_t\mathcal{R}_k^\psi\,G_\psi^{-1}\Gamma_{\bar{\chi}h}\,|\phi_m\rangle \, .
\end{split}
\ee
After this reordering the structure of $D_3^{\chi\chi h}$ becomes identical to the one found in $D_3^{hh\chi}$. The evaluation of $D_3^{\chi\chi h}$ then proceeds along the same lines as the one of $D_3^{hh\chi}$, replacing 
\begin{equation}\label{substitution rules}
	\begin{array}{ll}
	W_h(\Delta) \mapsto G_h^{-1}(\Delta) \, ,\qquad & \qquad G_\psi^{-1}(\ds) \mapsto G_\psi^{-2}(\ds)\partial_t\mathcal{R}_k^\psi(\ds) \, , \\[1.5ex]
	W_\chi(\ds^2) \mapsto V_\chi(\ds^2) \, , \qquad & \qquad \widetilde{W}_\chi(\ds^2) \mapsto \widetilde{V}_\chi(\ds^2) \, , 
	\end{array}
\end{equation}
where
\begin{equation}
	\begin{split}
		V_\chi(\ds^2) \equiv \frac{\ds^2 + \big(R_k^\psi\big)^2 + 2\bar{\alpha}R\,R_k^\psi}{\big[-\ds^2 + \big(R_k^\psi\big)^2 + 2\bar{\alpha}R\,R_k^\psi\big]^2}\partial_tR_k^\psi,\\
		\widetilde{V}_\chi(\ds^2) \equiv \frac{2(\bar{\alpha}R + R_k^\psi)}{\big[-\ds^2 + \big(R_k^\psi\big)^2 + 2\bar{\alpha}R\,R_k^\psi\big]^2}\partial_tR_k^\psi.
	\end{split}
\end{equation}
Making these substitutions in \eqref{trace rep scalar bosonic loop diagram} and taking into account the relative minus sign then yields the contribution of the fermion loop diagram
\be\label{Dcch-final}
\begin{split}
D_3^{\chi\chi h}\Big|_{R\,\bar{\theta}\gamma^5\theta} = & \frac{G\,k}{(1-2\lambda)\pi}\Bigg\{-\frac{39}{256}+(-\frac{27}{32}+\frac{27}{128}\pi)\beta\\
&+\alpha\,\Big[\frac{21}{16}-\frac{27}{64}\pi -\frac{3}{4}\beta\Big]
+\alpha^2\Big[\frac{89}{8}-\frac{105}{32}\pi+(\frac{135}{8}-\frac{135}{32}\pi)\beta\Big]+\alpha^3\frac{27}{20}\Bigg\}.
\end{split}
\ee

The evaluation of the Feynman diagrams including the traceless part of the graviton fluctuations proceeds along the same lines. In order to simplify the computation we perform a transverse-traceless decomposition \cite{York:1973ia,Lauscher:2002sq,Benedetti:2010nr,Groh:2011dw} of $\widehat{h}_{\mu\nu}$ 
on the spherically symmetric background, setting 
\begin{equation}
    \widehat{h}_{\mu\nu} = h^{TT}_{\mu\nu} + D_\mu\xi_\nu + D_\nu\xi_\mu + \big(D_\mu D_\nu + \frac{1}{4}g_{\mu\nu}\,\Delta\big)\sigma \, , 
\end{equation}
followed by the field redefinition
\be\label{TTredef}
 \xi_\mu \mapsto \frac{1}{\sqrt{2}}\big[\Delta - \frac{1}{4}R\big]^{-1/2}\,\xi_\mu, \quad \sigma \mapsto \big[\frac{3}{4}\Delta^2-\frac{1}{4}R\Delta\big]^{-1/2}\,\sigma \, . 
\ee
The component fields, given by a transverse-traceless symmetric tensor $h^{TT}_{\mu\nu}$, a transverse vector $\xi_\mu$, and a scalar $\sigma$, satisfy the differential constraints
\begin{equation}
    g^{\mu\nu}h^{TT}_{\mu\nu} = 0, \quad D^\mu h^{TT}_{\mu\nu} = 0, \quad D^\mu\xi_\mu = 0.
\end{equation}
The redefinition \eqref{TTredef} ensures that the decomposition does not give rise to non-trivial (operator-dependent) Jacobians. 

Following the algorithm described above allows to reduce the operator traces appearing in $D_3^{\hat{h}\hat{h}\chi}$ to traces whose arguments depend on the Laplacian $\Delta$ only. The expression analogous to \eqref{trace rep scalar bosonic loop diagram} reads 
\begin{equation}\label{trace rep hhat}
\begin{split}
   & D_3^{\hat{h}\hat{h}\chi}\Big|_{\bar{\theta}\gamma^5\theta} = \\ & \; \; -\frac{1}{32}\,\Tr_{TV}\big[W_{\hat{h}}\,\Delta\,W_\chi(-\Delta)\big]
    %\\     &
    +\frac{1}{128}\,\Tr_{TV}\big[W_{\hat{h}}\,R\,W_\chi(-\Delta)\big]
    %\\     &
    +\frac{1}{96}\,\Tr_{TV}\big[W_{\hat{h}}\,\Delta\,R\,W^\prime_\chi(-\Delta)\big]
    \\ & \; \; -\frac{1}{768}\,\Tr_{TV}\big[W_{\hat{h}}\,\Delta^2\,R\,W^{\prime\prime}_\chi(-\Delta)\big]
    % \\     &
    +\frac{5}{384}\,\Tr_{TV}\big[W^\prime_{\hat{h}}\,\Delta\,R\,W_\chi(-\Delta)\big]
    %\\    &
    -\frac{3}{64}\,\Tr_{0}\big[W_{\hat{h}}\,\Delta\,W_\chi(-\Delta)\big]
    \\ & \; \;
    +\frac{1}{64}\,\Tr_{0}\big[W_{\hat{h}}\,R\,W_\chi(-\Delta)\big]
    %\\     &
    +\frac{1}{256}\,\Tr_{0}\big[W_{\hat{h}}\,\Delta\,R\,W^\prime_\chi(-\Delta)\big]
    %\\&
    -\frac{1}{512}\,\Tr_{0}\big[W_{\hat{h}}\,\Delta^2\,R\,W^{\prime\prime}_\chi(-\Delta)\big]
    \\ & \; \;
    +\frac{1}{32}\,\Tr_{0}\big[W^\prime_{\hat{h}}\,\Delta\,R\,W_\chi(-\Delta)\big]
    +\frac{3}{8}\bar{\alpha}\,\Tr_{0}\big[W_{\hat{h}}\,\Delta^2\,\widetilde{W}_\chi(-\Delta)\big]
    -\frac{1}{8}\bar{\alpha}\,\Tr_{0}\big[W_{\hat{h}}\,\Delta\,R\,\widetilde{W}_\chi(-\Delta)\big]
    \\ & \; \;
    -\frac{3}{32}\bar{\alpha}\,\Tr_{0}\big[W_{\hat{h}}\,\Delta^2\,R\widetilde{W}^\prime_\chi(-\Delta)\big]
    +\frac{1}{64}\bar{\alpha}\,\Tr_{0}\big[W_{\hat{h}}\,\Delta^3\,R\,\widetilde{W}^{\prime\prime}_\chi(-\Delta)\big]
    -\frac{1}{4}\bar{\alpha}\,\,\Tr_{0}\big[W^\prime_{\hat{h}}\,\Delta^2\,R\,\widetilde{W}_\chi(-\Delta)\big]
    \\ & \; \;
    +\frac{3}{4}\bar{\alpha}^2\,\Tr_{0}\big[W_{\hat{h}}\,\Delta^2\,W_\chi(-\Delta)\big]
    -\frac{1}{4}\bar{\alpha}^2\,\Tr_{0}\big[W_{\hat{h}}\,\Delta\,R\,W_\chi(-\Delta)\big]
    -\frac{1}{4}\bar{\alpha}^2\,\Tr_{0}\big[W_{\hat{h}}\,\Delta^2\,R\,W^{\prime}_\chi(-\Delta)\big]
    \\ & \; \;
    +\frac{1}{32}\bar{\alpha}^2\,\Tr_{0}\big[W_{\hat{h}}\,\Delta^3\,R\,W^{\prime\prime}_\chi(-\Delta)\big]
    -\frac{1}{2}\bar{\alpha}^2\,\Tr_{0}\big[W^\prime_{\hat{h}}\,\Delta^2\,R\,W_\chi(-\Delta)\big].
\end{split}
\end{equation}
Here the subscripts $TV$ and $0$ indicate that the corresponding traces are over the space of transverse vectors and scalars, respectively. Applying \eqref{eq:heat}, one then finds
\be\label{D3ththc}
\begin{split}
 D_3^{\hat{h}\hat{h}\chi}\Big|_{R\,\bar{\theta}\gamma^5\theta} = &\frac{G\,k}{(1-2\lambda)^2\pi}\Bigg\{\frac{13}{256}+(\frac{129}{1280}-\frac{33}{1024}\pi)\eta_N + \frac{1}{1-2\lambda}\big[\frac{7}{20}-\frac{3}{32}\pi + (-\frac{179}{1120}+\frac{3}{64}\pi)\eta_N\big]\\
 & \qquad \qquad \quad + (\frac{9}{16}-\frac{9}{64}\pi)\beta + (-\frac{9}{128}+\frac{9}{256}\pi)\beta\,\eta_N\Big]\\
 & + \alpha\,\Big[-\frac{37}{48}+\frac{55}{128}\pi+(\frac{1379}{1920}-\frac{67}{256}\pi)\eta_N + \frac{1}{1-2\lambda}\big[-\frac{7}{30}-\frac{1}{8}\pi + (-\frac{11}{420}+\frac{1}{32}\pi)\eta_N)\big]\\
 &\qquad \qquad \quad+\frac{1}{8}\beta + (\frac{39}{80}-\frac{21}{128}\pi)\beta\,\eta_N\Big]\\
 &+\alpha^2\Big[-\frac{707}{240}+\frac{13}{16}\pi + (\frac{201}{280}-\frac{7}{32}\pi)\eta_N + \frac{1}{1-2\lambda}\big[-\frac{17}{105}+(\frac{143}{630}-\frac{1}{16}\pi)\eta_N\big]\\
 &\qquad \qquad \quad+(-\frac{9}{4}+\frac{9}{16}\pi)\beta + (\frac{61}{40}-\frac{15}{32}\pi)\beta\,\eta_N\Big]\\
 &+\alpha^3\Big[\frac{17}{20} + (-\frac{79}{56}+\frac{27}{64}\pi)\eta_N\Big]\Bigg\}.
\end{split}
\ee
Finally, one can compute $D_3^{\chi\chi\hat{h}}\Big|_{\bar{\theta}\gamma^5\theta}$ by plugging the substitutions \eqref{substitution rules} into \eqref{trace rep hhat}, giving
\be\label{D3ccth}
\begin{split}
D_3^{\chi\chi\hat{h}}\Big|_{R\,\bar{\theta}\gamma^5\theta} =\frac{G k}{(1-2\lambda)\pi}&\Bigg\{\frac{15}{256}+\frac{1}{1-2\lambda}\big[\frac{7}{16}-\frac{15}{128}\pi\big] + (\frac{28}{32}-\frac{27}{128}\pi)\beta\\
&+\alpha\,\Big[-\frac{119}{48}+\frac{55}{64}\pi + \frac{1}{1-2\lambda}\big[\frac{13}{20}-\frac{3}{16}\pi\big] + \frac{1}{4}\beta\Big]\\
&+\alpha^2\Big[-\frac{173}{24}+\frac{65}{32}\pi -\frac{17}{60} \frac{1}{1-2\lambda}+(-\frac{45}{8}+\frac{45}{32})\beta\Big] 
%\\ &
-\frac{9}{20}\alpha^3\Big]\Bigg\}.
\end{split}
\ee
For completeness we also give the expressions for the tadpole diagrams shown in Fig. \ \ref{fig.feyn}, which are evaluated rather straightforwardly, and read
\begin{equation}\label{Tadpole}
    \begin{split}
        &D_{\rm Tad}^h\Big|_{R\,\bar{\theta}\gamma^5\theta} = \frac{G k}{(1-2\lambda)^2\pi}\,\bigg\{\alpha\,\Big[\frac{1}{8}-\frac{1}{48}\eta_N\Big]\bigg\}, \\
        &D_{\rm Tad}^{\hat{h}}\Big|_{R\,\bar{\theta}\gamma^5\theta} = \frac{G k}{(1-2\lambda)^2\pi}\,\bigg\{\alpha\,\Big[\frac{27}{8}-\frac{9}{16}\eta_N+\frac{1}{1-2\lambda}\big[-2+\frac{1}{4}\eta_N\big]\Big]\bigg\}.
    \end{split}
\end{equation}
Together, the contributions \eqref{D3cch1}, \eqref{Dcch-final}, \eqref{D3ththc}, \eqref{D3ccth}, and \eqref{Tadpole} give rise to the beta function \eqref{betaalpha}. This closes our discussion on evaluating the functional renormalization group equation \eqref{FRGE} including fermions in a curved background.

%---------------------------------------------------------------
\acknowledgments{This article has been prepared for the Special Issue ``Asymptotic Safety in Quantum Gravity'' (Universe) edited by A.\ D.\ Pereira. We thank A.\ D.\ Pereira and M.\ Reuter for interesting discussions. J.\ W.\ acknowledges the China Scholarship Council (CSC) for financial support. 
}

\bibliographystyle{JHEP}
\bibliography{general_bib}

\end{document}